\documentclass[aps,prd,twocolumn,a4paper,floatfix,nofootbib,superscriptaddress,nofootinbib]{revtex4}
\usepackage{amsmath,amsfonts,amssymb}
\usepackage{graphicx}
\usepackage{units}
\usepackage{natbib}

\newcommand{\rth}{\mathrm{th}}
\newcommand{\re}{\mathrm{e}}
\newcommand{\rn}{\mathrm{n}}
\newcommand{\req}{\mathrm{eq}}
\newcommand{\rx}{\mathrm{x}}
\DeclareMathOperator{\sgn}{sgn}

\begin{document}

\title{Simulating bulk viscosity in neutron stars. II. Evolution in spherical symmetry}

\author{Giovanni \surname{Camelio}}
\affiliation{Nicolaus Copernicus Astronomical Center, Polish Academy of Science, Bartycka 18, 00-716 Warsaw, Poland}
\author{Lorenzo \surname{Gavassino}}
\affiliation{Nicolaus Copernicus Astronomical Center, Polish Academy of Science, Bartycka 18, 00-716 Warsaw, Poland}
\affiliation{Department of Mathematics, Vanderbilt University, Nashville, TN, USA}
\author{Marco \surname{Antonelli}}
\affiliation{CNRS/IN2P3, ENSICAEN, Laboratoire de Physique Corpusculaire, 14000 Caen, France}
\affiliation{Nicolaus Copernicus Astronomical Center, Polish Academy of Science, Bartycka 18, 00-716 Warsaw, Poland}
\author{Sebastiano \surname{Bernuzzi}}
\affiliation{Theoretisch-Physikalisches Institut, Friedrich-Schiller-Universit{\"a}t Jena, 07743, Jena, Germany}
\author{Brynmor \surname{Haskell}}
\affiliation{Nicolaus Copernicus Astronomical Center, Polish Academy of Science, Bartycka 18, 00-716 Warsaw, Poland}
\date{\today}

\begin{abstract}
Out-of-equilibrium reactions between different particle species are the main
processes contributing to bulk viscosity in neutron stars.
In this work, we numerically compare three different approaches to the modeling
of bulk viscosity: the multi-component fluid with reacting particle species and
two bulk stress formalism based on the M\"uller-Israel-Stewart theory, namely the
Hiscock-Lindblom and the Maxwell-Cattaneo models, whose flux-conservative
formulation in radial gauge-polar slicing coordinates and spherical symmetry is
derived in a companion paper.  To our knowledge, this is the first time that a
neutron star is simulated with the complete Hiscock-Lindblom model of bulk
viscosity.
We find that the Hiscock-Lindblom and Maxwell-Cattaneo models are good
approximations of the multi-component fluid for small perturbations and when the non-equilibrium equation of state of the fluid
depends on only one independent particle fraction. 
For more than one independent particle fraction and for large perturbations, the bulk stress approximation 
is still valid but less accurate.  In addition, we include the energy loss due to the luminosity
of the reactions in the bulk stress formulation. We find that the energy loss
due to bulk viscosity has a larger effect on the dynamics than the bulk stress
or the variation in particle composition \textit{per~se}.
The new one-dimensional, general-relativistic hydrodynamic code developed for
this work, \texttt{hydro-bulk-1D}, is publicly available.
\end{abstract}

\maketitle

\section{Introduction}
\label{sec:intro}

Recent advances in fields as diverse as gravitational wave astronomy and heavy
ion collisions have lead to a renewed interest in relativistic formulations of
dissipative hydrodynamics, as experiments and observations become sensitive
enough to probe these effects \cite{Romatchke2017,Florkowski2018,lhc2018}.

In neutron star simulations the effects of shear viscosity, which is expected
to play a strong role as magneto-hydrodynamic turbulence develops after a
neutron star merger, have been considered by few authors \citep{Kiuchi18,
Radice17, Shibata17, RadiceEjecta, Vigano20, Radice20, Shibata21}, while bulk
viscosity has received less attention. Nevertheless, recent studies suggest
that the thermodynamic conditions in a neutron star merger may be such that
bulk viscosity could be dynamically significant \citep{Haensel_Bulk_Urca,
Yakovlev_review_2001, Schmitt_review_2018, AlfordRezzolla}, even if no clear
evidence of this effect has been found in simulations \citep{Perego19, Most21,
Hammond2021, Radice22}, see for example the introductory discussion in our
companion paper \citep{Camelio22a}.
 
Bulk viscosity is an out-of-equilibrium dissipative process, which in
neutron stars is mostly due to particle reactions, and in particular to
the so-called Urca processes \citep{Haensel_Bulk_Urca}.  Bulk viscosity in neutron stars may impact on several aspects of their
dynamics.  For example, it determines the damping time-scale of radial
oscillations which have been excited at the time of the proto-neutron star
formation from supernovae. Viscosity also defines the window for the
gravitational wave instabilities in rotating neutron stars \citep{cutler1987,
cutler1990ApJ, andersson_jones_2000} and thereby the theoretical upper limits
on their rotation rates \citep{lindblom_1986ApJ}.

Many calculations of the bulk-viscosity coefficients arising from different
kinds of reactions have been reported in literature, e.g. 
\citep{Haensel_Bulk_Urca,haensel_bulkI,haensel_bulkII,Gusakov_kantor_2008,Haensel_Bulk_hyperons,AlfordBulk,alford_bulk_trapped}. 
Recently, there has been a renewed interest in this problem also because bulk viscosity
can in principle leave an observable imprint on the gravitational wave signal emitted during
neutron star mergers, even if preliminary studies have found that any effect
is probably minor \citep{Perego19, Most21, Hammond2021, Radice22, Zappa23}.

In this work, we numerically compare three different approaches to the modeling
of bulk viscosity in neutron stars: the multi-component fluid with tracking of
the reacting particle species \cite{Sawyer1989}, and two bulk stress formalism based on the
M\"uller-Israel-Stewart theory \cite{Israel_Stewart_1979}, namely the Hiscock-Lindblom \cite{Hiscock1983} and the Maxwell-Cattaneo \cite{Zakari93}
models, whose flux-conservative formulation in radial gauge-polar slicing
coordinates and spherical symmetry is discussed in the companion paper
\cite{Camelio22a}.
To our knowledge, this is the first implementation of M\"uller-Israel-Stewart
theories in radial gauge-polar slicing coordinates in spherical symmetry and the first implementation
of the complete Hiscock-Lindblom model in the context of neutron stars.
We find that the bulk stress formalism (i.e., Hiscock-Lindblom and
Maxwell-Cattaneo) is a good approximation of the multi-component fluid for small
perturbations and for one independent particle fraction (namely, when we can select the electron fraction as the only independent fraction variable of the equation of state). For more than one independent fraction variable in the equation of state, the bulk stress
approximation is still valid but loses accuracy\footnote{
When the viscous degrees of freedom are doubled, the relaxation-time approximation may no longer be valid \cite{NoronhaOrigin2011}. Sometimes, this can lead to exotic dynamics, which cannot be described within the M\"uller-Israel-Stewart formalism \cite{Heller2014}.}.  
In addition, we include for the first time the energy loss due to the reaction luminosity in the bulk stress
formulations \cite{Camelio22a}, finding that its inclusion influences the dynamics
more than the bulk stress alone.

In order to perform this study, we developed `\texttt{hydro-bulk-1D}', a new one
dimensional general relativistic hydrodynamic code. We publicly release \citep{Camelio22code}
\texttt{hydro-bulk-1D} on \texttt{zenodo} under the MIT license to allow the community
to check and improve our results.

This paper is organized as follows. In Sec.~\ref{sec:model} we describe the
bulk viscous models and the hydrodynamic equations.
In Sec.~\ref{sec:micro} we describe the equation
of state and the reaction rates.  In Sec.~\ref{sec:code} we introduce the
code. In Sec.~\ref{sec:results} we compare the
different approaches to bulk viscosity.  We draw our conclusions in
Sec.~\ref{sec:conclusions}. We describe in detail the numerical implementation
of the code in Appendix~\ref{sec:implementation}, we perform a set of
standard tests in Appendix~\ref{sec:tests}, and we derive a formula for the
relativistic linearized damping time of an oscillation in Appendix~\ref{sec:tdamp}.

We adopt the $(-,+,+,+)$ metric signature and, unless otherwise specified,
we set $c=G=M_\odot=k_\mathrm{B}=1$, which are our code units.  In these units,
the rest mass saturation density is $\rho_\rn\simeq 4.34\times10^{-4}$,
one kilometer is $\unit[]{km}\simeq0.677$, and one millisecond is
$\unit[]{ms}\simeq203$.

\section{Hydrodynamic model}
\label{sec:model}

We briefly recall some defining properties of the three approaches to bulk viscosity that we implement in our numerical simulations.
More details are given in the companion paper \citep{Camelio22a}.
Then, we specify their hydrodynamic equations in radial gauge-polar slicing coordinates and spherical symmetry.

\subsection{Bulk viscosity: three approaches}
\label{ssec:bulk}

We consider 3 different approaches to bulk viscosity: (i) the multi-component
fluid with the tracking of the chemical reactions and particle
abundances, and two M\"uller-Israel-Stewart theories, namely (ii) the Hiscock-Lindblom theory, and (iii) the Maxwell-Cattaneo
theory, see \citet{Camelio22a}.

As long as the fluid elements are isotropic (e.g., in the absence of heat, superfluid or electric currents), the stress-energy tensor for the three models (i)-(iii) can be written as:
\begin{equation}
\label{eq:tmunu}
T^{\mu\nu}=(\epsilon + p)u^\mu u^\nu + pg^{\mu\nu},
\end{equation}
where $u^\mu$ is the 4-velocity of the fluid and $g^{\mu\nu}$ is the metric.
The physical meaning of the pressure $p$ depends on the model considered: for the
multi-component fluid (i) it is
$p=p(\rho,\epsilon,\{Y_i\}_i)$, while for the M\"uller-Israel-Stewart theories (ii)-(iii) it is
$p=p^\req(\rho,\epsilon)+\Pi$.  
Above, we call $\rho$ the rest mass density, $Y_i=\rho_i/\rho$
the particle fraction of species $i$ ($\rho_i$ is the rest mass density of
particle $i$), $\Pi$ the bulk stress, and the `eq' superscript refers to
quantities in beta-equilibrium.

Note that $Y_i$ and $\Pi$ are independent variables that have their own dynamics in models (i) and (ii)-(iii), respectively, and
in order to evolve the system we need to provide an equation for them.
For the multi-component fluid the evolution of each chemical fractions $Y_i$ is given by the 
corresponding continuity equation:
\begin{equation}
\label{eq:chemical}
\nabla_\mu \big( \rho Y_i u^\mu \big)=m_\rn\mathcal R_i,
\end{equation}
where $m_\rn$ is the neutron mass and $\mathcal R_i$ is the net reaction rate
of particle species $i$.
In the Hiscock-Lindblom model, the additional equation for the bulk stress is:
\begin{multline}
\label{eq:hiscock-lindbloom}
\nabla_\mu(\Pi u^\mu)=-\frac\Pi\tau -\left(\frac1\chi-\frac\Pi2\right)\nabla_\mu u^\mu\\
-\frac\Pi2 u^\mu\nabla_\mu\left(\log\frac{\chi}{T^\req}\right),
\end{multline}
where $\chi$ is the bulk viscous parameter, $\tau$ is the bulk viscous
timescale, and $T^\req$ is the temperature in
beta-equilibrium.  Finally, Maxwell-Cattaneo is a linearization
of Hiscock-Lindblom, and for it the
additional equation that describes the evolution of the bulk stress is:
\begin{equation}
\label{eq:maxwell-cattaneo}
\nabla_\mu(\Pi u^\mu)=-\frac\Pi\tau -\left(\frac1\chi-\Pi\right)\nabla_\mu u^\mu.
\end{equation}

\subsection{From the multi-component fluid to M\"uller-Israel-Stewart }

For small deviations from equilibrium, we can derive an equivalent
M\"uller-Israel-Stewart description of bulk viscosity from the multi-component fluid one,
thanks to the relations \cite{Gavassino21bulk, GavassinoRadiazio2020, Camelio22a}:
\begin{align}
\label{eq:def-zeta}
\zeta={}&n^4\Xi^{ab}\frac{\partial Y_a^\req}{\partial n}\bigg|_s\frac{\partial Y_b^\req}{\partial n}\bigg|_s,\\
\label{eq:def-xi}
\Xi_{ab}={}&\left.\frac{\partial \mathcal R_a(\{\mathbb A^j=0\}_{\forall j})}{\partial \mathbb A^b}\right|_{\rho,s,\{\mathbb A^i\}_{i\neq b}},\\
\label{eq:a2pi}
\Pi={}&\mathbb A^i\left.\frac{\partial Y_i^\req}{\partial n}\right|_{s}n^2,
\end{align}
where $n=\rho/m_\rn$ is the baryon number density, $s$ the entropy per baryon, 
$\mathbb A^i$ the affinity of particle $i$, and the $a,b$ indices refer
to the particle species (e.g., $\{\re,\mu\}$), are summed with the Einstein
rule, and are raised and lowered with matrix inversion
($\Xi_{ab}\Xi^{bc}=\mathbb I$).  The bulk viscous timescale $\tau$ and the bulk
viscous parameter $\chi$ are, strictly speaking, defined only for one
independent species,
\begin{align}
\label{eq:tau-from-tracking}
\tau={}&-\frac{n}{\Xi}\left.\frac{\partial Y(\mathbb A=0)}{\partial \mathbb A}\right|_{\rho,s},\\
\chi={}&\frac\tau\zeta,
\end{align}
where $\Xi=\Xi_{11}$ and we dropped the species index because there is only one independent
species. However, we can extend this formulation to more than one independent species
requiring that the speed of sound of the multi-component 
fluid coincides with that of the M\"uller-Israel-Stewart theories for small perturbation from thermodynamic equilibrium \cite{Camelio22a}:
\begin{equation}
\label{eq:tau-multispecies}
\tau = \dfrac{n \Xi^{ab} \,  \dfrac{\partial Y_a^\req}{\partial \rho} \bigg|_s
\,  \dfrac{\partial Y_b^\req}{\partial \rho} \bigg|_s}{ m_\rn \dfrac{\partial^2
u}{\partial Y_c \partial Y_d} \bigg|_{\rho,s} \,  \dfrac{\partial
Y_c^\req}{\partial \rho} \bigg|_s \, \dfrac{\partial Y_d^\req}{\partial \rho}
\bigg|_s} \, .
\end{equation}

\subsection{Including luminosity}

For all the three models (i)-(iii), the total rest mass density satisfies the continuity equation
\begin{equation}
\label{eq:continuity}
\nabla_\mu \big( \rho u^\mu \big) =0,
\end{equation}
and the stress-energy tensor evolution is given by the
the usual momentum and energy conservation equations\footnote{Here we
assume that neutrinos are emitted isotropically in the fluid rest frame and
immediately leave the star. See Appendix~A of \citet{OConnor10} for a more
general setting.}:
\begin{equation}
\label{eq:conservation}
\nabla_\mu\big( T^{\mu \nu}\big)=-\mathcal Q u^\nu,
\end{equation}
where $\mathcal Q$ is the total luminosity.
For the multi-component fluid, $\mathcal Q=\sum_j\mathcal Q_j$,
where $\mathcal Q_j$ is the luminosity of reaction $j$.
In M\"uller-Israel-Stewart theories, the energy luminosity
is usually neglected, $\mathcal Q=0$. However, in the companion paper \cite{Camelio22a} we show
that it is possible to account for the energy luminosity
also in this kind of theories by performing an expansion around equilibrium:
\begin{align}
\label{eq:q}
\mathcal Q(\rho,s,\Pi)={}&\mathcal Q^\req(\rho,s^\req)
+\frac{\partial \mathcal Q}{\partial \Pi}\Pi + \mathcal O(\Pi^2),\\
\label{eq:dqdpi}
\frac{\partial \mathcal Q}{\partial \Pi}={}&
\frac1{n^2}\left(\left.\frac{\partial Y^\req}{\partial n}\right|_s\right)^{-1}
\left.\frac{\partial \mathcal Q(\mathbb A=0)}{\partial \mathbb A}\right|_{\rho,s},
\end{align}
where, for simplicity, the above formula refers to the particular case in which
there is a single independent fraction $Y$ in the non-equilibrium equation of
state of the corresponding multi-component fluid, $\epsilon = \epsilon(\rho, s,
Y)$.

\subsection{Hydrodynamic equations in radial gauge, polar slicing coordinates in spherical symmetry}
\label{ssec:hydro}

We adopt the radial gauge, polar slicing coordinates in spherical symmetry
(i.e.~Schwarzschild), whose metric is:
\begin{equation}
\label{eq:metric}
\mathrm dl^2= 
-\alpha^2(r,t)\mathrm dt^2 + X^2(r,t) \mathrm dr^2
+r^2\mathrm d\Omega^2,
\end{equation}
where $l$ is the proper time, $t$ and $r$ are respectively the time and radial
coordinates, $\mathrm d\Omega$ is the angular element, $\alpha$ is the
lapse function:
\begin{equation}
\label{eq:alpha}
\alpha(r,t)=\exp\big(\phi(r,t)\big),
\end{equation}
the metric function $X$ is given by:
\begin{equation}
\label{eq:x}
X(r,t)=\left(1 - \frac{2m(r,t)}r\right)^{-1/2},
\end{equation}
$m$ is the gravitational mass:
\begin{equation}
\label{eq:m}
m(r,t)=4\pi\int_0^r\big((\epsilon+ p) W^2 -  p\big)x^2\mathrm dx,
\end{equation}
$\phi$ is the general relativistic equivalent of the Newtonian gravitational field:
\begin{equation}
\label{eq:phi}
\phi(r,t)=\int_0^rX^2\left(\frac{m}{x^2} + 4\pi
x\big((\epsilon+ p) W^2v^2 + p\big)\right)\mathrm dx + \phi_0,
\end{equation}
$W$ is the Lorentz factor:
\begin{equation}
\label{eq:w}
W=\left(1 - v^2\right)^{-1/2},
\end{equation}
$v$ is the velocity of the fluid, and the integration constant $\phi_0$
is determined by the condition $\phi\to 0$ as $r\to \infty$,
from which we obtain:
\begin{equation}
\label{eq:phi0}
\phi(r>R,t)=\frac12\log\left(1 - \frac{2m(R,t)}r\right),
\end{equation}
where $R$ is the stellar radius.
Finally, the total baryon mass of the star is:
\begin{equation}
\label{eq:mb}
M_\mathrm{b}=4\pi\int_0^RX\rho Wx^2\mathrm dx.
\end{equation}

In radial gauge, polar slicing coordinates in spherical symmetry, the continuity equations reads \cite{Romero96, OConnor10}:
\begin{align}
\label{eq:continuity-rho}
\partial_t D
+{}&\frac1{r^2}\partial_r\left(\frac{\alpha r^2}X Dv\right)=0,\\
\label{eq:def-d}
D={}& X\rho W,
\end{align}
the momentum conservation equation reads \cite{Romero96, OConnor10}:
\begin{align}
\label{eq:conservation-momentum}
\partial_t S^r{}&{}
+\frac1{r^2}\partial_r\left(\frac{\alpha r^2}X (S^rv + p)\right)=- \alpha W v {\mathcal Q}\notag\\
{}&{}-\epsilon\alpha X\left(8\pi rp + \frac{m}{r^2}\right)
+\alpha X p\frac{m}{r^2} + \frac{2\alpha p}{Xr},\\
\label{eq:def-sr}
S^r{}&{}=(\epsilon + p)W^2v,
\end{align}
the energy conservation equation reads \cite{Romero96, OConnor10}:
\begin{align}
\label{eq:conservation-energy}
\partial_t \tau_\epsilon
+{}&\frac1{r^2}\partial_r\left(\frac{\alpha r^2}X \big(S^r - Dv\big)\right)= -\alpha W \mathcal Q,\\
\label{eq:def-tau}
\tau_\epsilon={}&(\epsilon + p)W^2 - p - D,
\end{align}
the particle continuity equation reads \cite{OConnor10}:
\begin{align}
\label{eq:continuity-yi}
\partial_t (DY_i)
+\frac1{r^2}\partial_r\left(\frac{\alpha r^2}X DY_i v\right)={}&\alpha Xm\mathcal R_i,
\end{align}
the bulk stress equation in the Hiscock-Lindblom theory reads \cite{Camelio22a}:
\begin{multline}
\label{eq:pi-evol-hl}
\partial_t(XW\Pi) + \frac1{r^2}\partial_r\left(r^2\alpha Wv\Pi\right) = -\frac{\alpha X\Pi}{\tau}\\
-\left(\frac1\chi-\frac{\Pi}2\right)\left(\partial_t (XW) +\frac1{r^2}\partial_r(r^2\alpha Wv)\right)\\
-\frac{\Pi W}2\left(X\partial_t\log\frac\chi{T^\req}+\alpha v\partial_r\log\frac\chi{T^\req}\right),
\end{multline}
and the bulk stress equation in the Maxwell-Cattaneo theory reads \cite{Camelio22a}:
\begin{multline}
\label{eq:pi-evol}
\partial_t(XW\Pi) + \frac1{r^2}\partial_r\left(r^2\alpha Wv\Pi\right) = -\frac{\alpha X\Pi}{\tau}\\
-\left(\frac1\chi-\Pi\right)\left(\partial_t (XW) +\frac1{r^2}\partial_r(r^2\alpha Wv)\right),
\end{multline}
where as above it is $p=p(\rho,\epsilon,\{Y_i\}_i)$ for the multi-component
 fluid and $p=p^\req(\rho,\epsilon)+\Pi$ for Hiscock-Lindblom and
Maxwell-Cattaneo.

\section{Microphysics}
\label{sec:micro}

Here we summarize the microscopic input needed for our numerical simulations. 
For a more detailed discussion, see the companion paper \citep{Camelio22a}.

\subsection{Reaction rates and luminosity}

We consider a neutrino-less fluid of neutrons `n', protons `p', electrons `e', and muons `$\mu$',
that undergoes direct beta reactions (i.e., direct Urca reactions):
\begin{align}
\label{eq:beta-minus-e}
\beta^-_\mathrm{e}:  {}&\quad \mathrm n \to \mathrm p + \mathrm e^- + \bar\nu_\mathrm{e},\\
\label{eq:beta-plus-e}
\beta^+_\mathrm{e}:  {}&\quad \mathrm p + \mathrm e^- \to \mathrm n + \nu_\mathrm{e},\\
\label{eq:beta-minus-mu}
\beta^-_\mu:{}&\quad \mathrm n \to \mathrm p + \mu^- + \bar\nu_\mu,\\
\label{eq:beta-plus-mu}
\beta^+_\mu:{}&\quad \mathrm p + \mu^- \to \mathrm n + \nu_\mu.
\end{align}
The net number reaction rate for the particle $i=\{\re,\mu\}$, which is responsible for the
change in the particle fraction [see Eq.~\eqref{eq:continuity-yi}], due to neutron
decay [Eqs.~\eqref{eq:beta-minus-e} and \eqref{eq:beta-minus-mu}] and lepton
capture [Eqs.~\eqref{eq:beta-plus-e} and \eqref{eq:beta-plus-mu}] and linearized around equilibrium in
the affinity $\mathbb A^i$, is \cite{Camelio22a}:
\begin{equation}
\label{eq:R}
\mathcal R_i= \frac{8.86\times 10^{31}}{\unit{cm^{3}\,s}}
\sqrt[3]{\frac{Y^\req_i\rho}{\rho_\rn}} \left(\frac{T^\req}{\unit[10^9]{K}}\right)^5\frac{17\pi^4}{30}\frac{\mathbb A^i}{k_\mathrm{B}T}.
\end{equation}
We chose to linearize the reaction rates in
order to simplify the implementation of the implicit step (See Appendix~\ref{sec:implementation}).
The reaction luminosity linearized around equilibrium in $\Delta
Y_i=Y_i-Y_i^\req$ [which in our case is linear in the affinity $\mathbb A^i$,
see Eq.~\eqref{eq:ai-dyi}], which is responsible for the change in the energy
and momentum of the fluid [see Eqs.~\eqref{eq:conservation-momentum} and
\eqref{eq:conservation-energy}], is \cite{Camelio22a}:
\begin{equation}
\label{eq:Q}
\mathcal Q_i= \frac{1.22\times10^{25}}{\unit{erg^{-1}\,cm^{3}\,s}}
\sqrt[3]{\frac{Y^\req_i\rho}{\rho_\rn}} \left(\frac{T^\req}{\unit[10^9]{K}}\right)^6\frac{457\pi^6}{1260}
\left(1 + \frac{\Delta Y_i}{3Y^\req_i}\right),
\end{equation}
where $\rho_\rn$ is the rest mass density at nuclear saturation.

Note that a more realistic set of reactions for the stellar conditions of the
models considered in this paper would have included the modified beta reactions
(i.e.~modified Urca, see Sec.~2.2 of \citet{Haensel92}).  Since we are not
interested in performing a quantitative study of the neutron star evolution,
but rather in comparing the multi-fluid formulation and the bulk stress
approximations for bulk viscosity, we focus on the results of simulations
including only un-suppressed direct Urca reactions.  However, we perform
some simulations using the modified Urca number ($\mathcal R_i^\mathrm{m}$) and
energy ($\mathcal Q_i^\mathrm{m}$) rates \cite{Camelio22a} instead of
the direct Urca ones,
\begin{align}
\label{eq:R-murca}
\mathcal R_i^\mathrm{m}\simeq{}& \frac{5.91\times 10^{23}}{\unit{cm^{3}\,s}}
\sqrt[3]{\frac{Y^\req_i\rho}{\rho_n}}\left(\frac{T^\req}{\unit[10^9]{K}}\right)^7
\frac{367\pi^6}{63} \frac{\mathbb A^i}{k_\mathrm{B}T^\req},\\
\label{eq:Q-murca}
\mathcal Q_i^\mathrm{m}\simeq{}& \frac{8.15\times10^{16}}{\unit{erg^{-1}\,cm^{3}\,s}}
\sqrt[3]{\frac{Y^\req_i\rho}{\rho_n}} \left(\frac{T^\req}{\unit[10^9]{K}}\right)^8
\frac{11513\pi^8}{2520} \left(1 + \frac{\Delta Y_i}{3Y^\req_i}\right).
\end{align}
We comment on these in Sec.~\ref{ssec:results:murca}.

\subsection{Analytic non-equilibrium equation of state}

We assume an analytic equation of state (EOS) for matter out of beta-equilibrium, defined by:
\begin{align}
\label{eq:eos}
u={}& k_0\rho + k_\rth s^2\rho^{\Gamma_\rth-1}
+ \sum_i k_i\Delta Y_i^2,\\
\Delta Y_i={}&{}Y_i - Y_i^0 \frac{\rho}{\rho_\rn},
\end{align}
where $i$ is the index of the independent particle fractions (e.g., $i=e,\mu$ in out-of-equilibrium npe$\mu$ matter). 
In the above expression, $u=u(\rho,s,\{Y_i\}_i)=\epsilon/\rho-1$ is the specific (per unit mass)
internal energy, $k_0,k_\rth,k_i$ are constant positive coefficients,
$\Gamma_\rth$ is the thermal polytropic exponent,
and $Y_i^0$ the equilibrium net particle fraction at nuclear saturation.
This EOS is a simple extension of a cold $\Gamma=2$ polytropic EOS; the
thermal component was already introduced in \citet{Camelio19}.
The dependence on $\rho$ of the cold and particle components of the EOS
was chosen \citep{Camelio22a} such that the pressure and the speed of sound are positive if
$Y_i\in[0,1]$ and
\begin{equation}
\label{eq:positivity-condition}
k_0>2{\sum_ik_i Y_i^0}/{\rho_\rn}.
\end{equation}

\begin{figure}[!ht]
\includegraphics[width=\columnwidth]{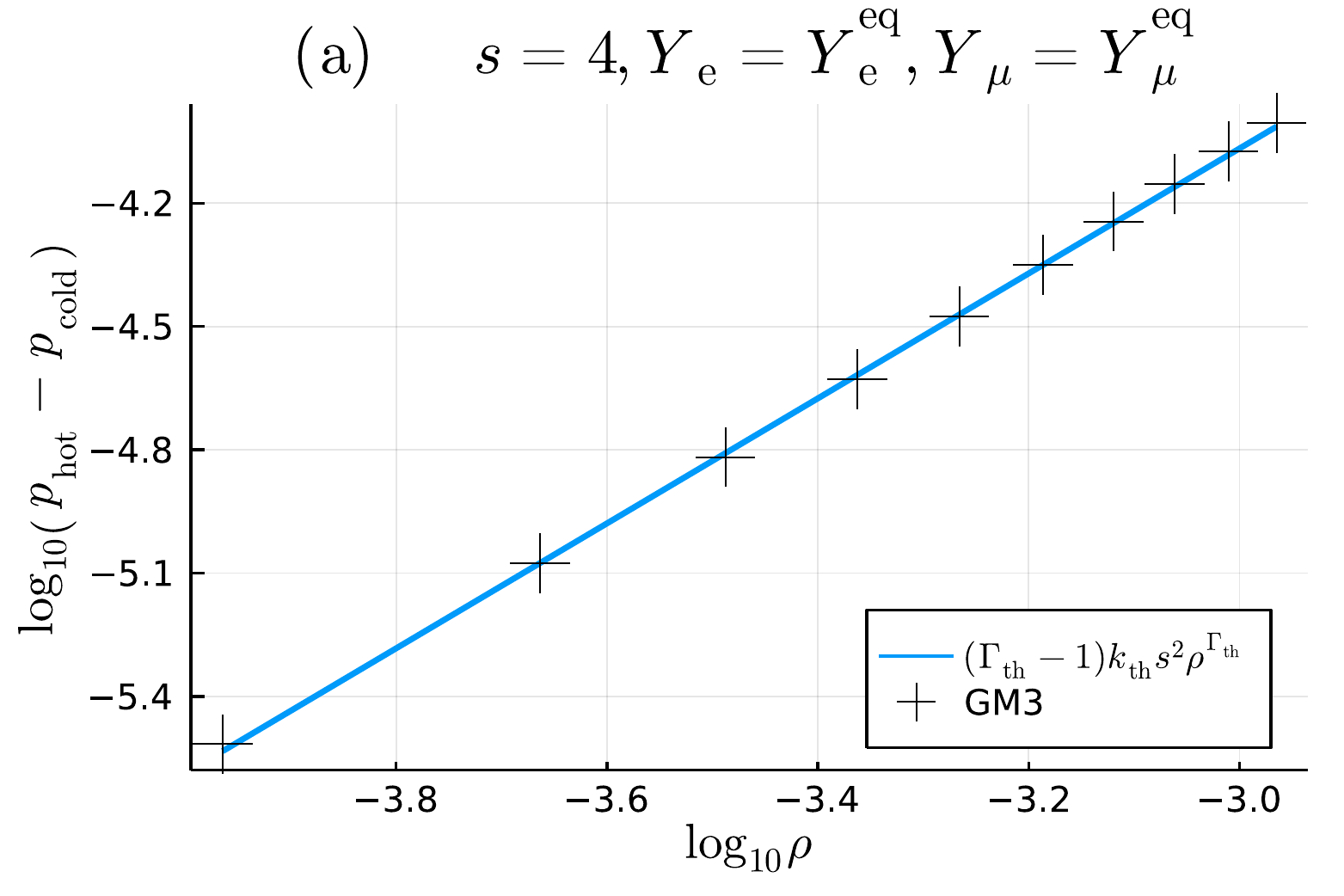}
\includegraphics[width=\columnwidth]{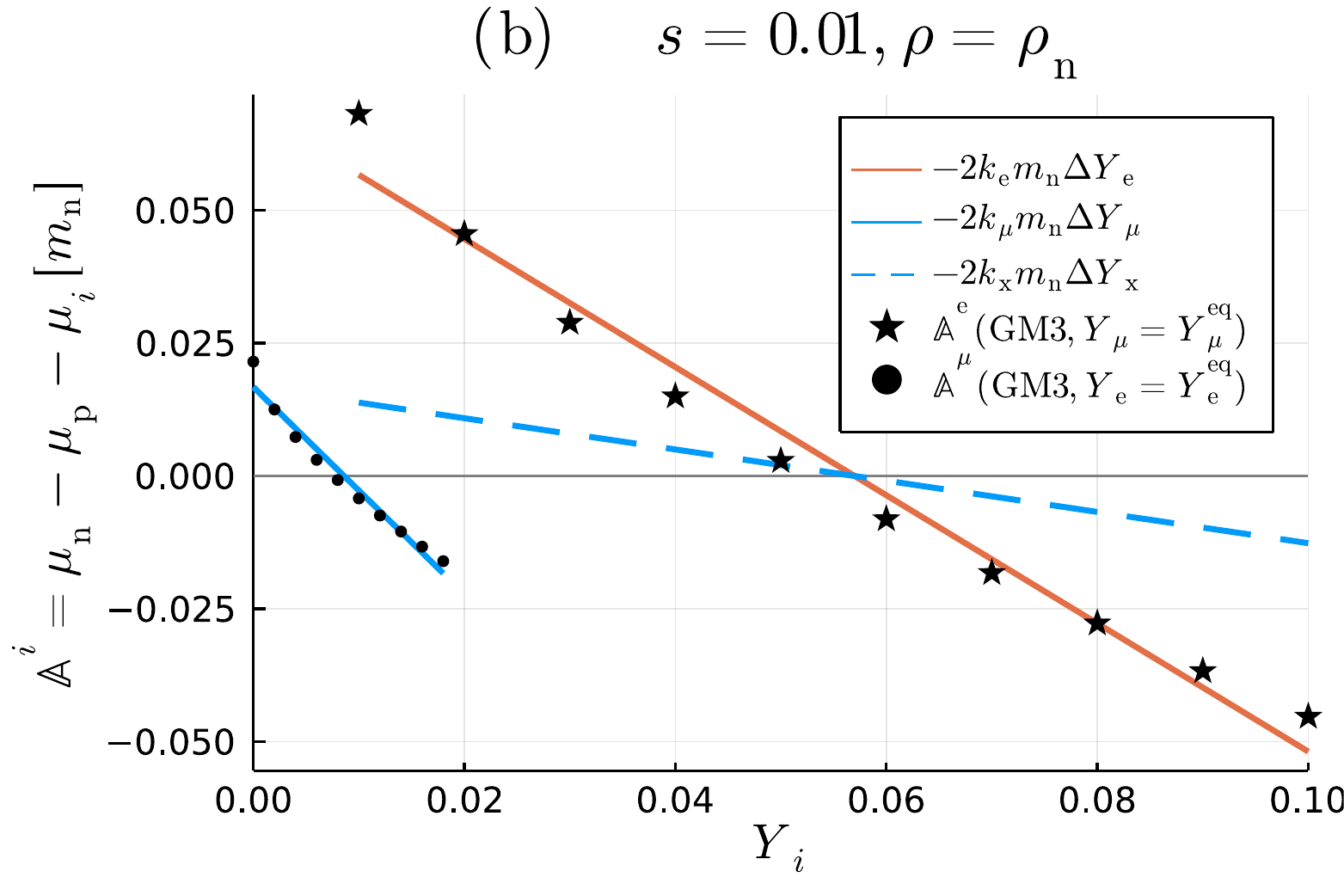}
\includegraphics[width=\columnwidth]{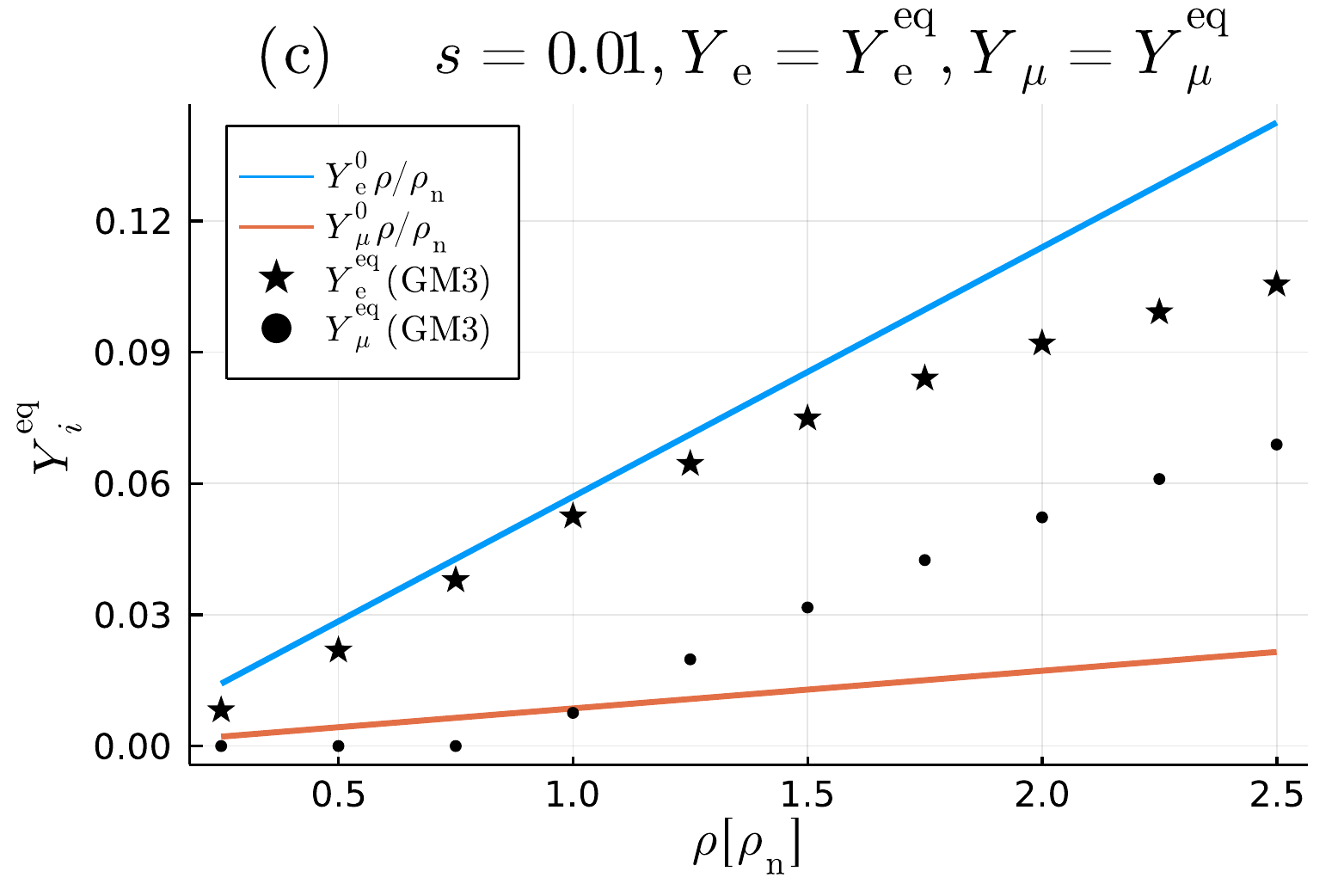}
\caption{EOS used in this paper and GM3~EOS. From top to bottom:
(a) thermal contribution to the equilibrium pressure (for $s=4$),
(b) dependence of the particle affinity ($\mathbb
A^i=\mu_\rn-\mu_\mathrm{p}-\mu_i$, where $\mu_{\{\rn,\mathrm{p},i\}}$ are the
chemical potentials) on the particle fraction $Y_i$, (c) dependence of the
equilibrium particle fraction on the rest mass density.
The value of the muon parameters obtained from the fit are $k_\mu=0.974, Y_\mu^0=0.00860$,
while the other parameters are reported in Table~\ref{tab:eos}.}
\label{fig:eos}
\end{figure}

Our EOS has the advantages to be simple and analytic, and at the same time it
reproduces some behaviors of a realistic EOS, and in particular: (i) the cold
EOS at equilibrium is a polytrope, (ii) the temperature goes to zero as the
entropy goes to zero, and (iii) the free parameters
$\Gamma_\rth,k_\rth,k_\re,Y_\re^0$ have been chosen in order to fit the GM3
EOS \cite{Glendenning85, Glendenning91}, see Table~\ref{tab:eos} and plots (a) and (b)
of Fig.~\ref{fig:eos}.

However, we are not trying to accurately reproduce an existing EOS, as for example the
polytropic parameters for the cold GM3 EOS are $\Gamma=2.88,
k_0=2.78\times10^4$ while we are setting $\Gamma=2, k_0=200$ in order for
Eq.~\eqref{eq:positivity-condition} to be true.

Given our analytical model for the EOS in \eqref{eq:eos}, the equilibrium value of the particle fractions 
(obtained in the limit $\mathbb A^e=\mathbb A^\mu=0$) is:
\begin{equation}
Y^\req_i(\rho)=Y^0_i\frac{\rho}{\rho_n} \qquad \quad (i=e,\mu) \, .
\end{equation}
Unfortunately, this expression does not reproduce the correct trend, see plot (c) of Fig.~\ref{fig:eos}.
In particular, the equilibrium fraction of muons $Y^\req_\mu$ deviates noticeably from the
correct trend (see plot (c) of Fig.~\ref{fig:eos}), and as a consequence the
contribution of the muons to the dynamics is negligible.
Since we are not aiming for an accurate physical description but
rather to clarify the role of bulk viscosity in simulations, we decided to artificially
enhance the role of muons in the dynamics by modifying their parameters, such that
condition~\eqref{eq:positivity-condition} is respected.
In order to do so, we substitute the muons `$\mu$' with an artificially modified muon particle
that we call `x' to avoid confusion, such that:
\begin{align}
Y_\rx^0 k_\rx={}& Y_\mu^0 k_\mu,\\
Y_\rx^0 = {}& Y_\re^0.
\end{align}
In Table~\ref{tab:eos} we report the EOSs used in this paper for the different
models described in Sec.~\ref{sec:results}. The EOS for the model `PF' (perfect fluid)
has both electrons and modified muons at equilibrium, the EOS
for model `MF' (multi-component fluid) has electrons out of equilibrium and modified muons
at equilibrium, and the EOS for model `MF-x' has both electrons and modified muons
out of equilibrium.

The other thermodynamic quantities can be derived from Eq.~\eqref{eq:eos}
and the first law of thermodynamics \cite{Camelio22a}:
\begin{align}
p={}&k_0\rho^2+ (\Gamma_\rth-1)k_\rth s^2\rho^{\Gamma_\rth} \notag\\
&{}- 2\frac{\rho^2}{\rho_\rn}\sum_i k_i Y_i^0 \Delta Y_i,\\
T={}& 2m_\rn k_\rth s \rho^{\Gamma_\rth-1},\\
\label{eq:ai-dyi}
\mathbb A^i={}&{}-2m_\rn k_i\Delta Y_i,\\
\label{eq:cs-uv}
c_\mathrm{s,uv}^2={}&\frac{\left.\frac{\partial p^\req}{\partial\rho}\right|_s
-2\frac{\rho}{\rho_\rn}\sum_ik_iY_i^0\left(2\Delta Y_i- Y_i^0\frac{\rho}{\rho_\rn}\right)}
{\left.\frac{\partial\epsilon^\req}{\partial\rho}\right|_s
+\sum_i k_i \Delta Y_i\left(\Delta Y_i - 2 Y_i^0 \frac{\rho}{\rho_\rn}\right)},\\
c_\mathrm{s,ir}^2={}&\left.\frac{\partial p^\req}{\partial\rho}\right|_s\div
\left.\frac{\partial\epsilon^\req}{\partial\rho}\right|_s,\\
\left.\frac{\partial p^\req}{\partial\rho}\right|_s={}&{}2k_0\rho+\Gamma_\rth(\Gamma_\rth-1)k_\rth s^2\rho^{\Gamma_\rth-1},\\
\left.\frac{\partial\epsilon^\req}{\partial\rho}\right|_s={}&{}1+2k_0\rho+\Gamma_\rth k_\rth s^2\rho^{\Gamma_\rth-1},
\end{align}
where $c_\mathrm{s,ir}$ and
$c_\mathrm{s,uv}$ are respectively the `infrared' (where the chemical composition is always at equilibrium \cite{Camelio22a}) and `ultraviolet' speeds of
sound \cite{Camelio22a} (where the thermodynamic derivatives are performed at constant composition \cite{Camelio22a}), and the `eq' superscript means that the quantity is
taken at equilibrium ($\{Y_i=Y_i^\req\}_i$).

For our choice of EOS and reaction rates, we have:
\begin{align}
\Xi_{ab}={}& \mathrm{diag}(\Xi_\re,\Xi_\rx),\\
\Xi_i={}& \frac{8.86\times10^{31}}{\unit{cm^{3}\,s}}\sqrt[3]{\frac{Y^0_i\rho^2}{\rho^2_n}}
\left(\frac{T}{\unit[10^9]{K}}\right)^5
\frac{17\pi^4}{30k_BT},\\
\zeta={}&n^2\frac{\rho^2}{\rho_n^2}\sum_i\frac{(Y^0_i)^2}{\Xi_i},\\
\tau={}&\frac{n}{2m_\rn}\times\sum_i{\frac{(Y_i^0)^2}{\Xi_i}}\div\sum_i{k_i(Y_i^0)^2},\\
\label{eq:effective-cs2}
c_{s,\mathrm{uv}}^{\req \, 2}={}& c_\mathrm{s,ir}^2 + \frac{1}{(\epsilon+p^\req)\chi},\\
\Pi={}&-2\rho\sum_i k_i Y^\req_i \Delta Y_i,
\end{align}
where $i=\{\re,\rx\}$.

\begin{table}
\centering
\caption{Parameters of the EOSs used in this paper.
In the last row we report if the parameter has been determined with a fit of the GM3~EOS
or if it has been chosen ad-hoc (see text).
The equilibrium temperature at saturation density and $s=4$ is
$T^\req(\rho_\rn,\unit[4]{k_\mathrm{B}})= \unit[56]{MeV/k_B}$ for all EOSs.}
\label{tab:eos}
\begin{tabular}{cccccccc}
model & $k_0$ & $\Gamma_\rth$ & $k_\rth$ & $k_\mathrm{e}$ & $Y_\mathrm{e}^0$ & $k_\rx$ & $Y_\rx^0$ \\
\hline
PF   & 200    & 1.52 & 0.374 & -      & -      & -      & - \\
MF   & 200    & 1.52 & 0.374 & 0.603  & 0.0570 & -      & - \\
MF-x & 200    & 1.52 & 0.374 & 0.603  & 0.0570 & 0.147  & 0.0570 \\
\hline
     & ad-hoc & fit  & fit   & fit    & fit    & ad-hoc & ad-hoc \\
\end{tabular}
\end{table}

\section{Numerical method}
\label{sec:code}

We implemented a new Eulerian, one-dimensional, general-relativistic
hydrodynamic code called \texttt{hydro-bulk-1D} in the \texttt{C} programming language
and release it \citep{Camelio22code} on \texttt{zenodo} under the MIT license, in such a way that the
community can check our results and improve on them.

The code uses finite-differences with the method of lines on an evenly spaced
grid.  The time evolution is performed with a 3rd order IMplicit-EXplicit
Runge-Kutta (IMEX-RK) solver \cite{Pareschi10} and a Courant-Friedrichs-Lewy factor CFL = 0.5.  The spatial
fluxes are obtained with the LLF (Local Lax-Friedrichs) approximate Riemann solver \cite{Lax54} with 2nd order
(piecewise linear) spatial reconstruction of the primitive variables and the
MINMOD slope limiter.  The gravitational mass is computed by
integrating Eqs.~\eqref{eq:m} and \eqref{eq:phi} with the trapezoid method.
We use as primitive variables $\rho$, $Wv$, $u$ with the
addition of $Y_\mathrm{e}$ and eventually $Y_\mathrm{x}$ (in the case of the
multi-component fluid), or $\Pi$ (in the case of Hiscock-Lindblom and
Maxwell-Cattaneo).  We perform the conservative to primitive inversion using
the Brent method to find the root of the conserved momentum, using $Wv$ as
independent variable. When the density is smaller than the threshold value
$\rho_\mathrm{thr}=100\rho_\mathrm{min}$, we set it as an atmosphere at
$\rho_\mathrm{min}=10^{-20}$.
In order to avoid division by zero, we set a floor for the temperature at
$T_\mathrm{min}=10^{-61}$ (in code units) in the derivative in the source of
the Hiscock-Lindblom equation [Eq.~\eqref{eq:pi-evol-hl}].  It is also possible to set in the code
a lower order time integration (2nd or 1st) and spatial reconstruction
(1st, i.e.~piecewise constant), the MCLIM slope limiter \cite{vanLeer77}, and the Harten-Lax-vanLeer (HLL)
Riemann-Solver \cite{Harten83}.

We adopt an implicit scheme to evolve the equations, as done also by \citet{MostNoronha21}, because the timescale of
the reactions can become shorter than the timestep, making an explicit method
unstable, see also \citep{celora22arxiv} for a recent discussion. 
Note that bulk stress formulations (namely Hiscock-Lindblom or
Maxwell-Cattaneo) do not solve this issue since the bulk viscous timescale
$\tau$ is related to the reaction timescale.

We describe in more detail the code in Appendix~\ref{sec:implementation} and
we test it in Appendix~\ref{sec:tests}.

\section{Results}
\label{sec:results}

In this section, we compare the bulk viscous evolution performed with different
models.  For each model, we consider both small deviations (neutron star
oscillations) and large deviations (neutron star migration from unstable to
stable branch) from hydrostatic equilibrium. 

\begin{figure}
\includegraphics[width=\columnwidth]{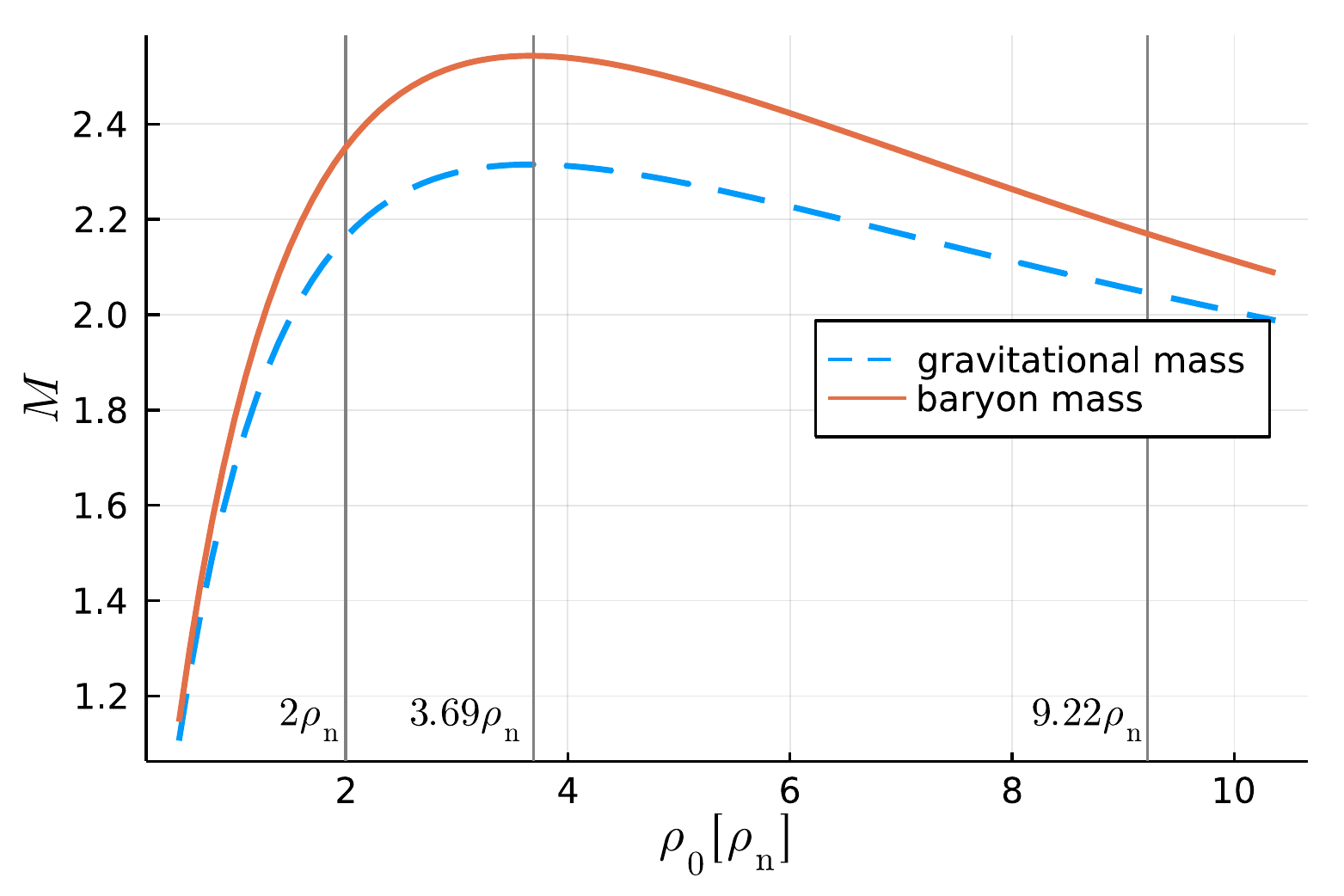}
\caption{Central rest mass density vs gravitational and baryon mass for a cold neutron
star in chemical equilibrium with the EOS considered in Sec.~\ref{sec:micro} ($\Gamma=2, k_0=200$).
The vertical lines refer to the oscillating and migration configurations considered in the
text (note that the addition of the uniform entropy per baryon $s=0.04$ does not
noticeably change the plot), and to the maximal gravitational mass configuration.
The latter separates the stable (left) and the unstable (right) branches and corresponds to
$\rho_0=3.69\rho_\rn$, $R=\unit[15.9]{km}$, $M=\unit[2.32]{M_\odot}$,
and $M_\mathrm{b}=\unit[2.54]{M_\odot}$.
The radius of the configuration with $M=\unit[1.4]{M_\odot}$ for this EOS is $R_{1.4M_\odot}=\unit[22.5]{km}$.}
\label{fig:m-rho}
\end{figure}

For the oscillation case, we choose as initial condition the stable and isentropic Tolman-Oppenheimer-Volkoff (TOV)
model in beta-equilibrium with central density $\rho_0=2\rho_\rn$.
The initial uniform entropy per baryon $s=0.04$ is such that the dissipative
timescale in the center is approximately of the order of 1~ms (see discussion in
Appendix~\ref{sec:tdamp}),
which is also the order of magnitude of the oscillation period.
For this model, $R=\unit[18.7]{km}$, $M=\unit[2.16]{M_\odot}$, and
$M_\mathrm{b}=\unit[2.35]{M_\odot}$.
The other settings specific for the oscillation case are $R_\mathrm{max}=13$,
$N=800$, and $v_\mathrm{pert}=5\times10^{-3}$ ($v_\mathrm{pert}$ determines the
amplitude of the initial velocity perturbation), see Appendix~\ref{sec:implementation} for details.
For the migration case, we choose as initial condition the unstable cold ($s=0$) TOV
model in beta-equilibrium with central density $\rho_0=4\times10^{-3}\simeq9.22\rho_\rn$, which
gives a star with $R=\unit[12.2]{km}$, $M=\unit[2.05]{M_\odot}$, and
$M_\mathrm{b}=\unit[2.17]{M_\odot}$.
The other settings specific for the migration model are $R_\mathrm{max}=80$,
$N= 8001$ ($\mathrm dr=0.01$), and $v_\mathrm{pert}=0$.
The other settings common to both models are described in Sec.~\ref{sec:code},
while the EOSs are described in Sec.~\ref{sec:micro}.
Unless otherwise specified, we consider only direct Urca reactions; modified
Urca reactions are consider only in Sec.~\ref{ssec:results:murca}.
In Fig.~\ref{fig:m-rho} we plot the dependence of the total baryon and gravitational masses
on the central density of the cold TOV model and we mark the central densities of
the oscillation and migration models.
We remark that, contrary to the oscillation and migration models considered in the
tests (see Appendix~\ref{sec:tests}), the two configurations do not have the
same baryon number, namely the end result of the migration is not equivalent to
the oscillation model.

For each case (oscillations and migration), we will consider 8 different models:
\begin{itemize}
\item PF: perfect fluid with all reactions at equilibrium (namely no viscosity and no
independent particle fractions),
\item MF: multi-component fluid with only electron beta reactions out of equilibrium
(namely the only independent particle fraction are the electrons) and no luminosity ($\mathcal Q=0$),
\item HL: Hiscock-Lindblom with only electron beta reactions out of equilibrium
(as MF) and no luminosity ($\mathcal Q=0$),
\item MC: Maxwell-Cattaneo with only electron beta reactions out of equilibrium
(as MF) and no luminosity ($\mathcal Q=0$),
\item MF-Q: multi-component fluid with luminosity and only electron beta reactions out of equilibrium
(as MF),
\item MC-Q: Maxwell-Cattaneo with luminosity and only electron beta reactions out of equilibrium
(as MF).
\item MF-x: multi-component fluid with electrons and modified muons `x' beta reactions out of equilibrium and no luminosity ($\mathcal Q=0$),
\item MC-x: Maxwell-Cattaneo with electrons and modified muons `x' beta reactions out of equilibrium (as MF-x) and no luminosity ($\mathcal Q=0$).
\end{itemize}

\begin{figure*}
\includegraphics[width=\columnwidth]{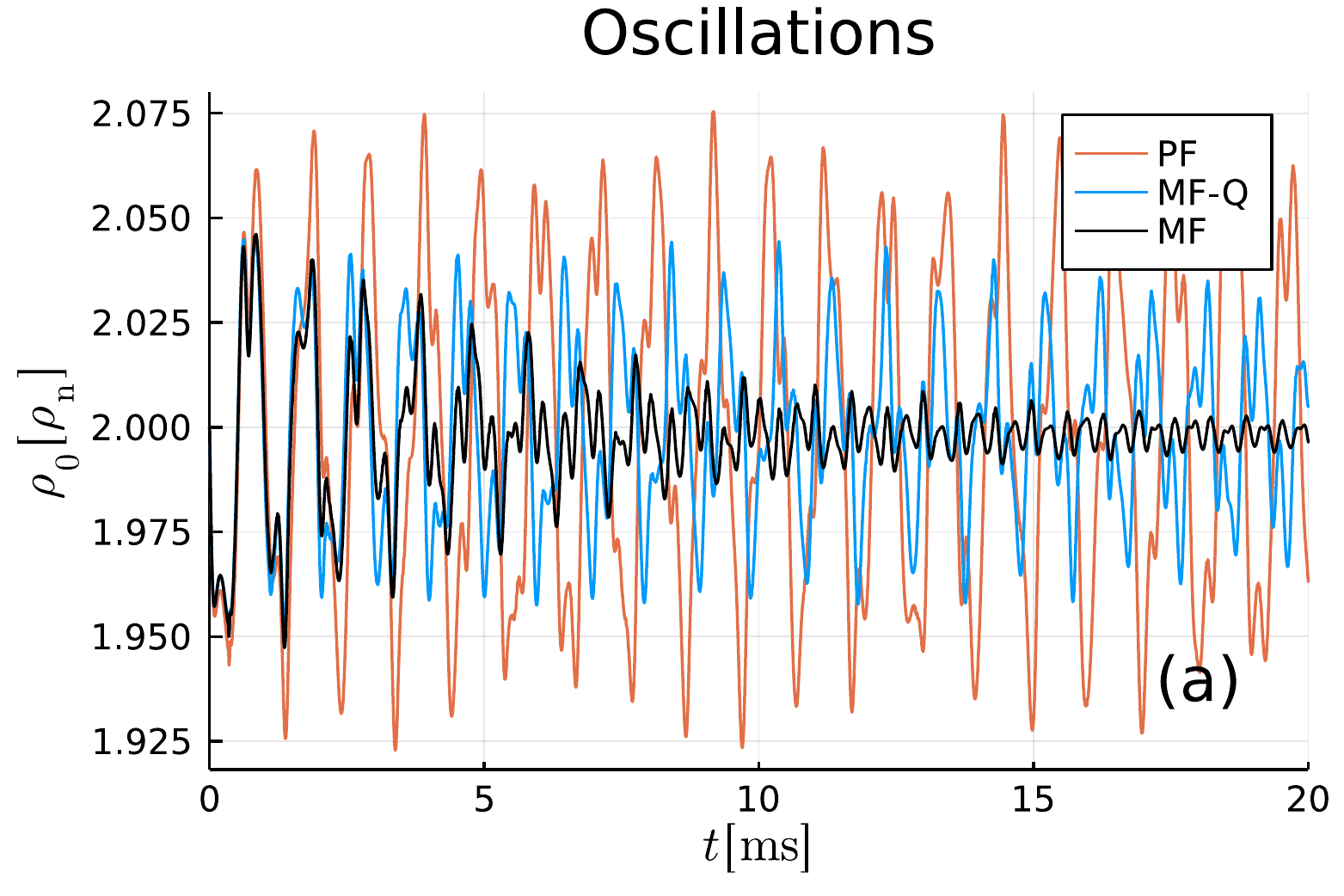}
\includegraphics[width=\columnwidth]{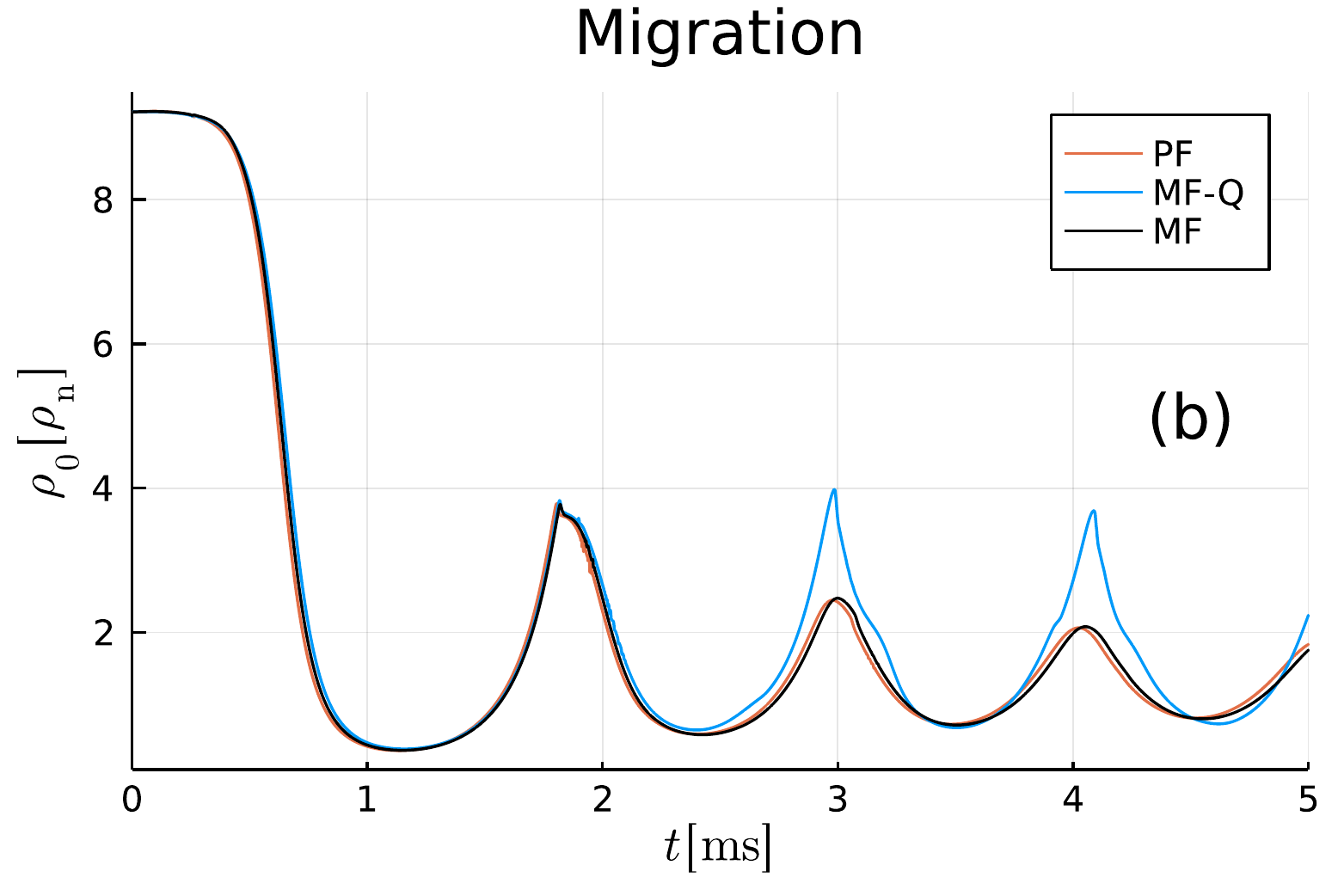}
\includegraphics[width=\columnwidth]{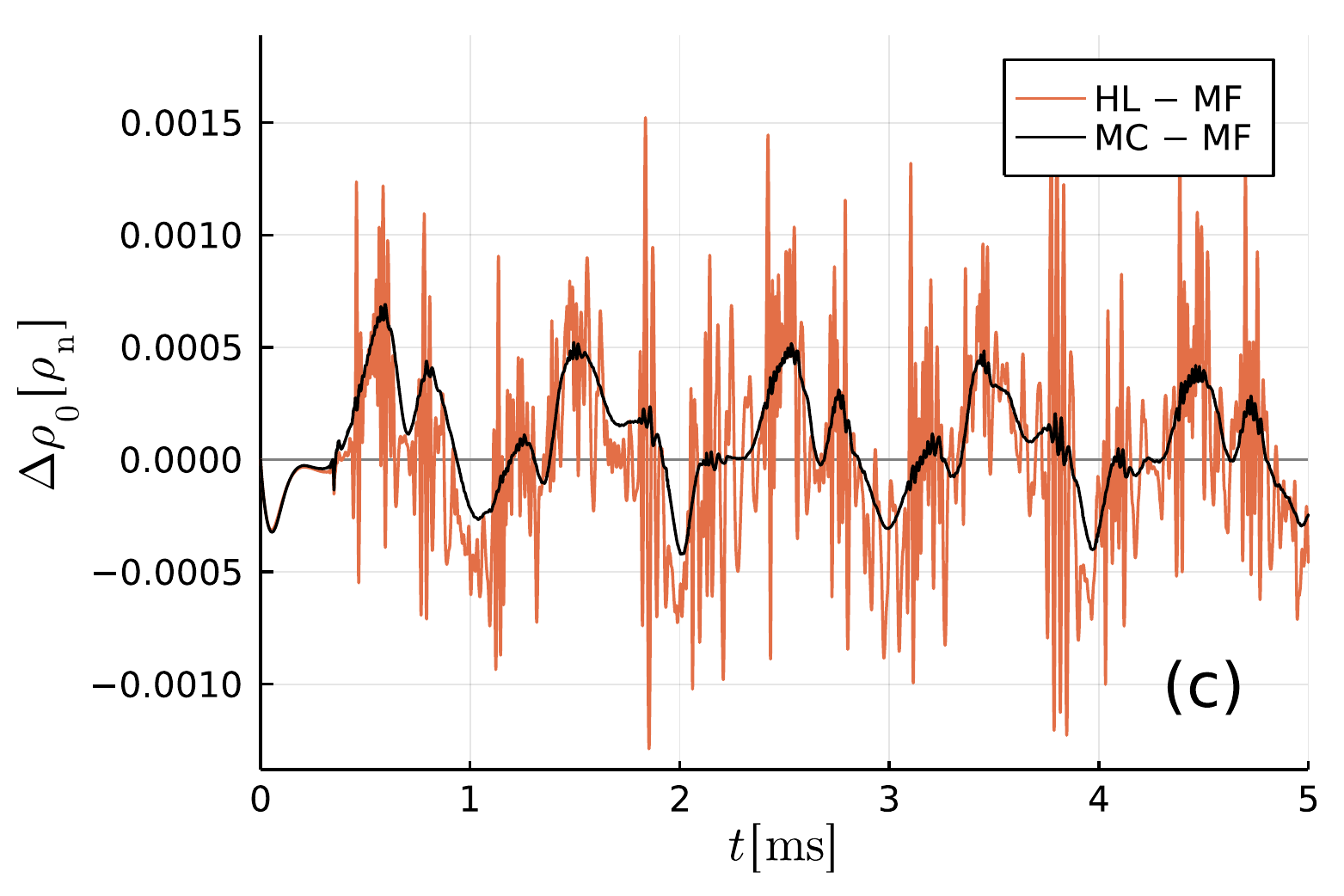}
\includegraphics[width=\columnwidth]{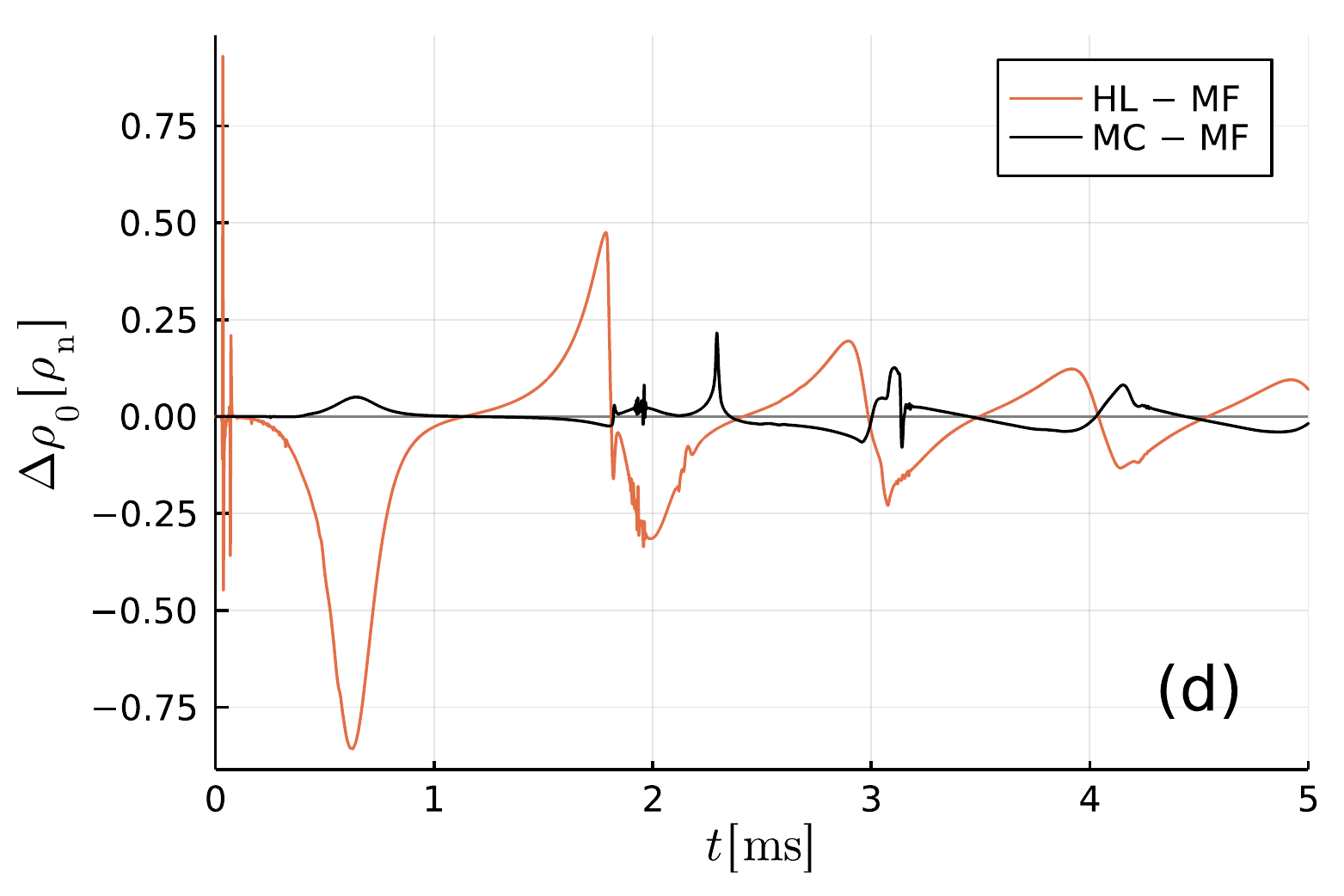}
\includegraphics[width=\columnwidth]{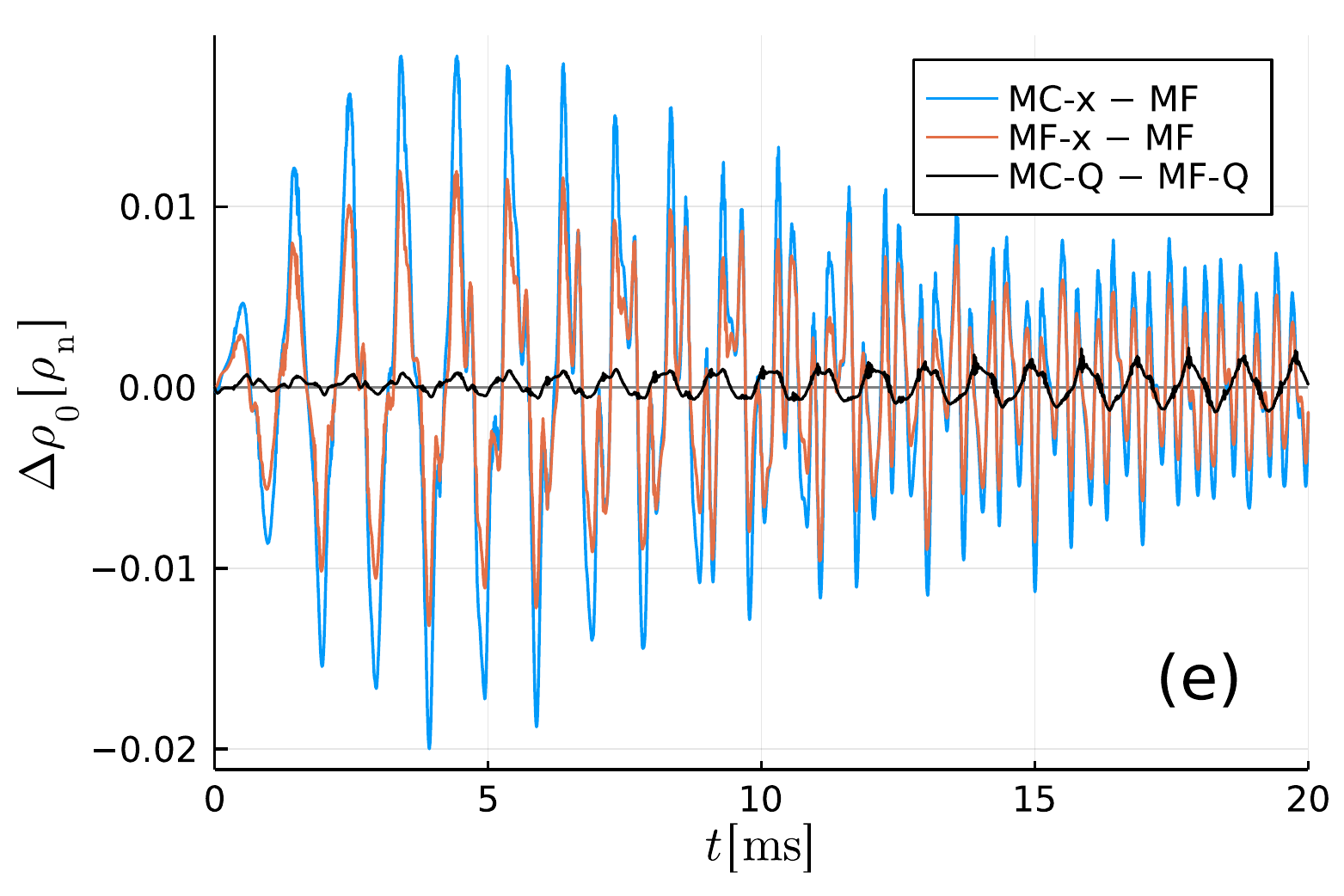}
\includegraphics[width=\columnwidth]{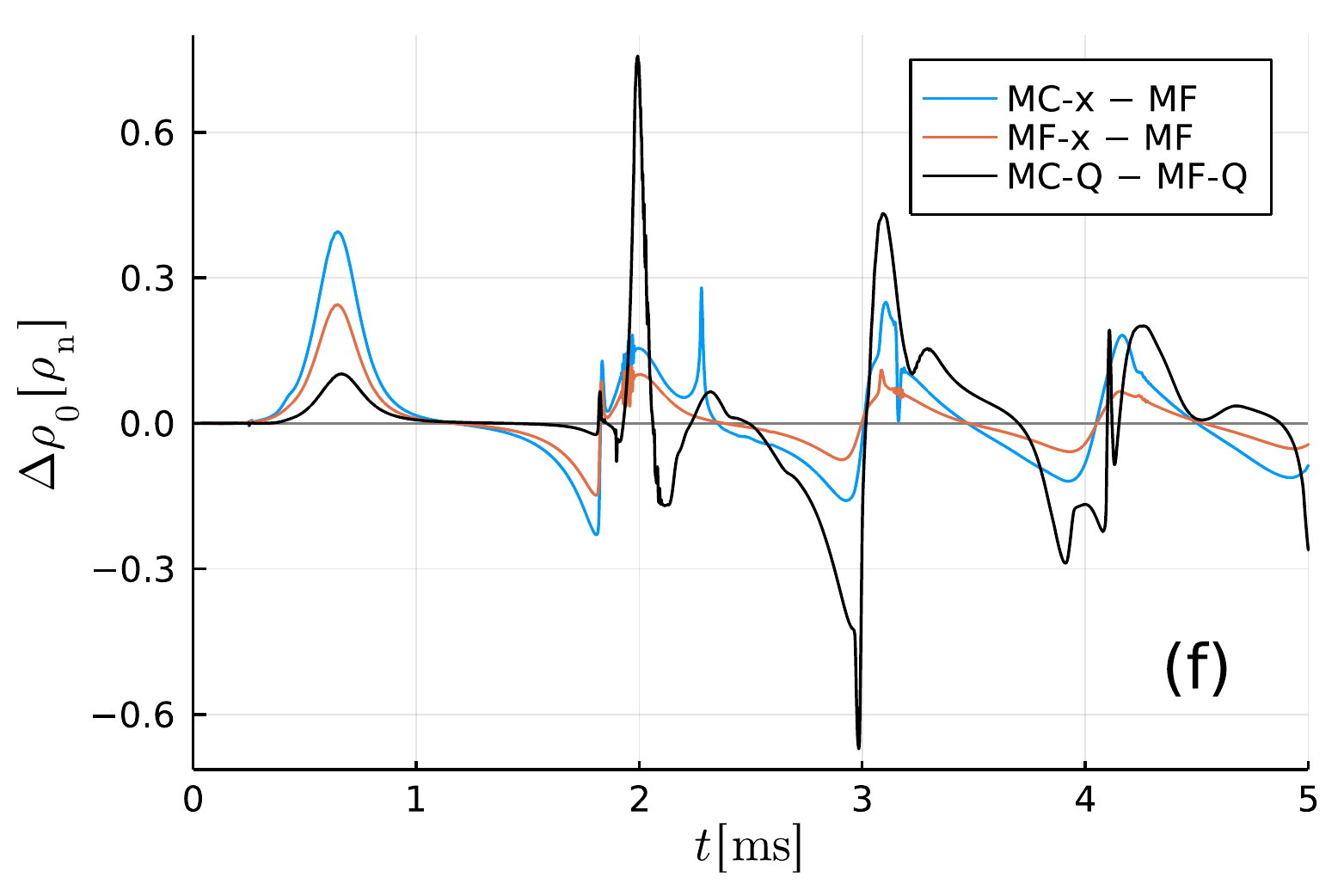}
\caption{Evolution of the central rest mass density (direct Urca reactions).
On the left [plots (a), (c), and (e)] we show the oscillation case, on the right [plots (b), (d), (f)] we show the migration case.
In the first row [plots (a) and (b)] we plot the central rest mass density for the PF, MF, and MF-Q models.
In the second and third row, we plot the difference of the central rest mass density between different models.
Plot~(c) shows a detail of the total evolution, up to 5~ms.}
\label{fig:rho0}
\end{figure*}

In Fig.~\ref{fig:rho0} we show the central rest mass density evolution of the
oscillation (left) and migration (right) models. In order to enhance the
differences between the models, we show the central rest mass density of models
PF, MF-Q, and MF in plots (a) and (b) and the difference between the central
rest mass densities of the other models and that of models MF and MF-Q in plots
(c)-(f).

\subsection{Comparison of the three approaches to bulk viscosity}
\label{ssec:results:models}

In this subsection, we compare the three dissipative models MF, HL, and MC, together with the additional model PF (no viscosity) that can be used as a benchmark reference.  
In all cases, we will assume that the modified muons `x' are in equilibrium, so that the only independent fraction that needs to be evolved is $Y_\re$. We also assume that no energy is emitted during the simulation ($\mathcal Q=0$).

\begin{figure}
\includegraphics[width=\columnwidth]{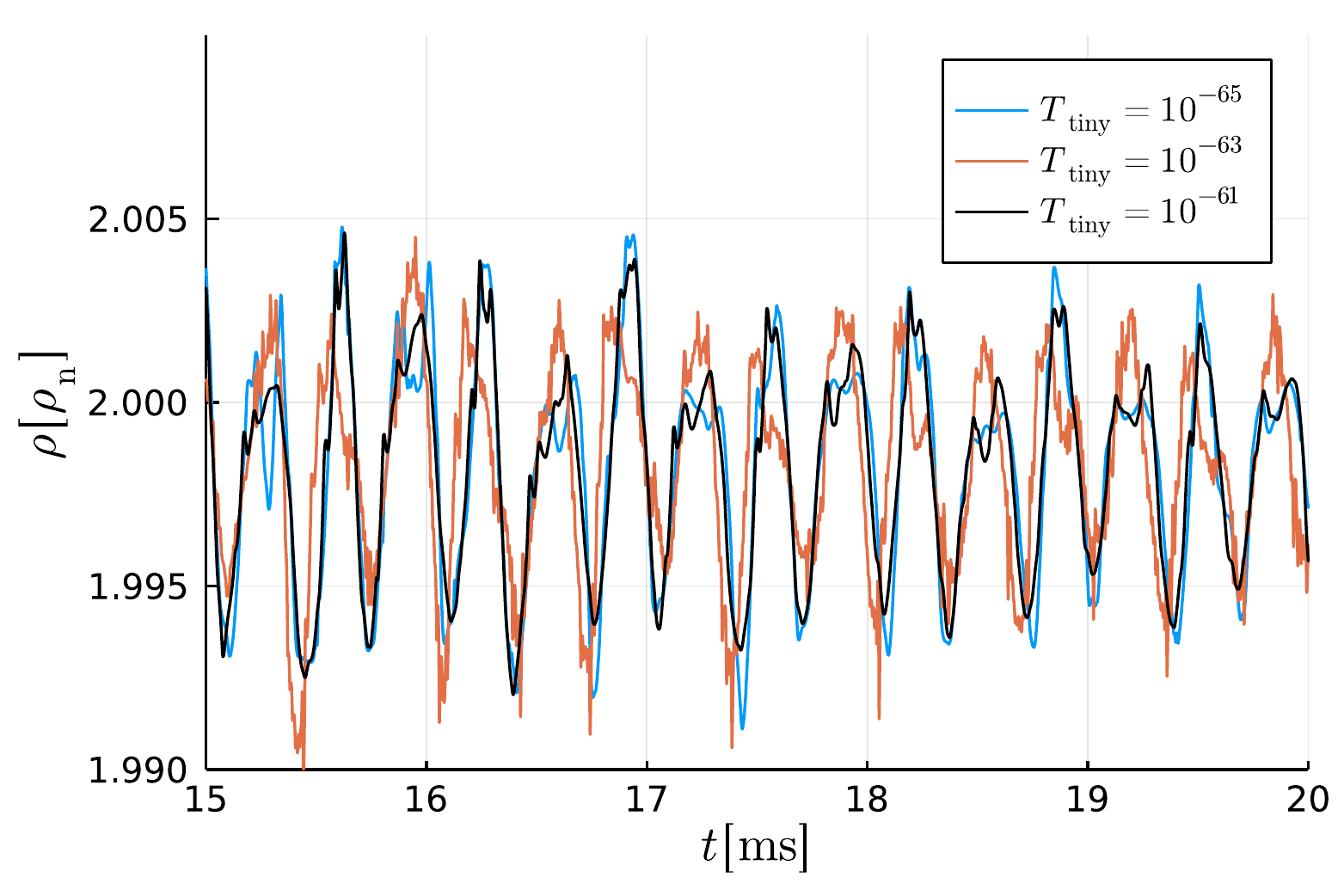}
\includegraphics[width=\columnwidth]{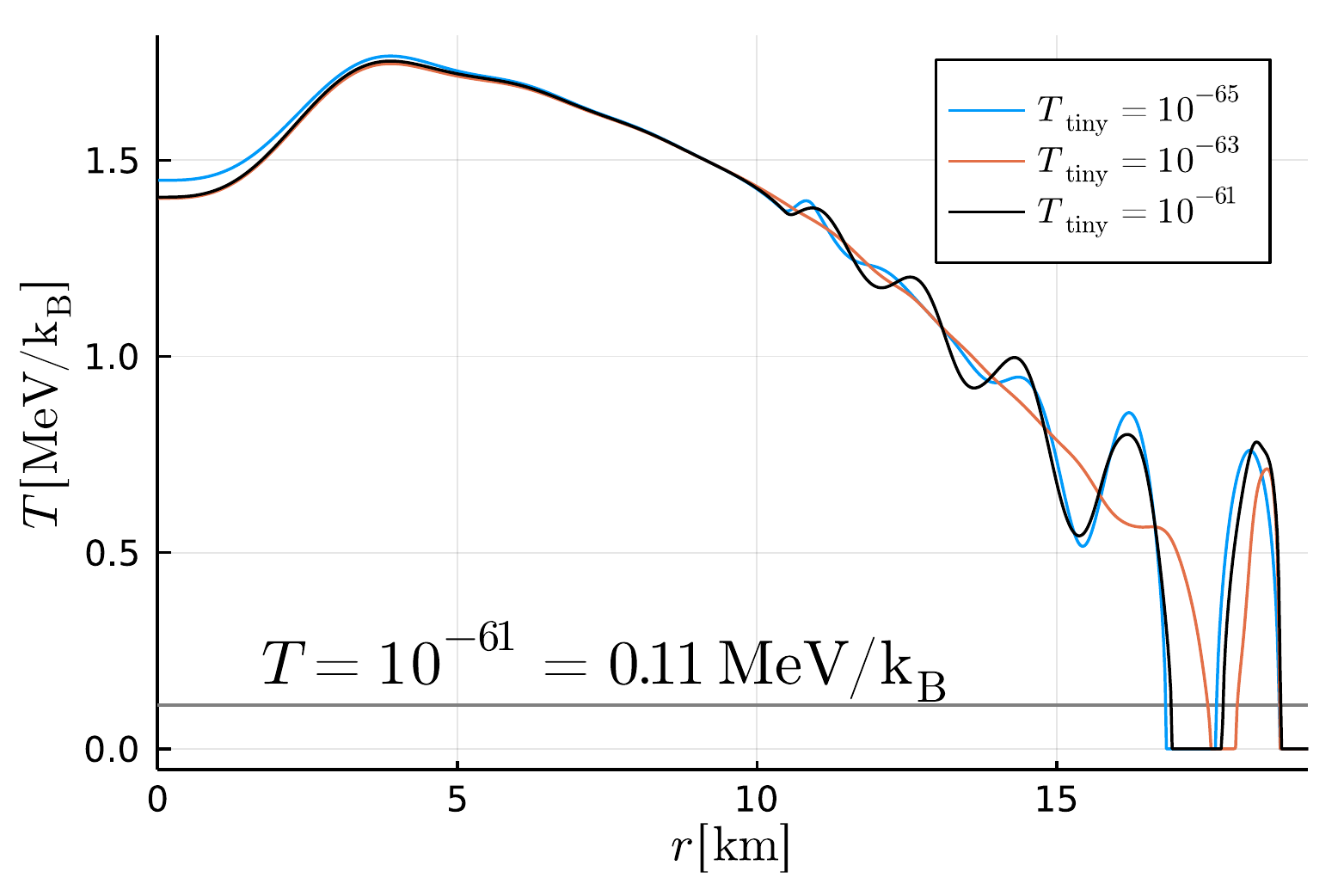}
\caption{Dependence of the simulation on the parameter $T_\mathrm{tiny}$ (direct Urca reactions).
On the top, central rest mass density from 15~ms to 20~ms.
On the bottom, thermal profile at 20~ms.}
\label{fig:tiny_t}
\end{figure}

In plots (a) and (c) of Fig.~\ref{fig:rho0} we show the central rest mass density for each model
for the oscillating case (i.e., small perturbations).  The oscillation of the central
density for the models with bulk 
viscosity (MF, MC, HL) is damped with time, while the oscillation of the non-viscous
model (PF) has no visible damping on the timescale of the simulation.
All three viscous
models evolve very similarly, which is a corroboration for the chemical
reactions-bulk viscosity correspondence \cite{Gavassino21bulk} and for the
code. However, the HL model is more noisy than MC and MF,
due to the additional derivative in the source, since close to the surface both
$\chi$ and $T^\req$ reach small values and the quantity $\log(\chi/T^\req)$ can
become problematic.
In order to check how the temperature floor $T_\mathrm{tiny}$ in the derivative in the source
of Eq.~\eqref{eq:pi-evol-hl} influences the evolution, we made some tests varying its value.
In Fig.~\ref{fig:tiny_t} we show the last 5~ms evolution of the central rest mass density (top)
and the final temperature profile (bottom) for the oscillating HL model with $T_\mathrm{tiny}=\{10^{-65},10^{-63},10^{-61}\}$.
Increasing the value of $T_\mathrm{tiny}$ reduces
the noise in the evolution of the central density because in this way the relative error
in the temperature profile is reduced too (note that the truncation error on the temperature is of the order of $10^{-66}$).
On the other hand, $T_\mathrm{tiny}$ should be small enough not to overcome the temperature profile.
We found that $T_\mathrm{tiny}=10^{-61}$ is a good compromise between reducing the noise and being able to resolve the thermal profile.

\begin{figure}
\includegraphics[width=\columnwidth]{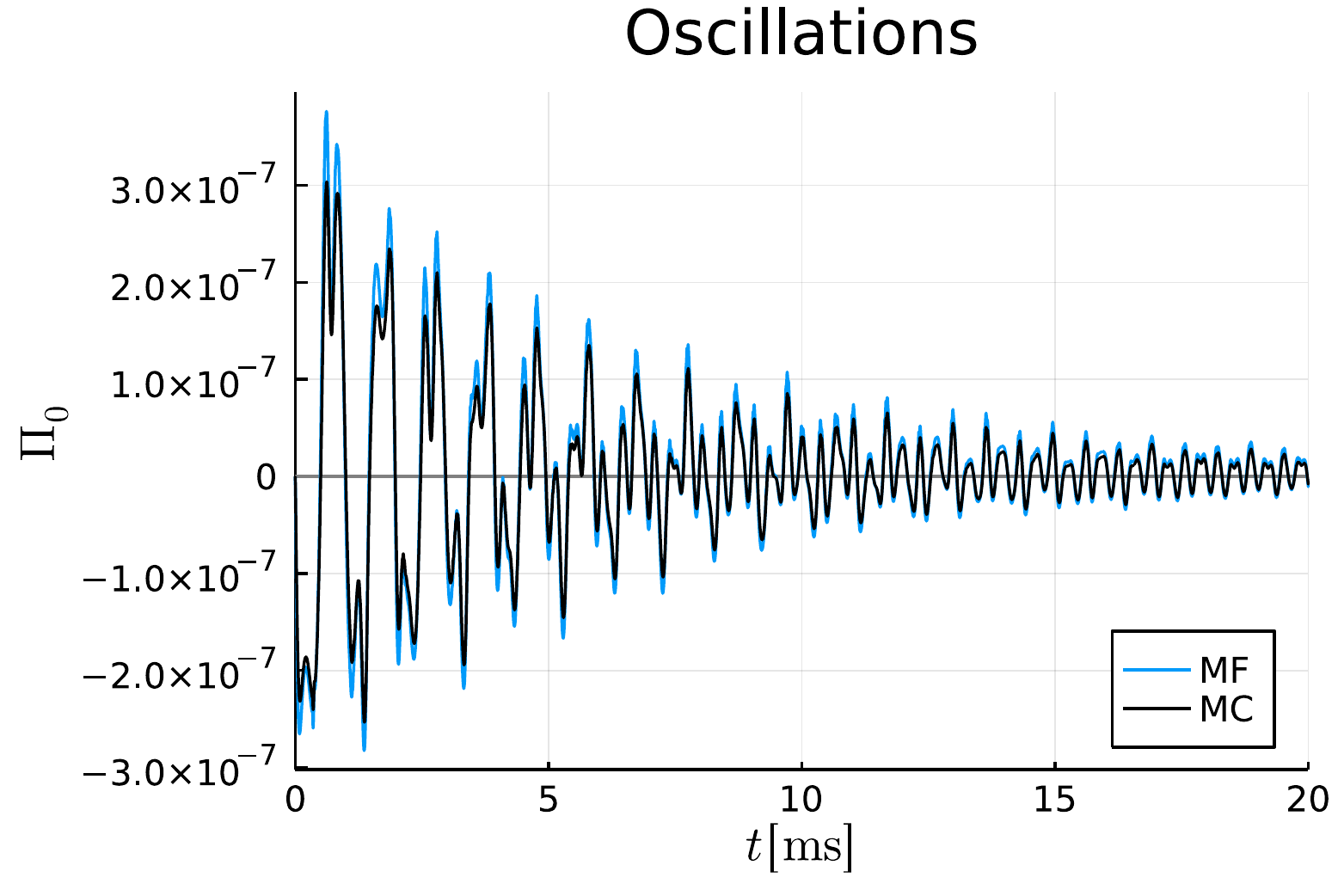}
\caption{Bulk stress $\Pi$ for the oscillating MF and MC models (direct Urca reactions).
The bulk stress for the MF model is obtained with Eq.~\eqref{eq:tau-from-tracking}.
The HL (not shown) and the MC models are very similar.}
\label{fig:osc-pi}
\end{figure}

In Fig.~\ref{fig:osc-pi} we plot the value of the bulk
stress $\Pi$ directly evolved with the MC model and that obtained with
Eq.~\eqref{eq:tau-from-tracking} from the evolution of the MF model, for the oscillation case.  The similarity of the bulk
stress obtained with two completely different models is an additional
corroboration of the correspondence between the bulk stress theories and the
multi-component fluid. 
As a comparison, the central pressure oscillates around
$p\simeq 1.5\times 10^{-3}$. In the final profile, the pressure is at least 9000
times larger than the absolute value of the bulk stress $|\Pi|$ in most of the star,
justifying the bulk stress approximation. However, close to the surface, the bulk stress
approaches the pressure and as a consequence the results here lose accuracy.

\begin{figure}
\includegraphics[width=\columnwidth]{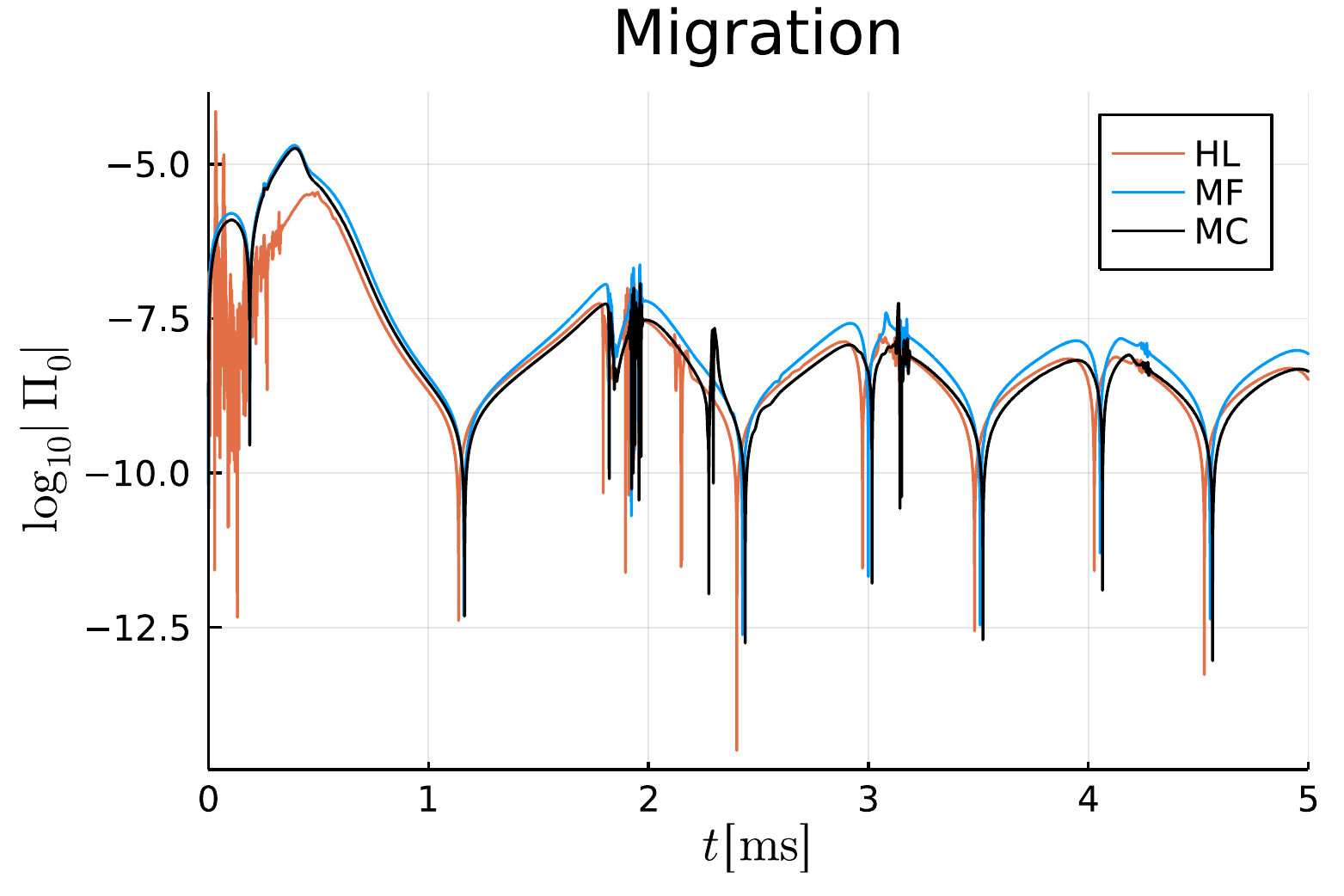}
\caption{Bulk stress $\Pi$ for the migration case (direct Urca reactions).
The bulk stress for the MF model is obtain with Eq.~\eqref{eq:tau-from-tracking}.}
\label{fig:mig-pi}
\end{figure}

For large perturbations (migration), the relative contribution due to bulk
viscosity to the dynamics is smaller, see plots (b) and (d) of Fig.~\ref{fig:rho0}. The
M\"uller-Israel-Stewart theories reproduce quite well the results of the
multi-component  fluid, but the approximation is less accurate than in the
oscillation case (as expected for larger deviations from equilibrium).
In Fig.~\ref{fig:mig-pi} we plot the absolute value of bulk stress $|\Pi|$ directly evolved
with the MC and HL models and that obtained with Eq.~\eqref{eq:tau-from-tracking}
from the evolution of the MF model, for the migration case.
As in the oscillation case, also for the migration the evolution with the Hiscock-Lindblom model
is more noisy than that with the Maxwell-Cattaneo model,
and the bulk stress derived from the
multi-component  fluid with Eq.~\eqref{eq:tau-from-tracking} is very
similar to the one evolved with the M\"uller-Israel-Stewart theories.
In the final profile, the pressure is at least 800 times larger than the
absolute value of the bulk stress $|\Pi|$ in most of the star, and hence the
bulk stress approximation is still valid.

\subsection{Switching on the luminosity}
\label{ssec:results:q}

We now include the luminosity in the evolution.
As in the previous subsection, the modified muons `x' are always in beta-equilibrium, so that we only have to evolve $Y_\re$.
In addition, the luminosity is only due to electron neutrinos.

\begin{figure*}
\includegraphics[width=\columnwidth]{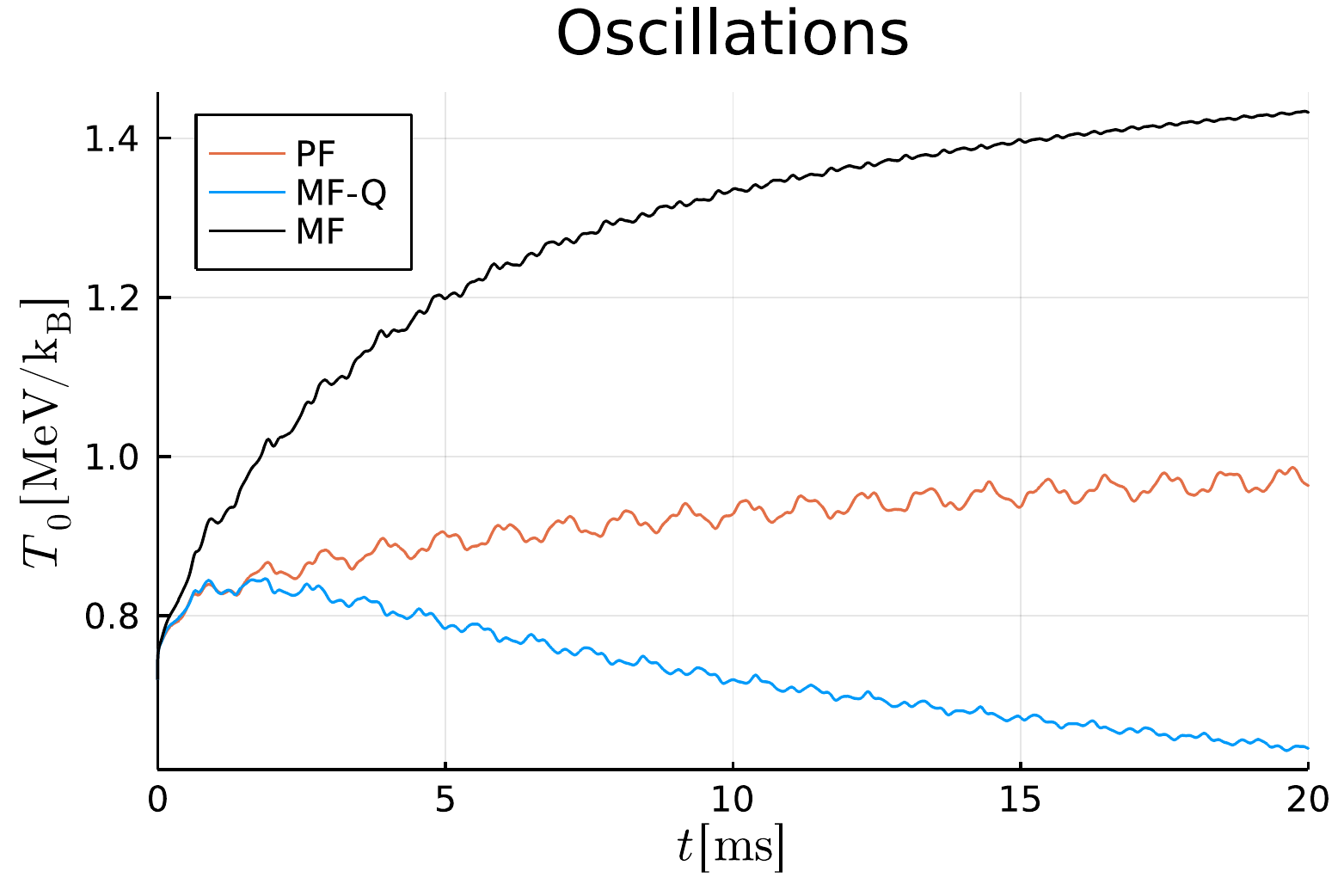}
\includegraphics[width=\columnwidth]{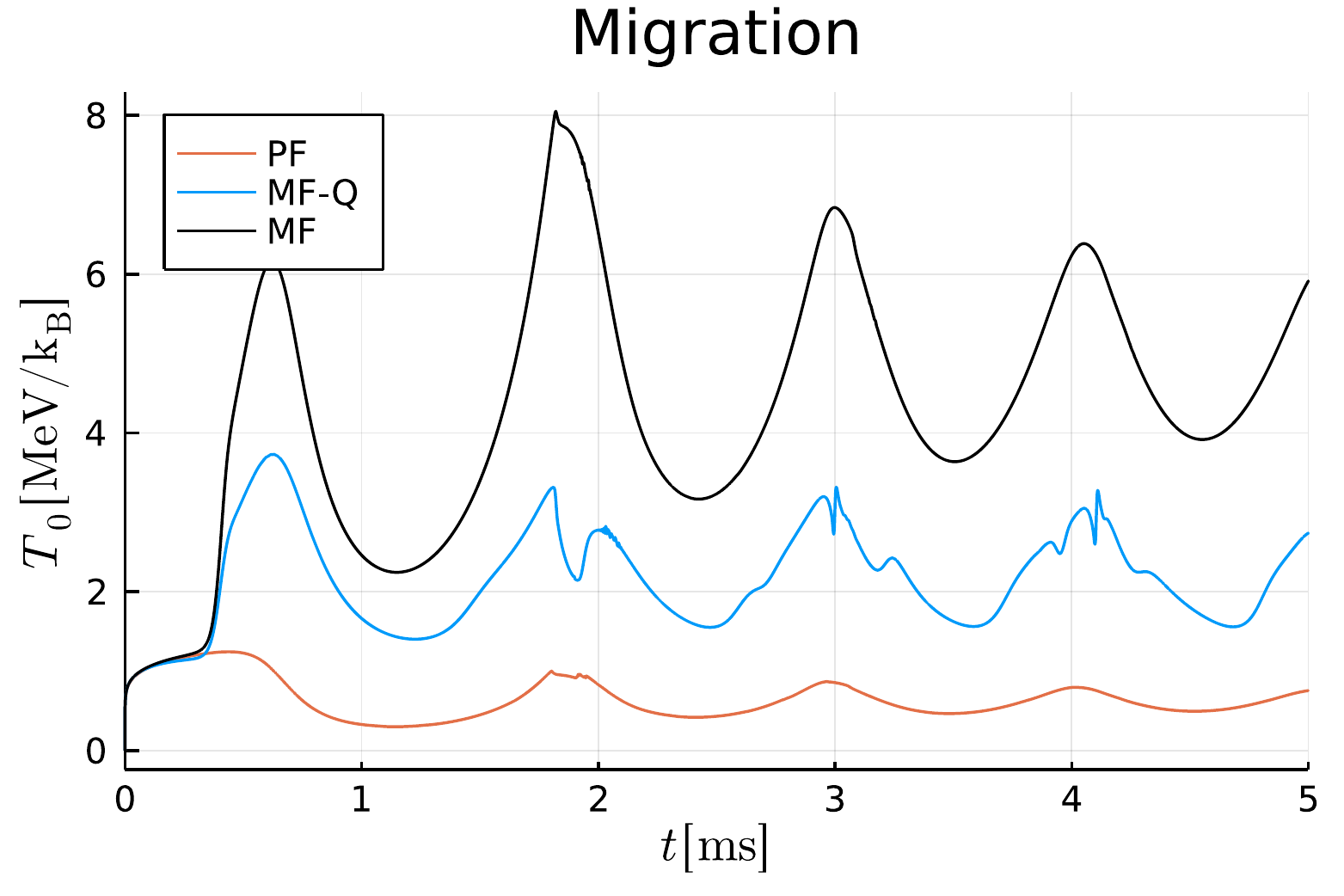}
\includegraphics[width=\columnwidth]{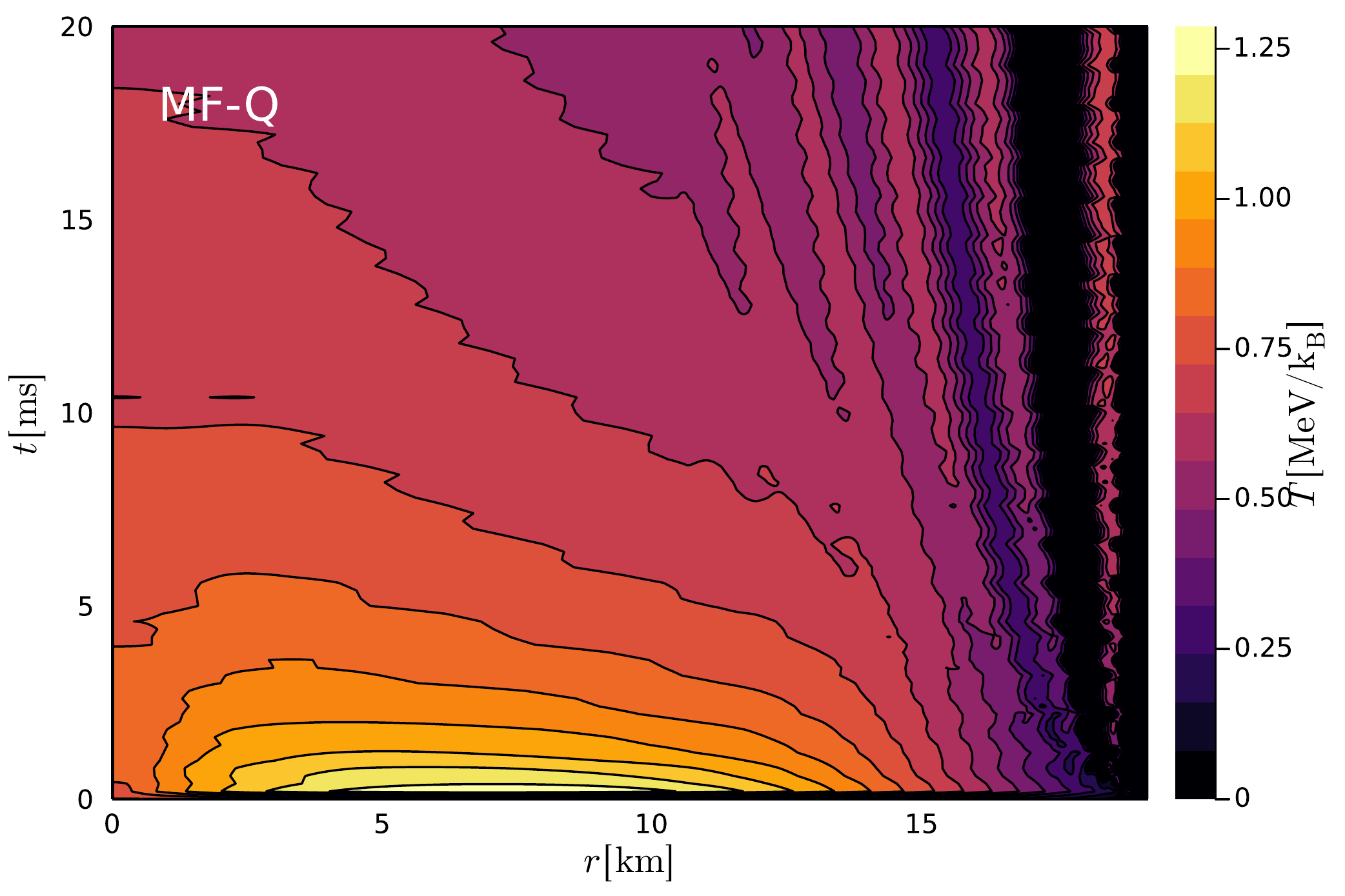}
\includegraphics[width=\columnwidth]{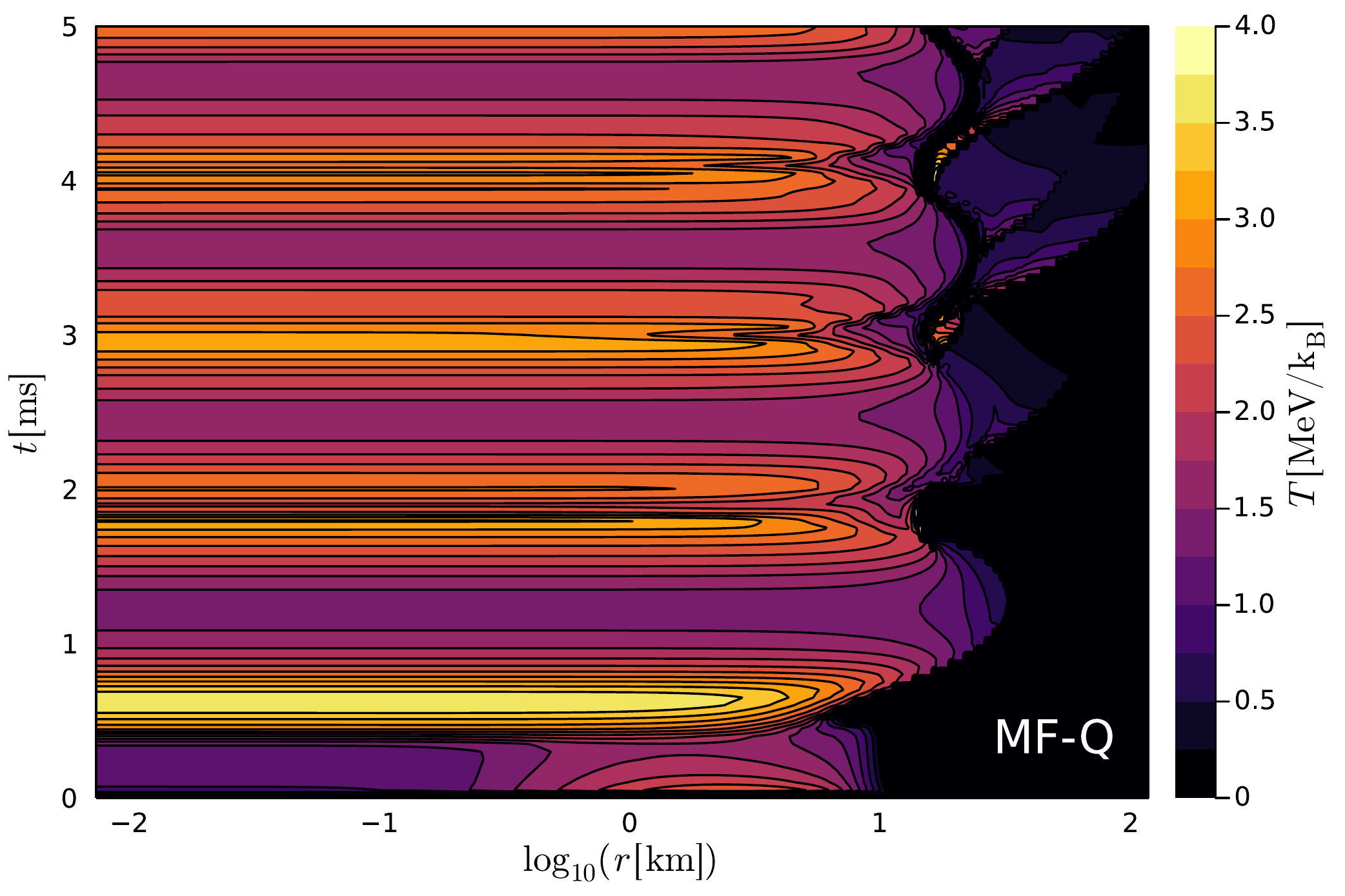}
\caption{
On the top, central temperature evolution for the PF, MF, and MF-Q models (direct Urca reactions).
On the bottom, temperature profile evolution for the MF-Q model (direct Urca reactions).
On the left, oscillation case, on the right, migration case.
}
\label{fig:temp}
\end{figure*}

As can be seen in plots (e) and (f) of Fig.~\ref{fig:rho0}, the central rest mass evolution
of the bulk stress model (MC-Q) reproduces very well the multi-component
fluid (MF-Q), both for the oscillations and the migration.
We also notice that, non-intuitively, the oscillations of the models with luminosity (MF-Q) are larger
than in those without (MF), see plots (a) and (b) of Fig.~\ref{fig:rho0}.
In order to understand this behavior, in Fig.~\ref{fig:temp} we plotted
the central temperature (top) and the time evolution of the temperature profile of the MF-Q model (bottom),
for the oscillation (left) and migration (right) cases.
The energy extraction occurring in the MF-Q model
reduces its temperature with respect to the MF model,
which in turn  implies that
also the number rates [which depend strongly on the temperature,
cf.~Eq.~\eqref{eq:R}] are reduced. As a consequence, the particle
fractions are allowed to oscillate farther from equilibrium than in the case
without luminosity, implying larger deviation from equilibrium of the other
thermodynamic quantities.

The PF model has no physical viscosity, but due to the always present numerical viscosity
there is anyway an increase in the temperature of the star, which is however smaller
than that due to the physical bulk viscosity of the MF model (cf.~top plots of Fig.~\ref{fig:temp}).
We remark that, while in a fluid in equilibrium (as in the PF model) the net number
reaction rate vanishes, there can still be luminosity due for example to
neutrino emission (see Eq.~\eqref{eq:Q}: $\mathcal Q_i\neq0$ even when $\Delta
Y_i=0$).
This is the reason why in the oscillating case the dissipating MF-Q model has a
temperature lower than that of the non-dissipating PF model: in the latter we
do not include the (physical) luminosity (which is present even at chemical
equilibrium), and therefore the heating due to the (nonphysical) numerical
viscosity cannot be dissipated.
On the other hand, the initial entropy of the oscillating MF-Q model was
chosen such that the viscosity and hydrodynamic timescales are of the same
order, and hence a more efficient emission is expected.
Compare this with the migrating case, which happens at temperatures far from
resonance between viscosity and hydrodynamics (top right plot of
Fig.~\ref{fig:temp}): as expected, the cooling due to density oscillation is
smaller than the dissipative heating, cf.~MF and MF-Q models.

Due to the many approximations made, and in particular to the use of direct
Urca instead of modified Urca reactions, the results presented should not be
taken as a quantitative description of what happens in a neutron star, but
rather as a qualitative indication of the importance of including luminosity in
the simulations.

\subsection{Two independent chemical fractions}

Finally, we relax the assumption that there is only one independent particle fraction
and we assume that there is no luminosity ($\mathcal Q=0$).
As mentioned in Sec.~\ref{sec:micro}, we modified the parameters of the muons
in the EOS in order to enhance their effects on the dynamics.  We refer to the
modified muons as particle `x'.

The difference between the evolution with independent $Y_\re,Y_\rx$ and the evolution 
with only one independent fraction ($Y_\re$) for the oscillations is shown in plot (e) of
Fig.~\ref{fig:rho0}.  We note that the addition of another degree
of freedom (the fraction $Y_\rx$) has a limited effect on the
dynamics (cf.~models MF-x, MF, and PF), probably due to the particular choice
of the parameters of the EOS.  More importantly, the bulk stress approximation
for two independent particle fractions
is less accurate in describing the bulk viscosity than for one independent particle fraction, cf.~plots (c) and (e) of
Fig.~\ref{fig:rho0}.  This is due to the fact that the bulk stress
correspondence with the multi-component fluid is valid for only one
additional independent particle fraction and small perturbations, while in the
MC-x model we are considering 2 species out of equilibrium.

For the migration, the results are similar (see plot (f) of Fig.~\ref{fig:rho0}),
but the bulk stress approximation MC-x of the multi-component fluid MF-x with 2
independent particle species is not appreciably worse than that for one
particle species out of equilibrium, cf.~with models MC and MF in plots (d)
and (f) of Fig.~\ref{fig:rho0}, probably due to the fact that the loss in accuracy due to
the larger deviations from equilibrium overcome that of considering more than
one independent particle fraction.

\subsection{Using modified Urca reactions}
\label{ssec:results:murca}

In low-density, low-temperature nuclear matter, direct Urca reactions are
kinematically suppressed and modified Urca reactions are the main channel for
dissipation.
For neutron stars in the conditions considered in this paper, direct Urca
reactions might not be the relevant rates; for this reason we performed some
simulations using the modified Urca reactions instead of the direct Urca ones.

\begin{figure*}
\includegraphics[width=\columnwidth]{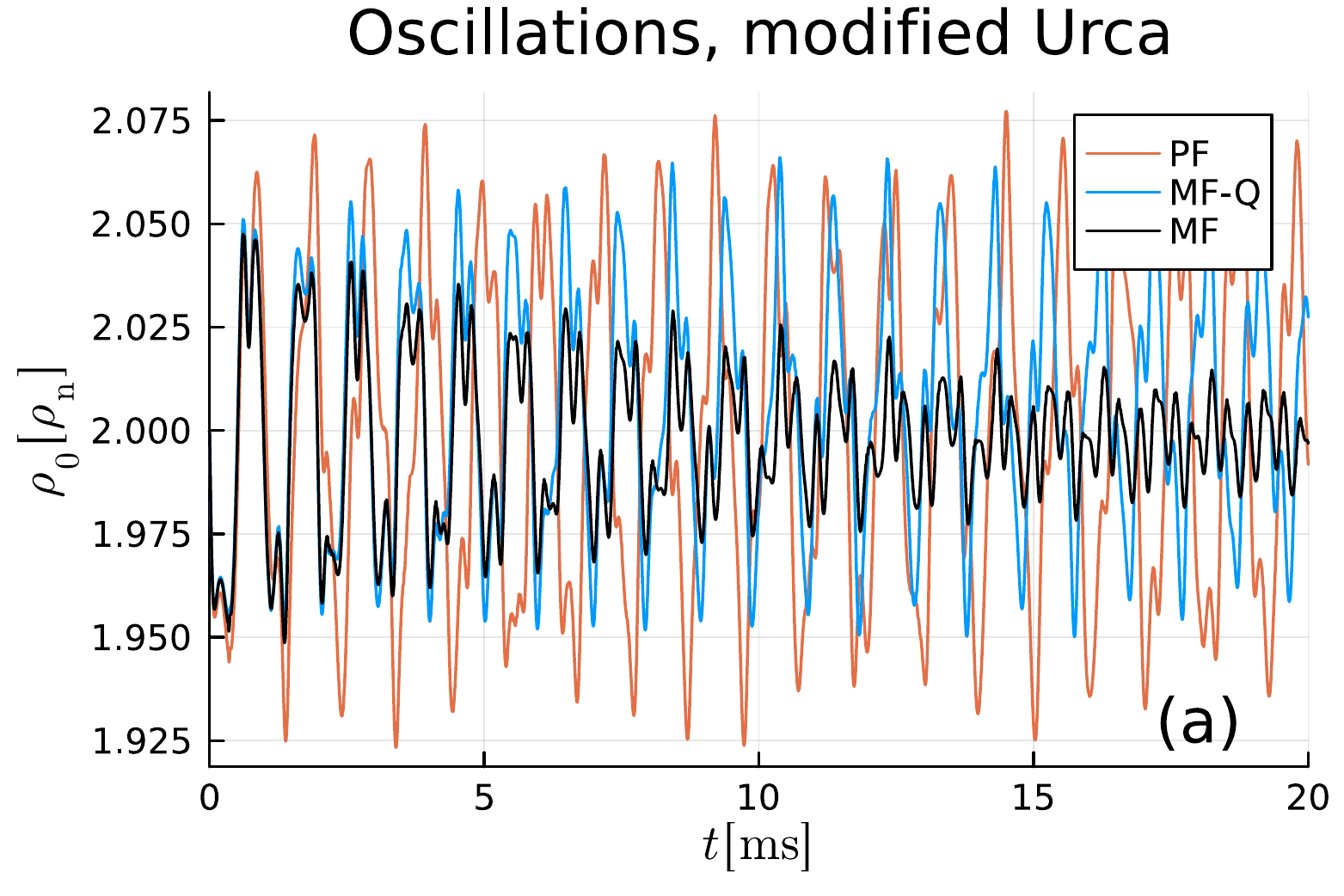}
\includegraphics[width=\columnwidth]{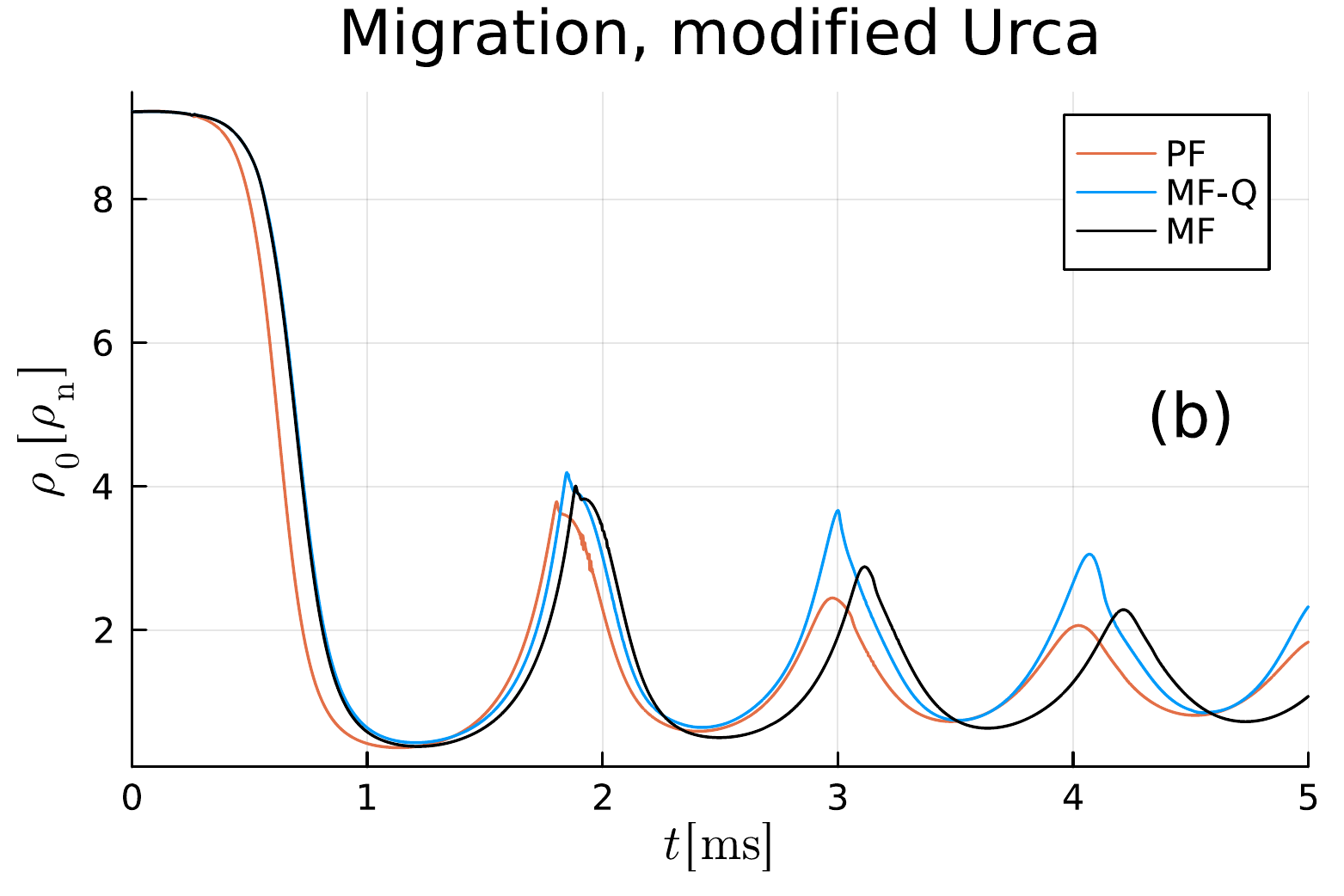}
\includegraphics[width=\columnwidth]{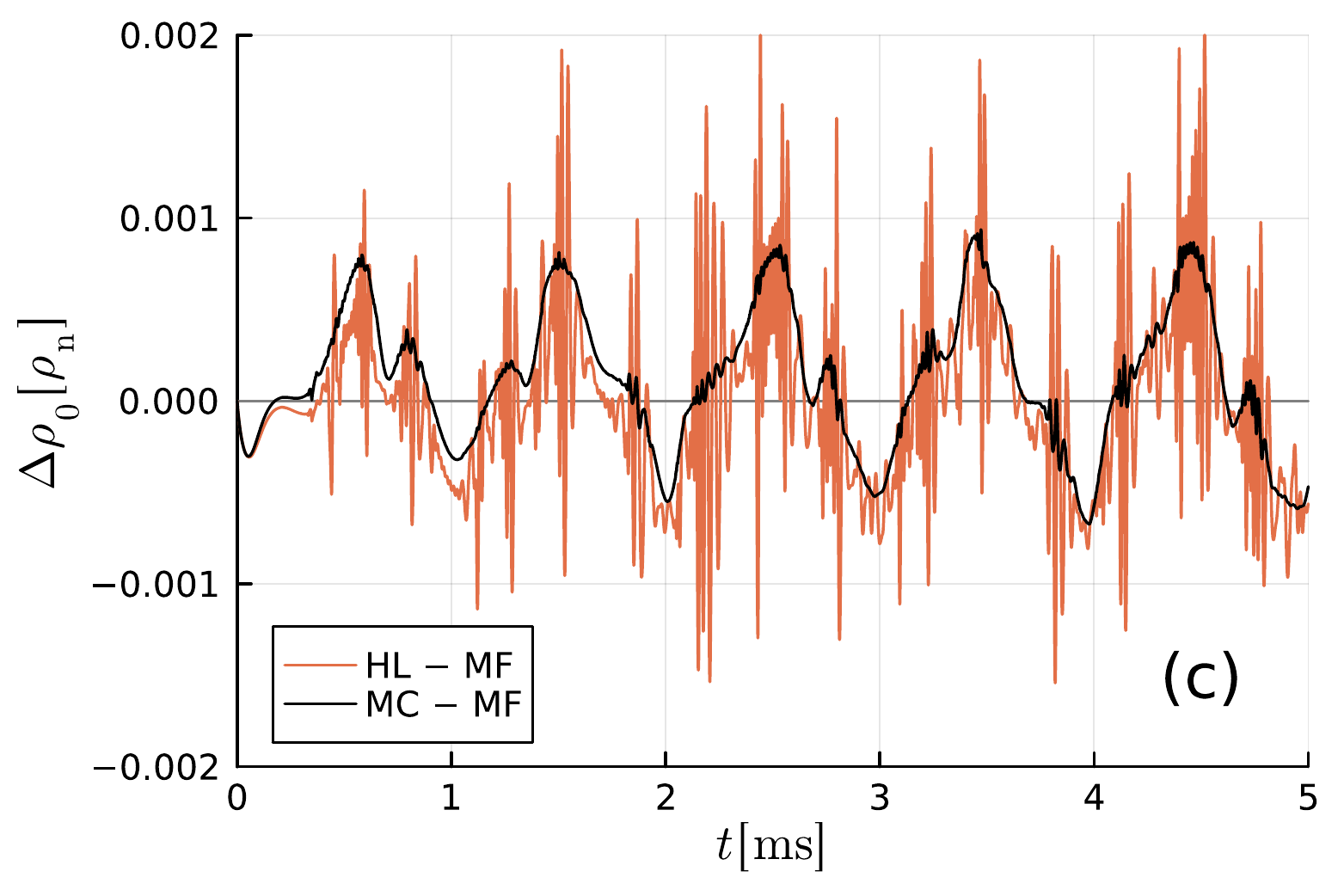}
\includegraphics[width=\columnwidth]{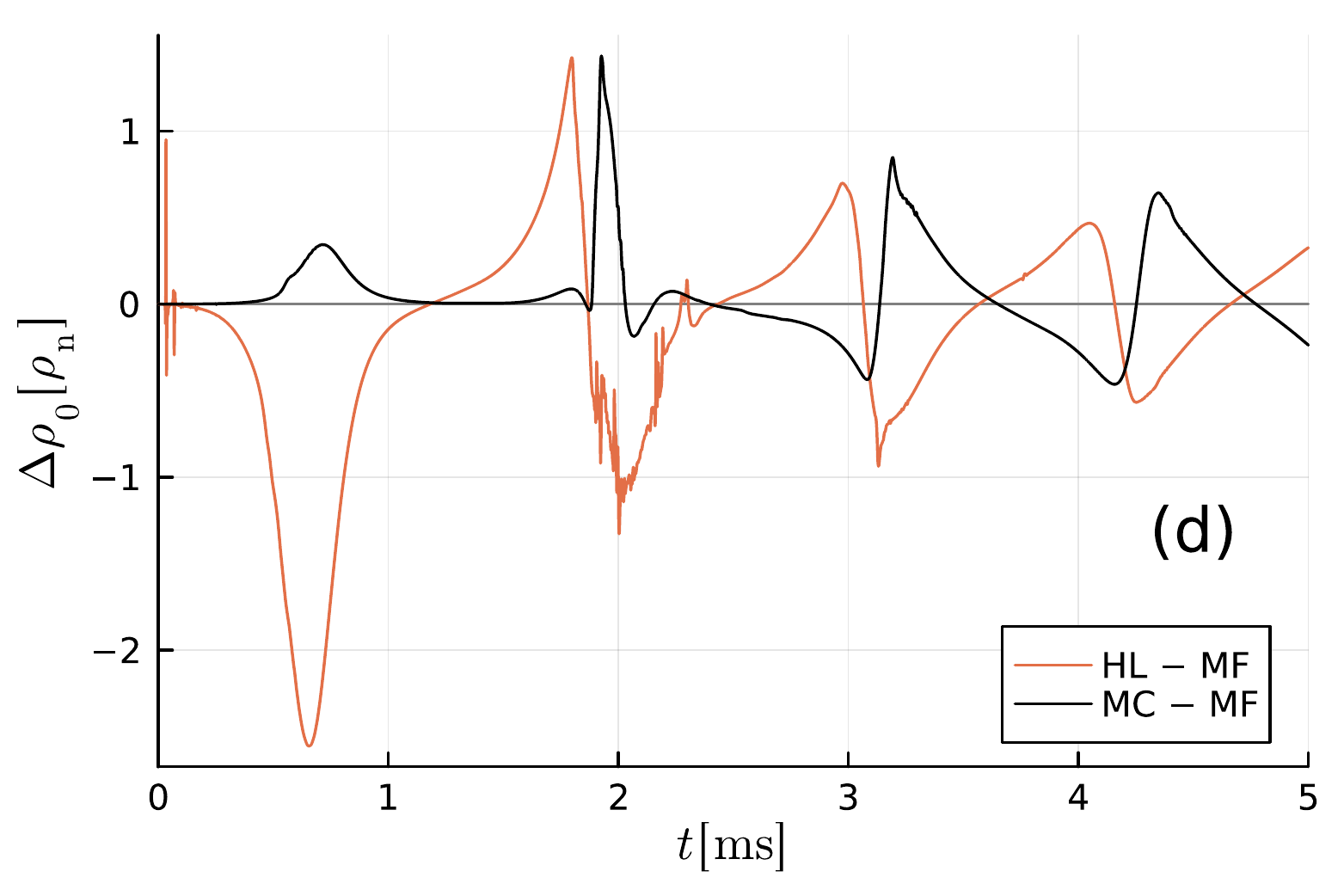}
\caption{Evolution with modified Urca reactions instead of direct Urca reactions.
On the left [Plots~(a) and (c)] we show the oscillation case, on the right [Plots~(b) and (d)] we show the migration case.
In the first row [Plots~(a) and (b)] we plot the central rest mass density for the PF, MF, and MF-Q models.
In the second row [Plots~(c) and (d)], we plot the difference of the central rest mass density between different models.
Plot~(c) shows a detail of the total evolution, up to 5~ms.
The damping timescale of the oscillations due to modified Urca reactions
is comparable to that due to direct Urca ones (cf.~Fig.~\ref{fig:rho0})
because we have artificially tuned the temperature,
cf.~Figs.~\ref{fig:temp-murca} and \ref{fig:temp} and see discussion in
Sec.~\ref{ssec:results:murca}.}
\label{fig:rho-murca}
\end{figure*}

\begin{figure}
\includegraphics[width=\columnwidth]{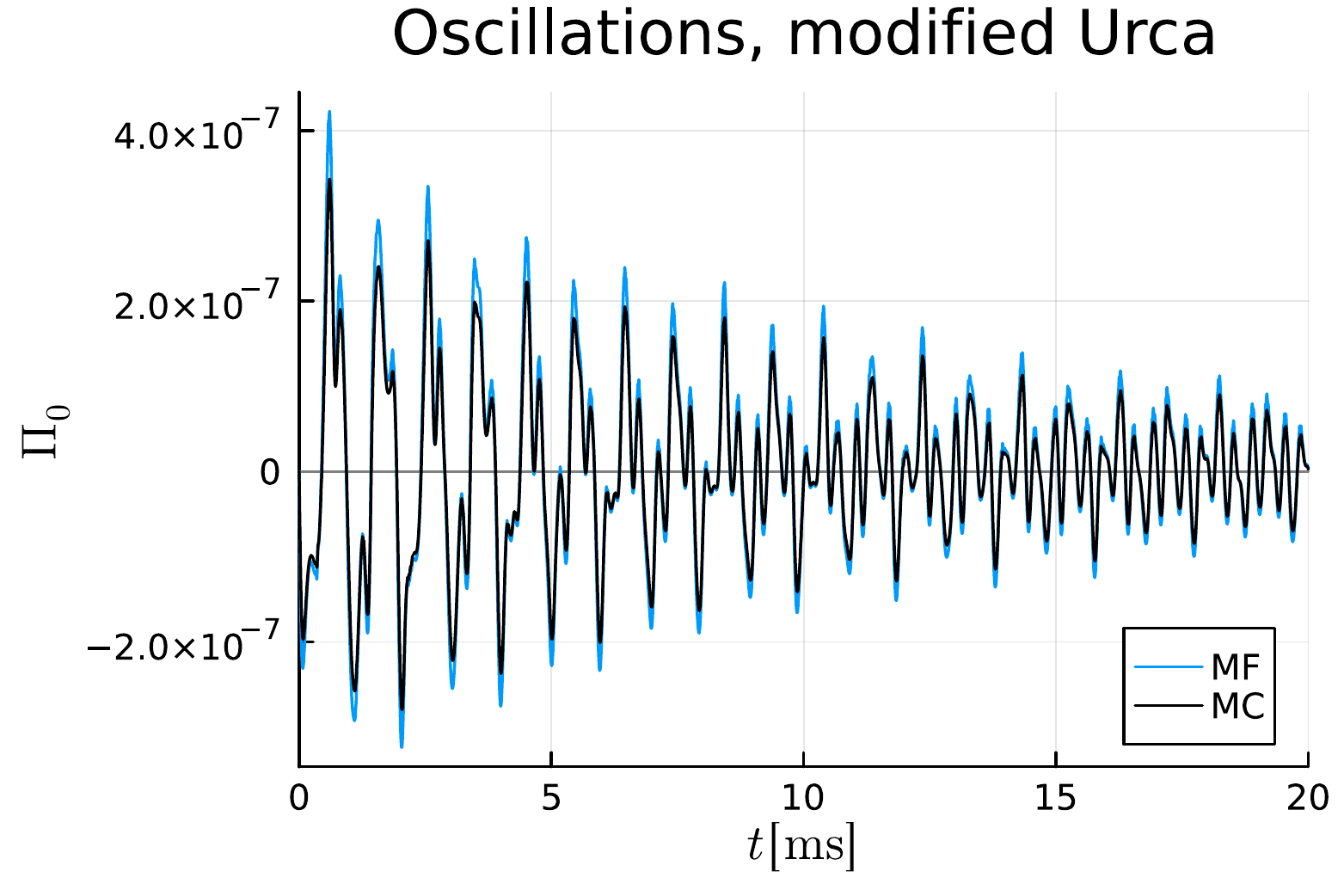}
\caption{Central bulk stress $\Pi$ for the oscillating MF and MC models
with modified Urca reactions instead of direct Urca reactions.
The bulk stress for the MF model is obtained with Eq.~\eqref{eq:tau-from-tracking}.
The HL (not shown) and MC models are very similar.}
\label{fig:pi-murca}
\end{figure}

\begin{figure}
\includegraphics[width=\columnwidth]{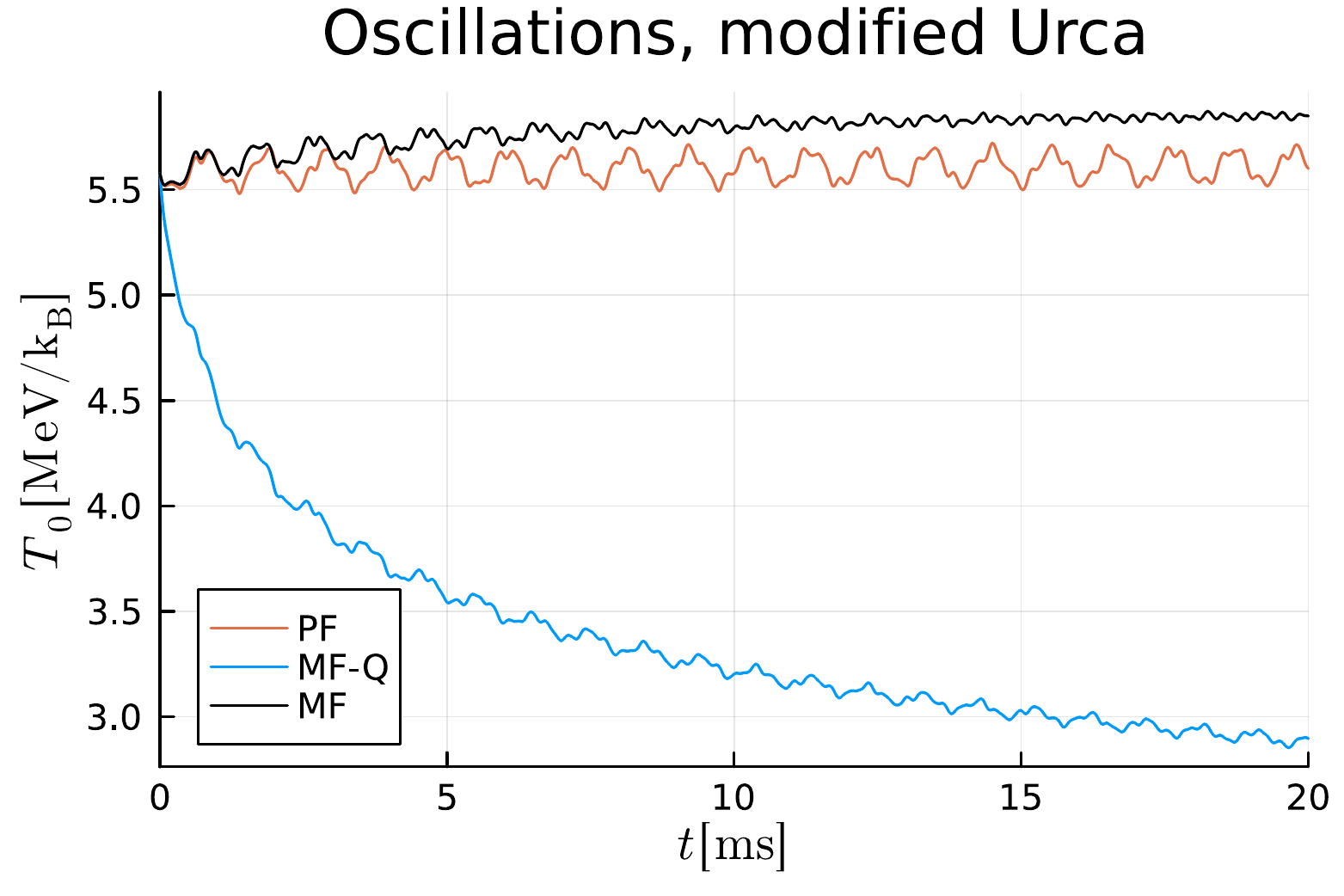}
\includegraphics[width=\columnwidth]{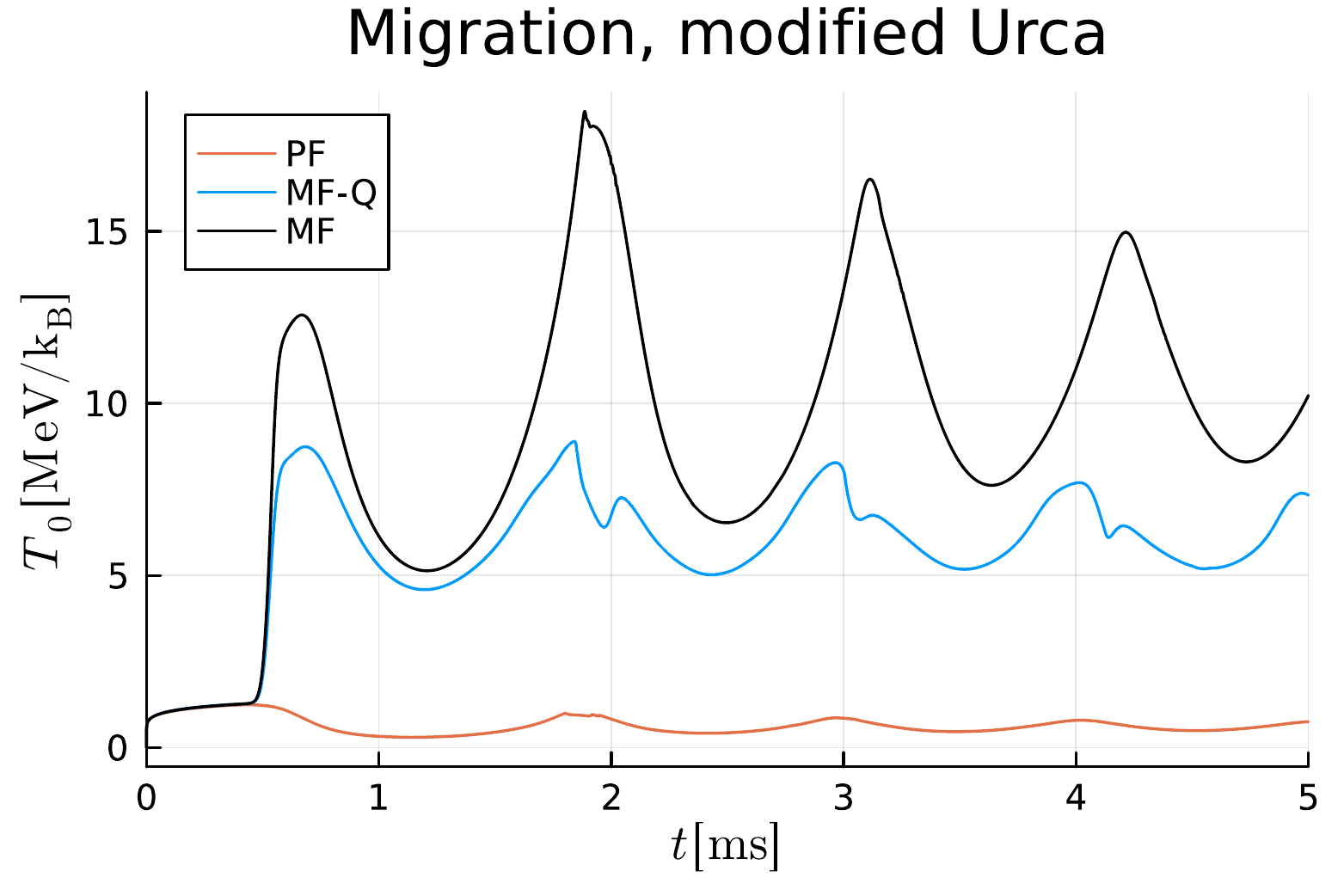}
\caption{Central temperature for the PF, MF, and MF-Q models
with modified Urca reactions instead of direct Urca reactions,
for the oscillation (top) and the migration (bottom) cases.}
\label{fig:temp-murca}
\end{figure}

A direct Urca reaction involves less particles (3 plus the escaping neutrino,
see Eqs.~(1) and (2) of \citet{Haensel92}) than a modified Urca reaction (which
involves 5 particles plus the escaping neutrino, see Eqs.~(10) and (11) of
\citet{Haensel92}).
As a consequence, direct Urca rates have a larger prefactor and depend on a
smaller power of the temperature than the modified ones,
cf.~Eqs.~\eqref{eq:R}-\eqref{eq:Q-murca}.
Since for the modified Urca reactions the prefactor is smaller, we need to set
a larger initial entropy (i.e., $s=0.31$) in the oscillation case in order
for the reaction timescale at the center to be approximately equal to the
oscillation period over $2\pi$ and hence maximize the dissipation (see
discussion in Appendix~\ref{sec:tdamp}).
Apart from this, we use the same initial configurations of the direct Urca
case.
Note that, in order to be able to compare equivalent configurations, we
performed for the oscillation case another perfect fluid (PF) simulation with
an initial uniform entropy of $s=0.31$, even if no direct and/or modified Urca
reaction is included in the PF model.
In Fig.~\ref{fig:rho-murca} we show the central rest mass density evolution for
the oscillation and migration cases,
in Fig.~\ref{fig:pi-murca} we show the central bulk stress evolution for the
oscillation case,
and in Fig.~\ref{fig:temp-murca} we show the central temperature evolution for
the oscillation and migration cases.

The perturbation in the oscillation case with modified Urca reaction is damped
on the timescale of the evolution, similarly to the direct Urca reactions case,
cf.~Plot~(a) in Figs.~\ref{fig:rho0} and \ref{fig:rho-murca} and
Figs.~\ref{fig:osc-pi} and \ref{fig:pi-murca} (see also discussion in
Appendix~\ref{sec:tdamp}).
However, this is only due to the fact that we have artificially
tuned the initial entropy in order to have the same relaxation time.
Therefore, one should keep in mind that the temperatures in the center differ
by a factor 4-5 (cf.~Figs.~\ref{fig:temp-murca} and \ref{fig:temp}).

We notice that in the oscillation case, the bulk stress approximation (MF and
HL models) is still very good (see Fig.~\ref{fig:pi-murca} and Plots~(c) of
Fig.~\ref{fig:rho-murca}), while in the migration case the approximation is
even less accurate than for the direct Urca case (compare Plots~(d) of
Figs.~\ref{fig:rho0} and \ref{fig:rho-murca}).
Moreover, the ratio between the bulk stress and the pressure in the migration
case with modified Urca reactions is larger than with direct Urca reactions,
and as a consequence the bulk stress approximation is less accurate.
This is likely due to the steeper dependence of the reaction rates on
temperature in the modified Urca case, which pushes the star farther from
equilibrium.
Note that the temperature in the migration case with modified Urca reactions is
larger than in the direct Urca case, cf.~top left plot in Fig.~\ref{fig:temp}
and bottom plot in Fig.~\ref{fig:temp-murca}.

Similarly to the simulations with direct Urca reactions, the model with
luminosity (MF-Q) with modified Urca reactions has larger oscillations (both in
the oscillation and the migration cases) than the model without (MF),
see Plots~(a) and (b) of Fig.~\ref{fig:rho-murca}.
This means that also for the case with modified Urca reactions, the main effect
of bulk viscosity on the evolution is indirect and caused by the extraction of energy
due to neutrinos leaving the star.

The temperature evolution is qualitatively similar between the direct and
modified Urca cases, cf.~top row of Fig.\ref{fig:temp} with
Fig.~\ref{fig:temp-murca}.
However, for the oscillation simulations the increase of entropy in the
modified Urca case is much smaller due to the fact that the entropy production
rate is inversely proportional to the temperature (Eq.~(8) of
\citet{Camelio22a}) and the initial temperature is larger than in the case with
direct Urca reactions (as discussed above).\\

The adopted reaction rates do not qualitatively influence the comparison between
different bulk viscosity formulations.
In other words, in the regime in which the M\"uller-Israel-Stewart
theories are valid (i.e., for small deviation from equilibrium),
both the direct and the modified Urca reaction rates are equivalent to the
multi-fluid formulation.
This is due to the fact that the formal equivalence between these formulations
does not depend on the details of the reaction rate \cite{Gavassino21bulk}.

\section{Conclusions}
\label{sec:conclusions}

In this paper we numerically implement the hydrodynamic equations derived in a
companion paper \cite{Camelio22a}.  To do so, we have developed a new
one-dimensional, general relativistic, hydrodynamic code called
\texttt{hydro-bulk-1D} \citep{Camelio22code}.  We use this code to compare different approaches to
the inclusion of bulk viscosity in the context of neutron stars, both for small
perturbations (oscillations) and large perturbations (migration from unstable
to stable branch).  In particular, we consider the bulk viscosity caused by
beta reactions and we adopt the following approaches: (i) the multi-component
fluid, namely the direct tracking of the chemical reactions and the particle
species, and two approximations based on the M\"uller-Israel-Stewart theory, namely (ii)
Hiscock-Lindblom, and (iii) Maxwell-Cattaneo, which is a linearization of
Hiscock-Lindblom.  In our knowledge, this is the first time that the
Hiscock-Lindblom formulation has been implemented in the context of neutron
stars, and the first time that M\"uller-Israel-Stewart theories are implemented in
radial gauge, polar slicing coordinates in spherical symmetry.  We evolve our
model with and without the energy loss due to the luminosity of the reactions
and with one (only electrons) or two (both electrons and `modified' muons)
species out of equilibrium. The inclusion of the luminosity in the
M\"uller-Israel-Stewart theory is a novel result of the companion paper
\cite{Camelio22a}, and this is the first implementation of such extension.
Finally, in Appendix~\ref{sec:tdamp} we derive a formula for the relativistic
linearized damping time of an oscillation valid in any regime (i.e., in the
frozen, quasi-stationary, and intermediate regimes) and discuss its
implications on our simulations.\\

We find that the correspondence between the multi-component description of the
fluid and the bulk stress one identified in \citet{Gavassino21bulk} is
accurate, both in the case of direct and modified Urca reactions.
The accuracy of the M\"uller-Israel-Stewart models decreases for
more than one independent particle species and for large perturbations (because
the mathematical correspondence is no longer rigorous), but they still
reproduce correctly the qualitative behavior of the system.  However, the
Hiscock-Lindblom model is more noisy than the Maxwell-Cattaneo one due to the
presence of an additional derivative in the source of Hiscock-Lindblom that
involves quantities that are vanishing at the surface.

Our simulations show that the largest effect of bulk viscosity in violent
(i.e., far from equilibrium, e.g.~in the migration case) neutron star dynamics
is indirect and due to the energy sink rather than to particle reactions
\textit{per~se}, because the inclusion of luminosity reduces the temperature of
the star, which in turn reduces the reaction rate, and as a consequence the
fluid is allowed to oscillate farther from equilibrium (we find this behavior
both in simulation with direct and modified Urca reactions).
In other words, it is important to correctly include reaction luminosity in the
simulations. This can be done also within the bulk stress approach, as shown in
this paper and its companion~\cite{Camelio22a}.\\

When the dominant contribution to bulk viscosity in neutron stars comes from chemical reactions, 
we find that the implementation of the multi-component fluid is more convenient than the bulk stress
models. In fact, it is directly and naturally linked to the physics of the system in consideration, but it is also
capable to accurately describe dissipation in systems with more than one independent chemical fraction.  
In addition, its numerical implementation is simpler than that of the
M\"uller-Israel-Stewart models (it has no derivatives in the source).
Moreover, contrary to the M\"uller-Israel-Stewart models that are perturbative, the chemical framework is
robust when the system is driven far from thermodynamic equilibrium \citep{Gavassino21bulk}.
Finally, numerical implementations of a multi-component fluid in the context of
neutron stars are much more common and better understood (e.g., supernova simulations)
than the ones adopting bulk stress theories.

When bulk viscosity is not directly due to a net imbalance in the direction of chemical reactions, 
we recommend the use of the Maxwell-Cattaneo model over the Hiscock-Lindblom one, since the latter has
an additional derivative in the source that worsen the numerical evolution.  
In any case, the ratio between the pressure and the bulk stress during the
evolution should be monitored, since M\"uller-Israel-Stewart theories assume the
bulk stress to be much smaller than the pressure.

An alternative approach and possible future line of research is to use the
correspondence between the multi-component fluid and the bulk
stress~\cite{Gavassino21bulk} to derive the equivalent reaction rates from
bulk stress parameters, in order to treat the case when bulk viscosity is not
due to the presence of non-equilibrium reactions, still using a multi-component fluid.
This approach is the inverse of the one adopted in this paper, and it is discussed in~\citet{Camelio22a}.

As mentioned above, the aim of this paper and its companion \cite{Camelio22a}
is to compare the multi-component fluid description and the bulk stress
approximation of bulk viscosity.  Other than that, the approximations made
allow only for a qualitative description of the neutron star evolution.  For a
more quantitative description it would be necessary to abandon spherical
symmetry, to use a tabulated EOS and accurate reaction rates (including
modified Urca and the gradual opening of the direct Urca reactions with density
and temperature), and to include trapped neutrinos, which are present at high
temperatures.

\begin{acknowledgments}
This work was supported by the Polish National Science Centre (NCN) grant number OPUS~2019/33/B/ST9/00942.
SB acknowledges support by the EU H2020 under ERC Starting Grant, no.~BinGraSp-714626.
LG is partially supported by a Vanderbilt's Seeding Success Grant.
The plots in this paper have been created using the \texttt{Julia} programming language \cite{bezanson2017julia}.
\end{acknowledgments}

\bibliographystyle{unsrtnat}
\bibliography{paper20230519.bbl}

\appendix

\section{Numerical implementation}
\label{sec:implementation}

In this appendix we discuss the \texttt{hydro-bulk-1D} code \citep{Camelio22code} in detail.

\subsection{Grid}
\label{ssec:grid}

We use a 1D, evenly spaced, staggered grid with 2 (for the piecewise linear
reconstruction) or 1 (for the piecewise constant reconstruction) ghost cells on
each side.  The position of each grid point is defined by
\begin{align}
r_i={}&\left(i-\frac12\right)\mathrm dr,\\
\mathrm dr={}&\frac{R_\mathrm{max}}{N-1},
\end{align}
where $R_\mathrm{max}$ is a parameter setting the physical dimension of the
grid, $i=1\ldots N$, and $N$ is the number of grid points (without ghost
cells).  The ghost cells are filled with symmetry conditions at the center and
by linear extrapolation of the primitive quantities at the external boundary.

\subsection{Spatial reconstruction}
\label{ssec:reconstruction}

According to the order of the spatial reconstruction,
the primitive variable $Q$ is obtained at the left and right
interfaces of the central position between grid points as:
\begin{itemize}
\item[1st:] piecewise constant:
\begin{align}
\label{eq:constant}
Q^R_{i-1/2}={}& Q_i + \mathcal O(\mathrm dr),\\
Q^L_{i+1/2}={}& Q_i + \mathcal O(\mathrm dr),
\end{align}
\item[2nd:] piecewise linear:
\begin{align}
\label{eq:linear}
Q^R_{i-1/2}={}& Q_i - \frac{r_i-r_{i-1}}2\bar Q'_i + \mathcal O(\mathrm dr^2)\\
Q^L_{i+1/2}={}& Q_i + \frac{r_{i+1}-r_i}2\bar Q'_i + \mathrm O(\mathrm dr^2)
\end{align}
where $\bar Q'_i$ is the limited slope. It is necessary to limit the slope to
avoid oscillatory solutions.  The slope limiter can be set to:
\begin{itemize}
\item MINMOD limiter:
\begin{equation}
\bar Q'_i=\begin{cases}
Q'_{-} & |Q'_{-}| \le |Q'_{+}| \mbox{ and } Q'_{-}\cdot Q'_{+} > 0,\\
Q'_{+} & |Q'_{+}| < |Q'_{-}| \mbox{ and } Q'_{-}\cdot Q'_{+} > 0,\\
0 & \mbox{otherwise},
\end{cases}
\end{equation}
\item MCLIM limiter (also called MC limiter) \cite{vanLeer77}:
\begin{equation}
\bar Q_i=\begin{cases}
\min(|Q'|,|2Q'_{-}|,|2Q'_{+}|)\sgn(Q') & Q'_{-}\cdot Q'_{+}>0,\\
0 & \mbox{otherwise},
\end{cases}
\end{equation}
\end{itemize}
where
\begin{align}
Q'={}&\frac{Q_{i+1}-Q_{i-1}}{r_{i+1}-r_{i-1}},\\
Q'_{-}={}&\frac{Q_i - Q_{i-1}}{r_i - r_{i-1}},\\
Q'_{+}={}&\frac{Q_{i+1} - Q_i}{r_{i+1} - r_i}.
\end{align}
\end{itemize}

The reconstructed state is used to solve the Riemann problem and to compute the
spatial derivatives in the source of Eqs.~\eqref{eq:pi-evol-hl} and
\eqref{eq:pi-evol}.

\subsection{Riemann solver}
\label{ssec:riemann}

The code can use either the HLL or LLF approximate Riemann solvers,
which compute the fluxes at the grid interfaces.
The HLL Riemann solver is the following:
\begin{equation}
\label{eq:hll}
F_{i-1/2}=\begin{cases}F_L&\lambda_- > 0,\\
\frac{\lambda_+F_L - \lambda_-F_R +\lambda_+\lambda_-(U_R-U_L)}{\lambda_+ - \lambda_-}&\lambda_-<0<\lambda_+,\\
F_R&\lambda_+<0,\end{cases}
\end{equation}
where $F$ are the fluxes, $U$ the conservative variables, the subscripts
$L,R$ refer to the left and right state, respectively, at the center of the
cell (namely at $i-1/2$), and the signal propagation speeds are:
\begin{align}
\lambda^+={}&\max\left(0,\frac{v_L + c_L}{1 + v_Lc_L},\frac{v_R + c_R}{1 + v_Rc_R}\right),\\
\lambda^-={}&\min\left(0,\frac{v_L - c_L}{1 - v_Lc_L},\frac{v_R - c_R}{1 - v_Rc_R}\right),
\end{align}
where $c$ is the `ultraviolet' speed of sound [Eqs.~\eqref{eq:cs-uv} and \eqref{eq:effective-cs2}].

The LLF Riemann solver is:
\begin{equation}
F_{i-1/2}=\frac12\Big(F_L+F_R-\max(|\lambda^+|, |\lambda^-|)(U_R - U_L)\Big).
\end{equation}

\subsection{Time evolution}
\label{ssec:runge-kutta}

The hydrodynamic equations can be written as:
\begin{equation}
\label{eq:evolution}
\frac{\mathrm dU}{\mathrm dt}= \mathcal E(U,t) + \mathcal I(U,t),
\end{equation}
where $\mathcal E(U,t)$ includes the contribution from the fluxes and the explicit sources,
while $\mathcal I(U,t)$ includes the contribution from the implicit sources.

The timestep $\mathrm dt$ is determined by:
\begin{equation}
\label{eq:courant}
\mathrm dt= \mathrm{CFL}\times\min_i\big(\mathrm dr_i/\max(|v_i|,|\lambda^+_i|,|\lambda^-_i|)\big),
\end{equation}
where CFL is the Courant-Friedrichs-Lewy factor.

To numerically integrate Eq.~\eqref{eq:evolution} at a given order, we use 
IMplicit-EXplicit Runge Kutta (IMEX~RK):

\begin{align}
\label{eq:imex1}
U^i=U(t)+{}&\mathrm dt\sum_{j=1}^{i-1} a_{ij}^{\mathcal E} \mathcal E(U^j,t + c_j^{\mathcal E} \mathrm dt) \notag\\
        +{}&\mathrm dt\sum_{j=1}^i a_{ij}^{\mathcal I} \mathcal I(U^j,t + c_j^{\mathcal I} \mathrm dt),\\
\label{eq:imex2}
U(t+\mathrm dt)=U(t)+{}&\mathrm dt\sum_{j=1}^n b_j^{\mathcal E} \mathcal E(U^j,t + c_j^{\mathcal E} \mathrm dt) \notag\\
                    +{}&\mathrm dt\sum_{j=1}^n b_j^{\mathcal I} \mathcal I(U^j,t + c_j^{\mathcal I} \mathrm dt),
\end{align}
where $a_{ij}, b_j, c_j$ are parameters given by the Butcher tableau, see Table~\ref{tab:tableau}.
\begin{table}
\caption{General form of the Butcher tableau.}
\begin{tabular}{c|c}
$c^\mathcal{E}_j$ & $a^\mathcal{E}_{ij}$ \\
\hline
& $b^\mathcal{E}_j$
\end{tabular}
\hspace{1cm}
\begin{tabular}{c|c}
$c^\mathcal{I}_j$ & $a^\mathcal{I}_{ij}$ \\
\hline
& $b^\mathcal{I}_j$
\end{tabular}
\label{tab:tableau}
\end{table}
We report the adopted parameters of the 2nd order IMEX~RK in Table~\ref{tab:2nd} and the 3rd order IMEX~RK in Table~\ref{tab:3rd}.
\begin{table}
\caption{Butcher tableau for IMEX~RK at 2nd order,
with $\gamma= 1 - 1/\sqrt 2$. Compare with Table~2 of \citet{Pareschi10}.}
\begin{tabular}{c|cc}
0 & 0 & 0 \\
1 & 1 & 0 \\
\hline
& 1/2 & 1/2
\end{tabular}
\hspace{1cm}
\begin{tabular}{c|cc}
$\gamma$ & $\gamma$ & 0 \\
$1-\gamma$ & $1-2\gamma$ & $\gamma$ \\
\hline
& 1/2 & 1/2
\end{tabular}
\label{tab:2nd}
\end{table}
\begin{table}
\caption{Butcher tableau for IMEX~RK at 3nd order,
with $\gamma= 1 - 1/\sqrt 2$. Compare with Table~5 of \citet{Pareschi10}.}
\begin{tabular}{c|ccc}
0 & 0 & 0 & 0 \\
1 & 1 & 0 & 0 \\
1/2 & 1/4 & 1/4 & 0 \\
\hline
& 1/6 & 1/6 & 2/3
\end{tabular}
\hspace{1cm}
\begin{tabular}{c|ccc}
$\gamma$ & $\gamma$ & 0 & 0 \\
$1-\gamma$ & $1-2\gamma$ & $\gamma$ & 0\\
1/2 & $1/2-\gamma$ & 0 & $\gamma$ \\
\hline
& 1/6 & 1/6 & 2/3
\end{tabular}
\label{tab:3rd}
\end{table}
When first order integration in time is set, we use an implicit-explicit Euler:
\begin{align}
U(t+\mathrm dt)= U(t) +{}& \mathrm dt \mathcal E(U(t),t)\notag\\
+{}& \mathrm dt \mathcal I(U(t+\mathrm dt),t+\mathrm dt).
\end{align}

\subsection{Conservative-to-primitive inversion}
\label{ssec:c2p}

We choose the quantities $\rho,Wv,u$ as primitive variables, with the addition of
$\{Y_i\}_i$ in the case of the multi-component fluid and of $\Pi$ in the case of M\"uller-Israel-Stewart theories.
For each grid point, we determine the value of the function $X$ from the conservative
variables, noting that the integrand of Eq.~\eqref{eq:m} is equal to $\tau+D$.
We then use the Brent method to find the root of the function
\begin{equation}
\label{eq:root-function}
f(z)= S^r - \big(\hat\epsilon + \hat p\big)z\hat W,
\end{equation}
where $z$ is the independent variable and
\begin{align}
\hat W={}& \sqrt{1 + z^2},\\
\hat\rho={}& \max\left(\rho_\mathrm{min},\frac{D}{X\hat W}\right),\\
\hat\epsilon={}& \max\left(\epsilon_\mathrm{min}, \tau + D - \frac{S^rz}{\hat W}\right),\\
\label{eq:root-p}
\hat p={}&\begin{cases}
p(\hat\rho,\hat \epsilon) & \mbox{PF},\\
p(\hat\rho,\hat \epsilon, \{\hat Y_i\}_i) & \mbox{MF},\\
p^\req(\hat\rho,\hat \epsilon) + \hat \Pi & \mbox{HL and MC},
\end{cases}
\end{align}
where $\hat Y_i$ and $\hat \Pi$ are\footnote{We also enforce $0\le\hat Y_i\le1$ and
$-p^\req\le\hat \Pi\le p^\req$.} computed implicitly in the conservative-to-primitive
inversion.
At the root of $f$, the quantities in
Eqs.~\eqref{eq:root-function}--\eqref{eq:root-p} correspond to the primitives,
namely we can drop the hat, and $z=Wv$.

The minimal rest mass density is set to $\rho_\mathrm{min}=10^{-20}$ and the
minimal energy density $\epsilon_\mathrm{min}$ is obtained from
$\rho_\mathrm{min}$ and the cold EOS.
Moreover, when the rest mass density obtained from the conservative-to-primitive
inversion or from the reconstruction is smaller than the threshold value
$\rho_\mathrm{thr}=100\rho_\mathrm{min}$, we set the primitive variables to their
minimal values $\rho=\rho_\mathrm{min},Wv=0,u=\epsilon_\mathrm{min}/\rho_\mathrm{min}-1,Y_i=0,\Pi=0$
(i.e., we set them to their atmosphere values).

The time derivatives in the source of Eqs.~\eqref{eq:pi-evol-hl} and
\eqref{eq:pi-evol} are performed implicitly by including the advanced time
quantity in the conservative-to-primitive inversion\footnote{Note that
\citet{Chabanov_2021} performed an equivalent time derivative in the source
term with a two-step predictor-corrector time evolution (Eq.~(106) of
\citet{Chabanov_2021}).}.
We computed implicitly also the terms containing the number reaction rates $\mathcal R_i$ in
Eq.~\eqref{eq:continuity-yi} and the viscous timescale $\tau$ and the bulk
stress $\Pi$ in the right-hand side of Eqs.~\eqref{eq:pi-evol-hl} and
\eqref{eq:pi-evol}.
Since the implicit terms are linear, a simple
inversion is enough to recover the conservative variables when solving for
Eq.~\eqref{eq:imex1}, while the implicit term is already known when solving for
Eq.~\eqref{eq:imex2}.  The spatial derivatives in the source of
Eqs.~\eqref{eq:pi-evol-hl} and \eqref{eq:pi-evol} are performed explicitly
(i.e., at the previous timestep), and are limited with the slope limiter.  In
order to avoid division by zero, we set a floor for the temperature at
$T_\mathrm{min}=10^{-61}$ (in code units) in the derivative in the source of
the Hiscock-Lindblom equation [Eq.~\eqref{eq:pi-evol-hl}], while $\chi$ is
already limited by our choice of $\rho_\mathrm{min}$.

\section{Tests}
\label{sec:tests}

In this section we consider a set of standard tests for hydrodynamic codes in
special and general relativity. Unless otherwise specified, we use the same
code settings described in Sec.~\ref{sec:code}.
In order to make a quantitative evaluation, we define the residual $\Delta f$
of a quantity $f$ as
\begin{equation}
\label{eq:residual}
\Delta f=\sum_{i=1}^N\frac{\left|f_i - \bar f_i\right|}N,
\end{equation}
where the index $i$ identifies the point in the grid of dimension $N$ and $\bar
f_i$ is either the exact value or the value at the highest available
resolution, according to the test.

In the following, we use 3rd order IMEX-RK with $\mathrm{CFL}=0.5$,
2nd order spatial reconstruction with MINMOD limiter, and the LLF Riemann
solver.

\subsection{Special-relativistic shocktube}
\label{ssec:shocktube}

The shocktube is a one dimensional problem where two different thermodynamic
states of a fluid are initially separated by a wall, which is lifted at $t=0$.
The hydrodynamic equations are:
\begin{align}
\label{eq:continuity-rho-cart}
\partial_t D
+\partial_x\left(Dv\right)={}&0,\\
\label{eq:conservation-momentum-cart}
\partial_t S^x
+\partial_x\left(S^xv + p\right)={}& 0,\\
\label{eq:conservation-energy-cart}
\partial_t \tau_\epsilon
+\partial_x\left(S^x - Dv\right)={}& 0,
\end{align}
where $D=\rho W,S^x=(\epsilon+p)W^2v$, and $\tau_\epsilon=(\epsilon+p)W^2-p-D$.

\begin{figure}
\includegraphics[width=\columnwidth]{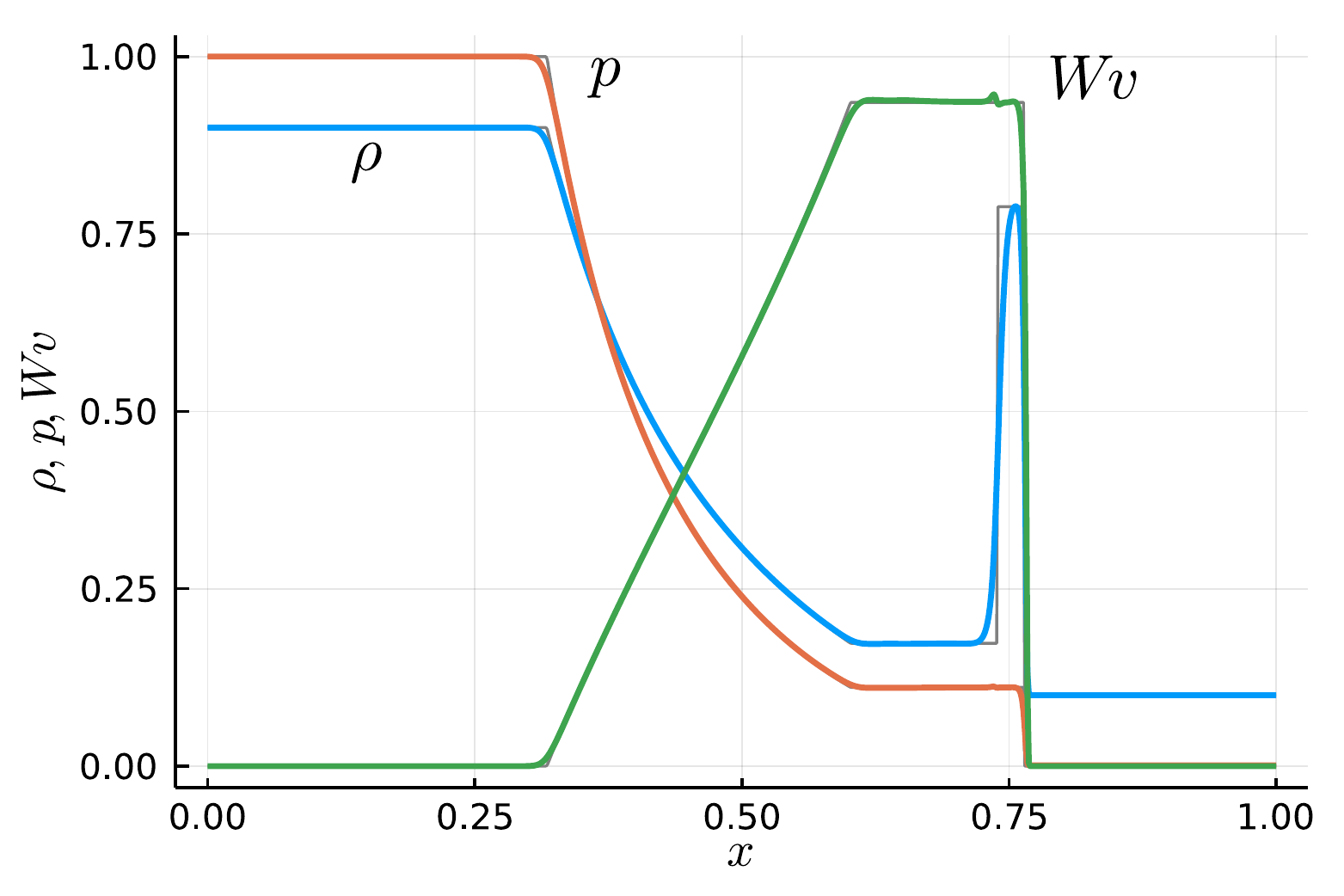}
\includegraphics[width=\columnwidth]{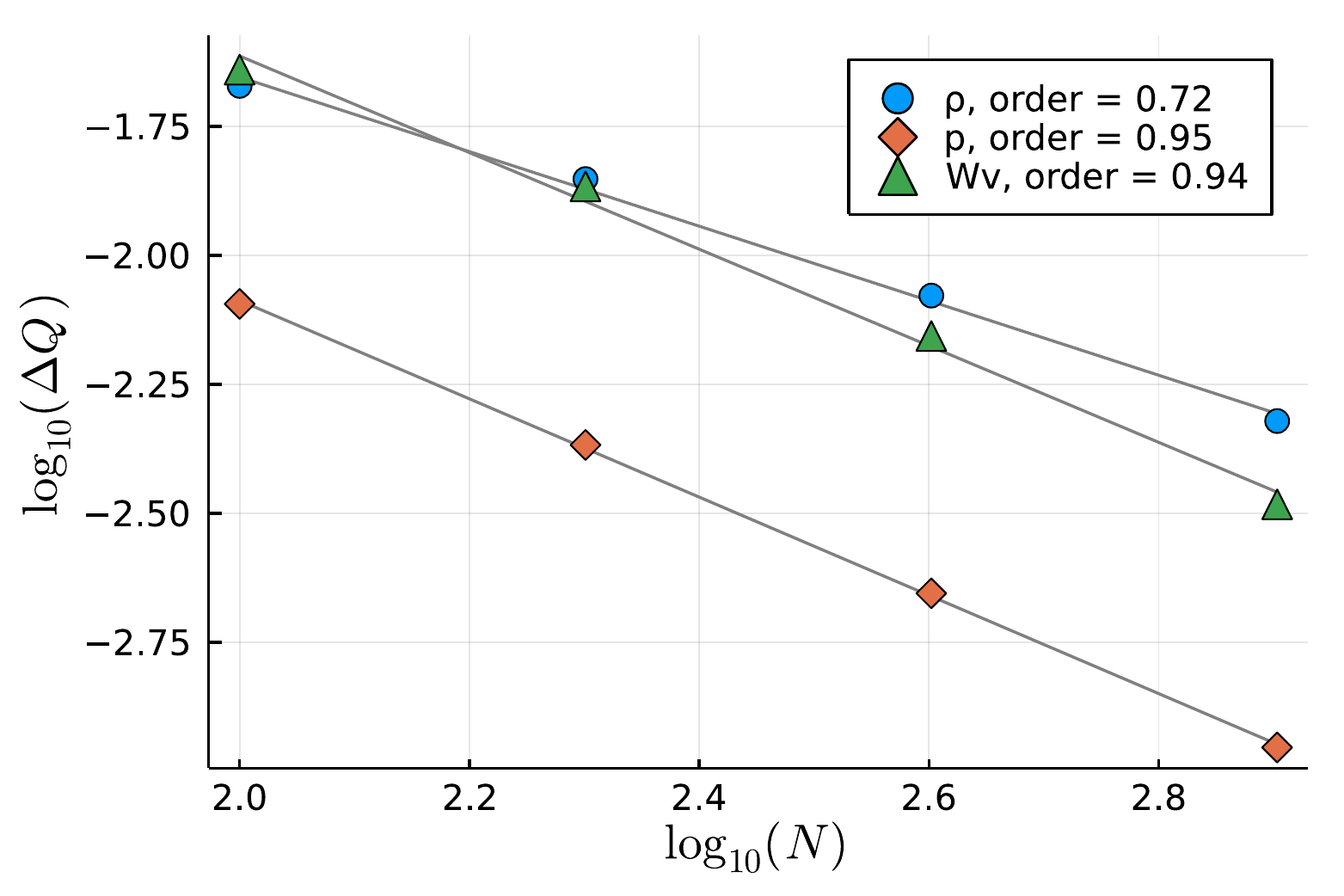}
\caption{Result of the relativistic shocktube test at $t=0.35$.  On the top, we
plot the rest mass density $\rho$ (light blue), pressure $p$ (orange), and
the Lorentz factor times the velocity $Wv$ (green), for the grid size $N=800$, against the exact solution (gray).  On the bottom, we
plot the residual of the same quantities with respect to the exact solution as a function of the grid size and
report the convergence order in the legend for each of them.}
\label{fig:shocktube}
\end{figure}

We use an evenly spaced Cartesian grid from $x=0$ to $x=1$ with varying size
$N=\{100,200,400,800\}$, constant values of the primitives in the ghost cells, and the ideal EOS:
\begin{equation}
\label{eq:eos-ideal}
p= (\Gamma - 1)\rho u,
\end{equation}
with $\Gamma=4/3$.  The fluid is initialized in a rarefaction-shock
configuration: $\rho(x<0.5)=0.9$, $\rho(x>0.5)=0.1$, $p(x<0.5)=1$,
$p(x>0.5)=0.001$, and $v=0$.
In Fig.~\ref{fig:shocktube} we plot the result of the test at $t=0.35$.  The
residuals with respect to the exact solution  decrease increasing the grid
size, as expected, which proves the convergence of the code. The order of
convergence is smaller than the nominal order of our code (which is second
order), because of the presence of the shock.

\subsection{Bjorken flow}
\label{ssec:bjorken}

The Bjorken flow describes the quark-gluon plasma in the center-of-mass
of an ultra-relativistic heavy-ion collision \citep{Bjorken83}.
This is a standard test for viscous codes describing heavy-ion collisions \cite{DelZanna13},
and it has been used also in an astrophysical context by \citet{Chabanov_2021}.
Only in this section, we use the Milne coordinates:
\begin{equation}
\label{eq:milne}
\mathrm dl^2= - \mathrm d t^2 + t^2\mathrm dx^2,
\end{equation}
where $\tilde t= t\cosh x$ and $\tilde x=t\sinh x$ are the
time and spatial coordinates in the Minkowski spacetime.
The determinant of the metric is $\sqrt{-g}= t$ and the only nonzero connection coefficients are 
$\Gamma^{t}_{xx}=t$ and
$\Gamma^{x}_{xt}=\Gamma^{x}_{tx}= t^{-1}$.
In the boosted system the velocity is null and therefore we have $v=0$, $W=1$, $u^\mu=(1,0)$, and
$T^{\mu\nu}=\mathrm{diag}\big(\epsilon,(p^\req+\Pi)/t^2\big)$.
Also, there are no spatial fluxes;
the hydrodynamic equations therefore read (see also Sec.~3.4 of \citet{DelZanna13}):
\begin{itemize}
\item continuity equation:
\begin{equation}
\partial_t(\rho t)=0,
\end{equation}
\item since there are no fluxes the momentum conservation equation is trivial,
\item energy conservation equation:
\begin{equation}
\partial_t\epsilon=-\frac{\epsilon+p^\req(\rho,\epsilon)+\Pi}t,
\end{equation}
\item bulk stress evolution in Hiscock-Lindblom theory:
\begin{equation}
\label{eq:pi-evol-hl-bjorken}
\partial_t (t\Pi)= -\frac{t\Pi}2\left(\frac2{\tau}-\frac1t+\partial_t\log\frac\chi{T^\req}\right)-\frac1\chi,
\end{equation}
\item bulk stress evolution in Maxwell-Cattaneo theory:
\begin{equation}
\label{eq:pi-evol-bjorken}
\partial_t (t\Pi)=-t\Pi\left(\frac1{\tau}-\frac1t\right) -\frac1\chi.
\end{equation}
\end{itemize}
For constant $\chi$ and $\tau$ the Maxwell-Cattaneo equation~\eqref{eq:pi-evol-bjorken} has the analytical solution \cite{DelZanna13}:
\begin{equation}
\label{eq:bjorken-flow}
\Pi(t)=\Pi(t_0)\mathrm e^{-\frac{t-t_0}{\tau}} + \frac{\zeta}{\tau}\mathrm e^{-t/\tau}
\big(\mathrm{Ei}(t_0/\tau) - \mathrm{Ei}(t/\tau)\big),
\end{equation}
where $t_0$ is the initial time and Ei the exponential integral function.

\begin{figure}
\includegraphics[width=\columnwidth]{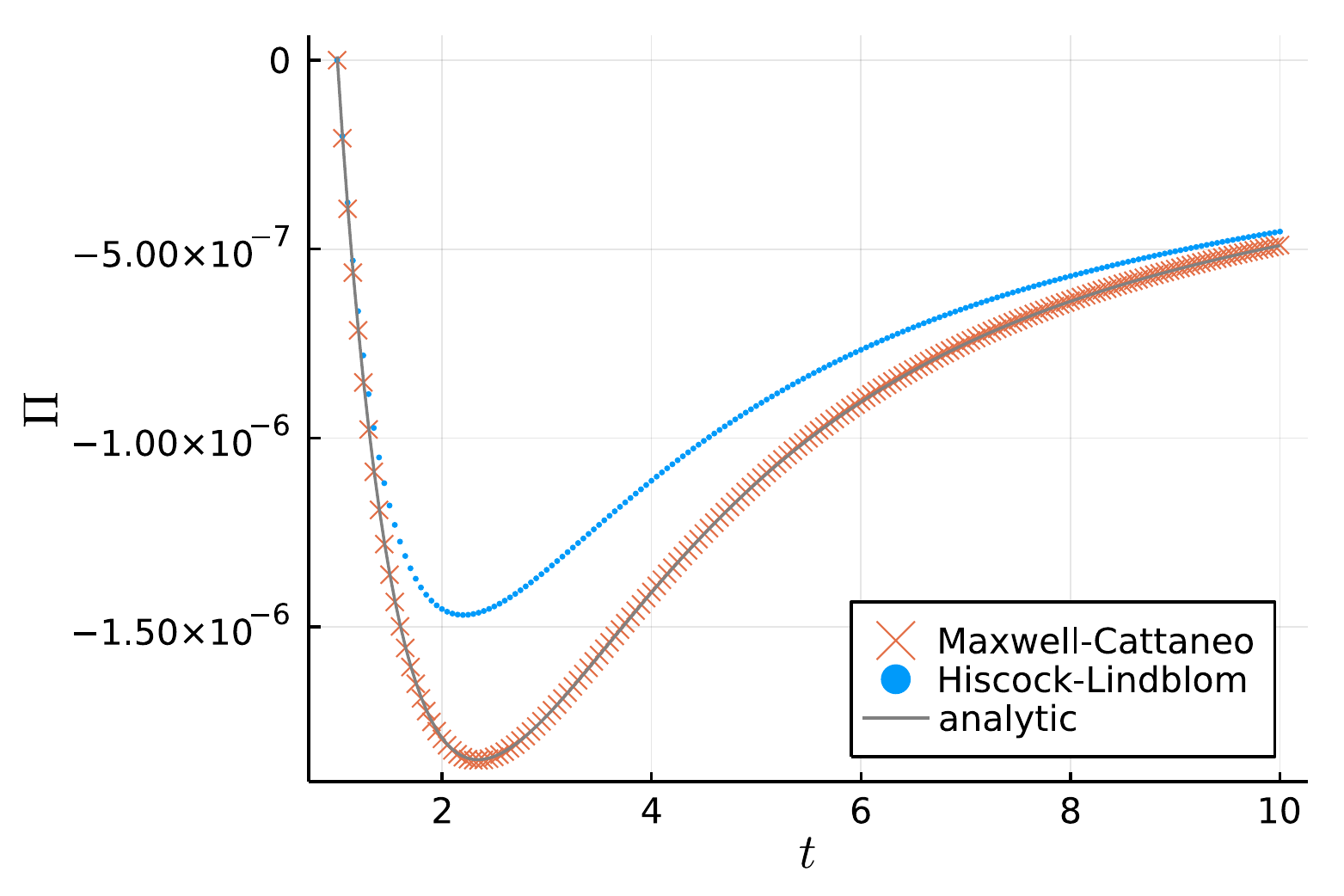}
\includegraphics[width=\columnwidth]{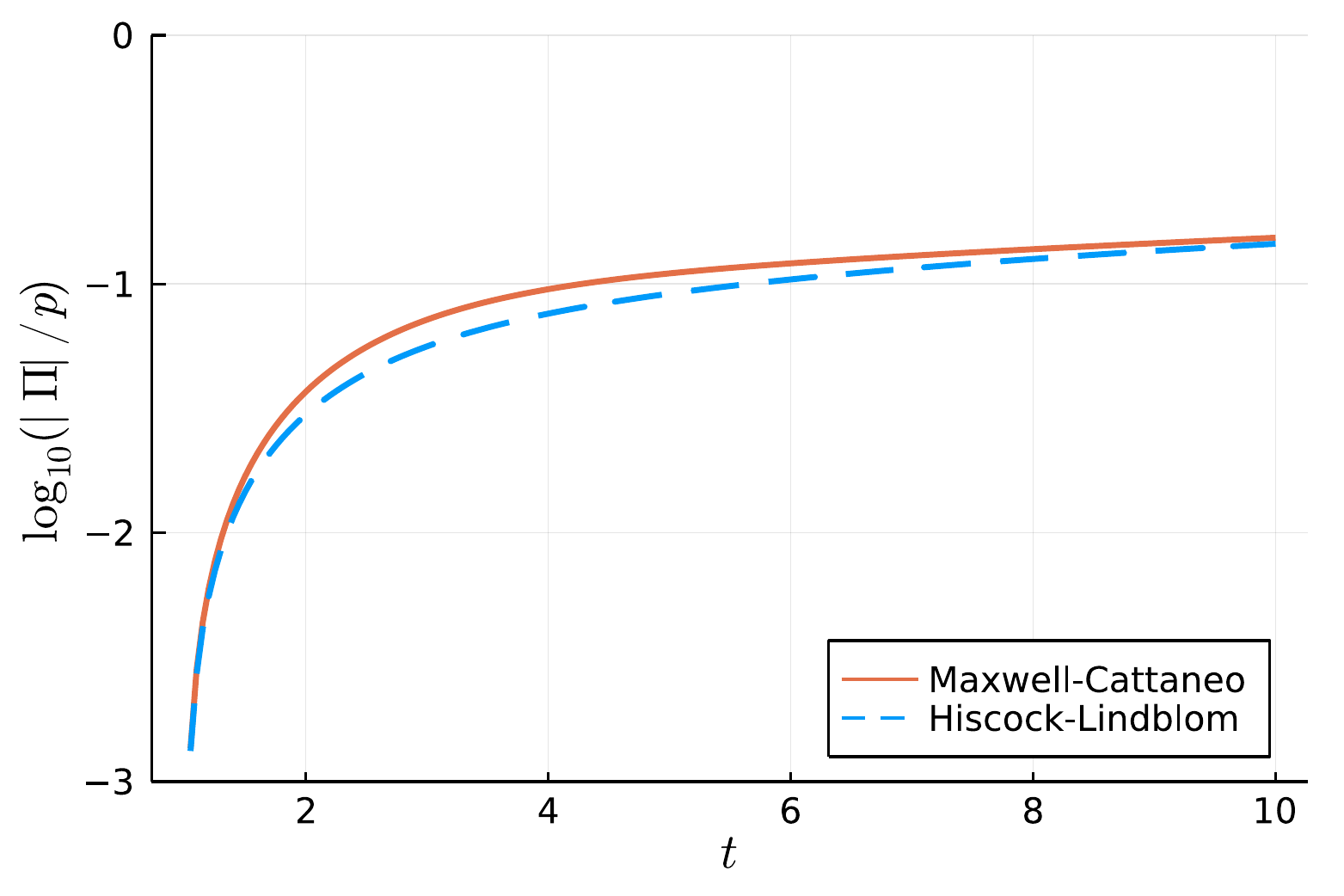}
\caption{Bjorken flow.
On the top, we plot the bulk stress $\Pi$, on the bottom, the logarithm of the
absolute value of the bulk stress over the pressure $|\Pi|/p$.
We compare the evolution of the Hiscock-Lindblom equation
(Eq.~\eqref{eq:pi-evol-hl-bjorken}, blue dots in the top figure and blue dashed line in the bottom figure) and the Maxwell-Cattaneo
equation (Eq.~\eqref{eq:pi-evol-bjorken}, orange crosses in the top figure and orange solid
line in the bottom figure), for which it does exist an analytic solution
(Eq.~\eqref{eq:bjorken-flow}, gray solid line in the top figure).}
\label{fig:bjorken}
\end{figure}

We use the following EOS \cite{Camelio19}:
\begin{equation}
\label{eq:eos-gam-gamth}
u(\rho,s)= k_0\rho^{\Gamma-1} + k_\rth s^2 \rho^{\Gamma_\rth - 1},
\end{equation}
with $\Gamma=2, \Gamma_\rth=1.75, k_0=100, k_\rth=1.5$.
The initial time is $t_0=1$, the initial rest mass density $\rho(t_0)=\rho_n$,
the initial entropy $s(t_0)=10$, the initial bulk stress $\Pi(t_0)=10^{-8}$, the
bulk parameter and bulk timescales are the constant values $\chi=100/\rho_\rn$ and
$\tau=1$, and the timestep is $\mathrm dt= \mathrm{CFL} \times 0.1$.
We evolve implicitly the terms containing the viscous timescale $\tau$, the bulk stress
$\Pi$, and the time derivative in the right-hand side of
Eqs.~\eqref{eq:pi-evol-hl-bjorken} and \eqref{eq:pi-evol-bjorken}.
In Fig.~\ref{fig:bjorken} we plot the results of the simulations.
The Maxwell-Cattaneo equation [Eq.~\eqref{eq:pi-evol-bjorken}]
reproduces very well the exact solution [Eq.~\eqref{eq:bjorken-flow}], while the
Hiscock-Lindblom equation [Eq.~\eqref{eq:pi-evol-hl-bjorken}]
has the same trend of the Maxwell-Cattaneo case but differs from it, because the
second order effects are important in the Bjorken flow.
The solution for the Hiscock-Lindblom equation is closer to local thermodynamic
equilibrium (i.e., the absolute value of the bulk stress is smaller) than the
Maxwell-Cattaneo equation. This is similar to what happens in viscous cosmological models (see Eq.~(5) of \citet{Maartens95}).  During the evolution, the
pressure remains greater of the magnitude of the bulk stress, but due to the
expansion of the Bjorken flow it decreases with time and $|\Pi|\to p$, at which
point the viscous approximation is no more valid.

\subsection{General-relativistic dust collapse}
\label{ssec:dust}

The dust collapse (also called Oppenheimer-Snyder collapse) is the
gravitational collapse in spherical coordinates of an initially homogeneous
pressure-less fluid.
The exact solution for the general relativistic
dust collapse in radial gauge, polar slicing coordinates was determined by \citet{Petrich86} (see also
the Appendix of \citet{Romero96}).

\begin{figure}
\includegraphics[width=\columnwidth]{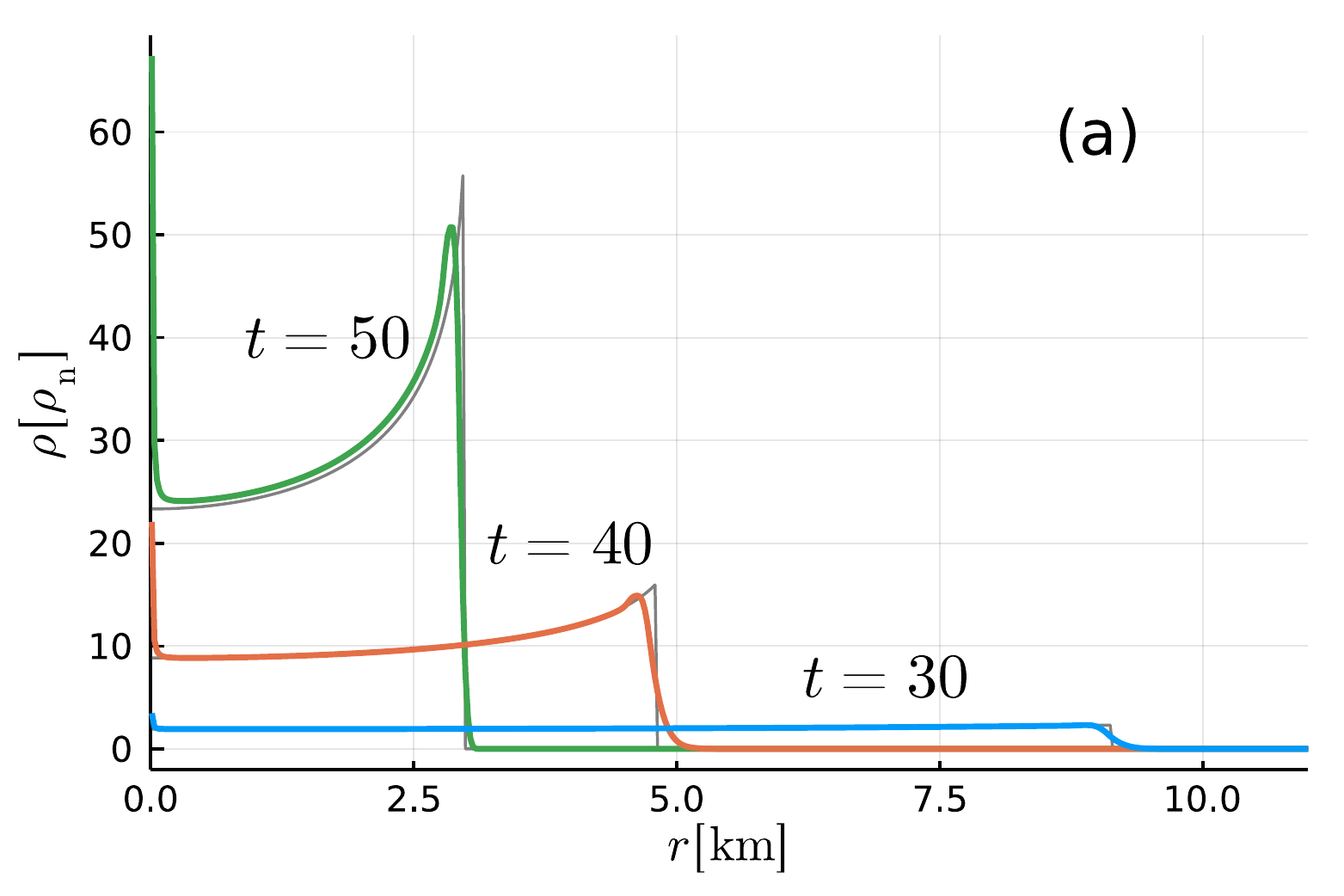}
\includegraphics[width=\columnwidth]{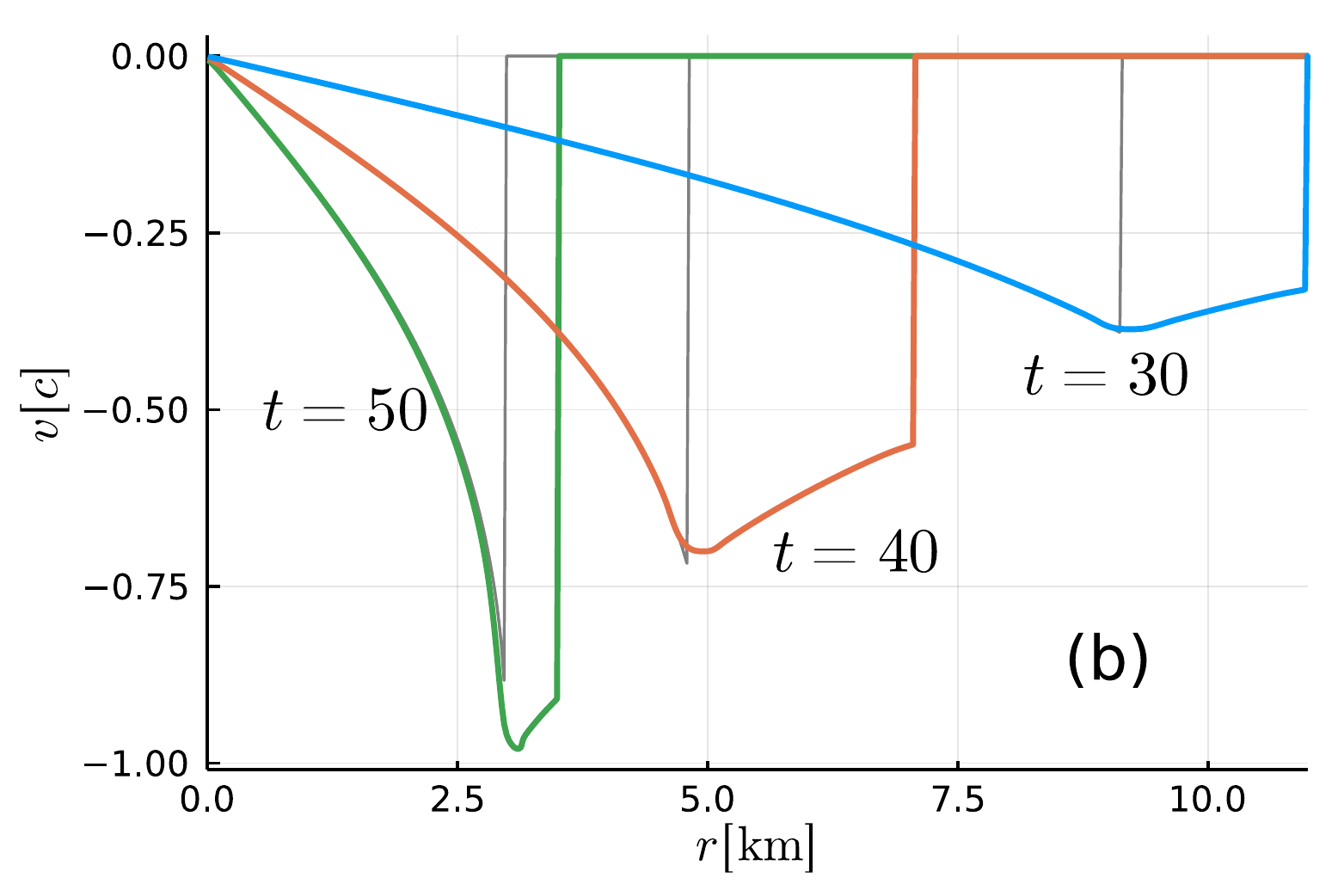}
\includegraphics[width=\columnwidth]{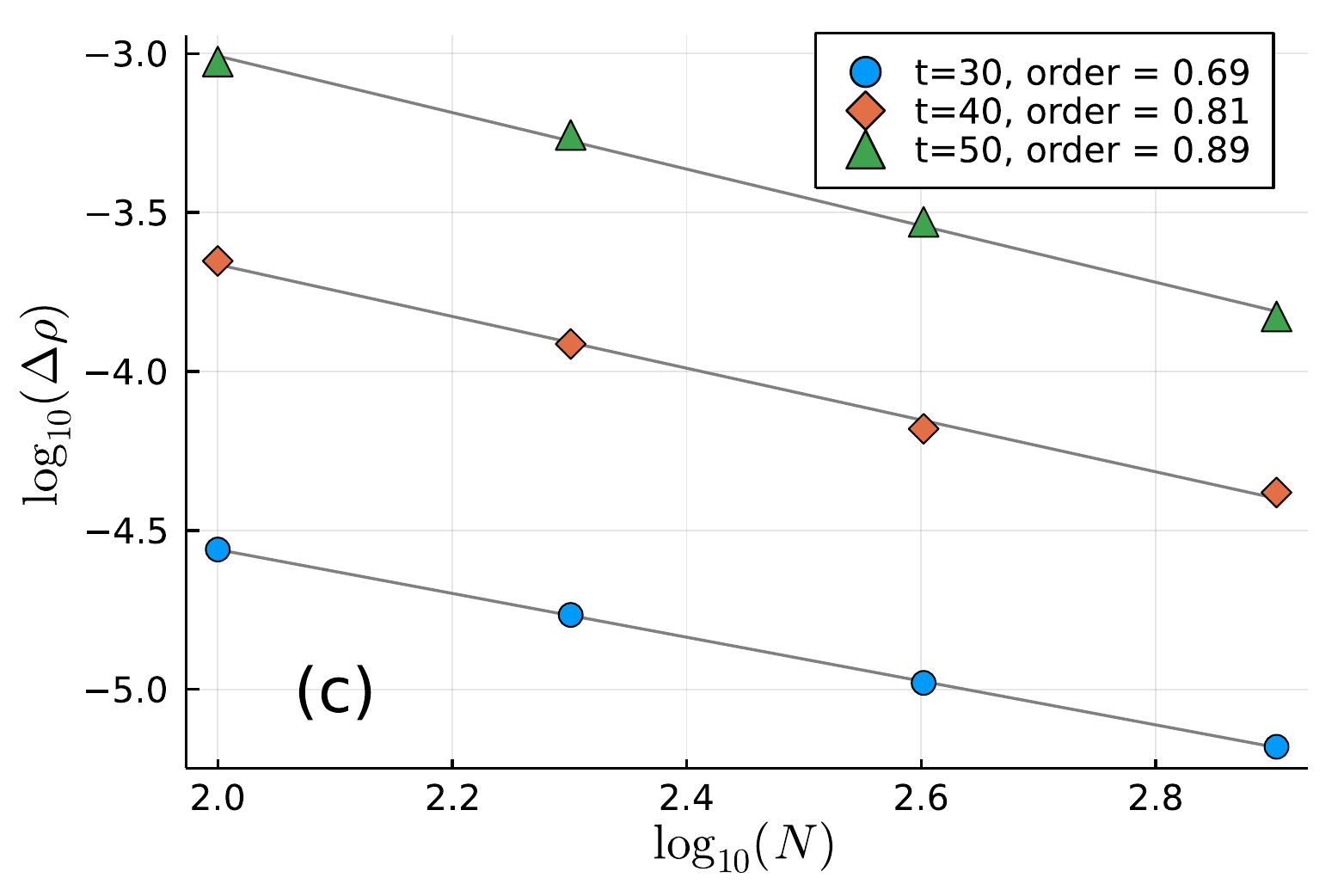}
\includegraphics[width=\columnwidth]{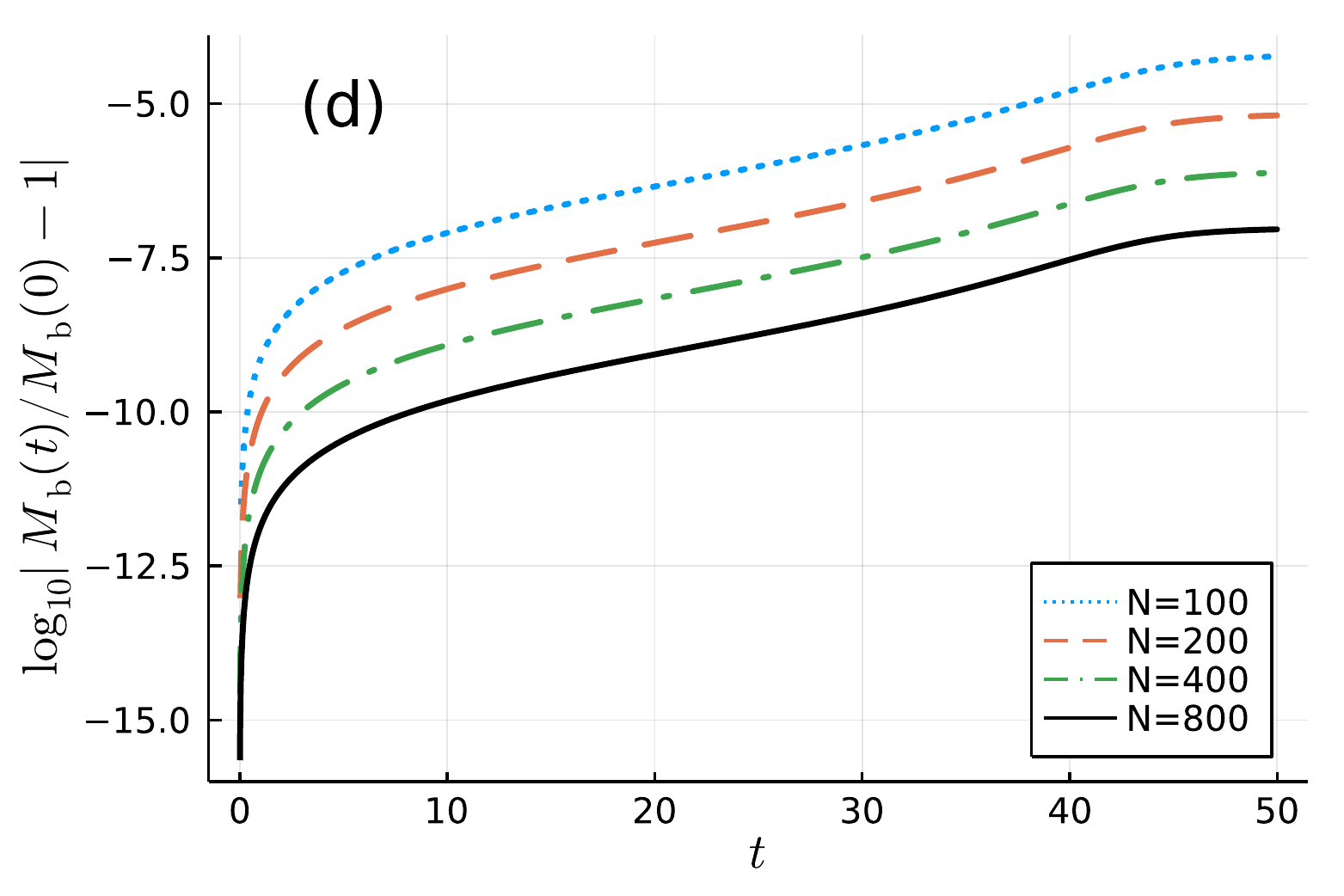}
\caption{Pressure-less dust collapse at times $t=30,40,50$. From top to the bottom: density (a) and velocity (b) profiles
at grid resolution $N=800$ against the exact solution (gray line);
residual of the rest mass density with respect to the exact solution (c), and
the total rest mass variation during the evolution (d).}
\label{fig:dust}
\end{figure}

We use the pressure-less EOS given by
\begin{equation}
\label{eq:eos-dust}
\epsilon(\rho)=\rho,
\end{equation}
and set $p=\mathcal Q=0$ in Eqs.~\eqref{eq:continuity-rho}--\eqref{eq:def-tau}.
In Fig.~\ref{fig:dust} we show the results at times $t=\{30,40,50\}$ for an
initially homogeneous configuration of total gravitational mass $M=1$ and initial radius
$R(t=0)=10$; we consider the different grid resolutions $N=\{100,200,400,800\}$,
the maximal grid radius is $R_\mathrm{max}=13$, and the timestep is $\mathrm dt=\mathrm{CFL}\times\mathrm dr$.
The simulation follows the exact solution for most of the evolution, but
(i) there is a buildup of material close to the origin, as noticed also
by \citet{OConnor10}, (ii) the surface of the collapsing cloud is smeared out (note that
the velocity tail corresponds to a vanishing density), as
noticed also by \citet{Romero96}, and (iii) at late times the simulation
starts diverging from the exact solution.
In any case, the code converges to the exact solution increasing the resolution
(see plots (c) and (d) of Fig.~\ref{fig:dust}),
even if the order of convergence is lower than 1.
These inaccuracies are due to the steep spatial and temporal gradients that
develop in the simulation.

\subsection{Neutron star oscillations}
\label{ssec:test-oscillations}

The hydrodynamic equation for this test are Eqs.~\eqref{eq:continuity-rho}--\eqref{eq:def-tau}
with $\mathcal Q=0$.
We use the ideal EOS [Eq.~\eqref{eq:eos-ideal}] with $\Gamma=2$.
The initial configuration is a cold spherical solution of the TOV
equations with central density $\rho_0=1.28\times10^{-3}$, initialized
with the cold EOS:
\begin{equation}
\label{eq:eos-cold}
p(\rho)= k_0\rho^\Gamma,
\end{equation}
with $k_0=100$, and we set an initial velocity perturbation $v(r)=v_\mathrm{pert}\times\sin(\pi r/R)$,
where $v_\mathrm{pert}=5\times 10^{-3}$ (unless otherwise specified) and $R$ is the stellar radius.
We consider 4 different resolutions for the grid: $N=\{100,200,400,800\}$, we set
$R_\mathrm{max}=11$, and we end the evolution at $t_\mathrm{end}=\unit[20]{ms}$.

\begin{figure}
\includegraphics[width=\columnwidth]{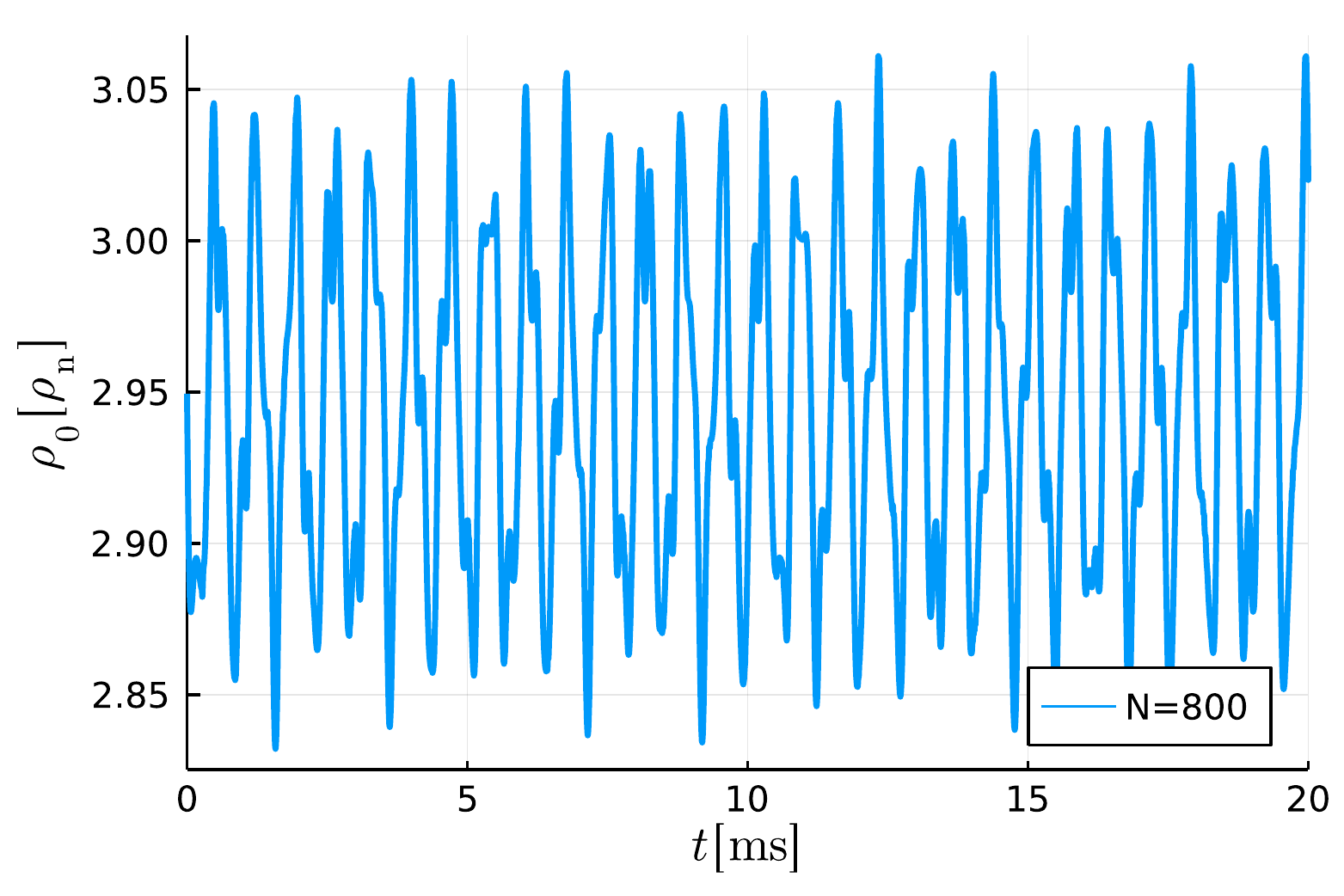}
\includegraphics[width=\columnwidth]{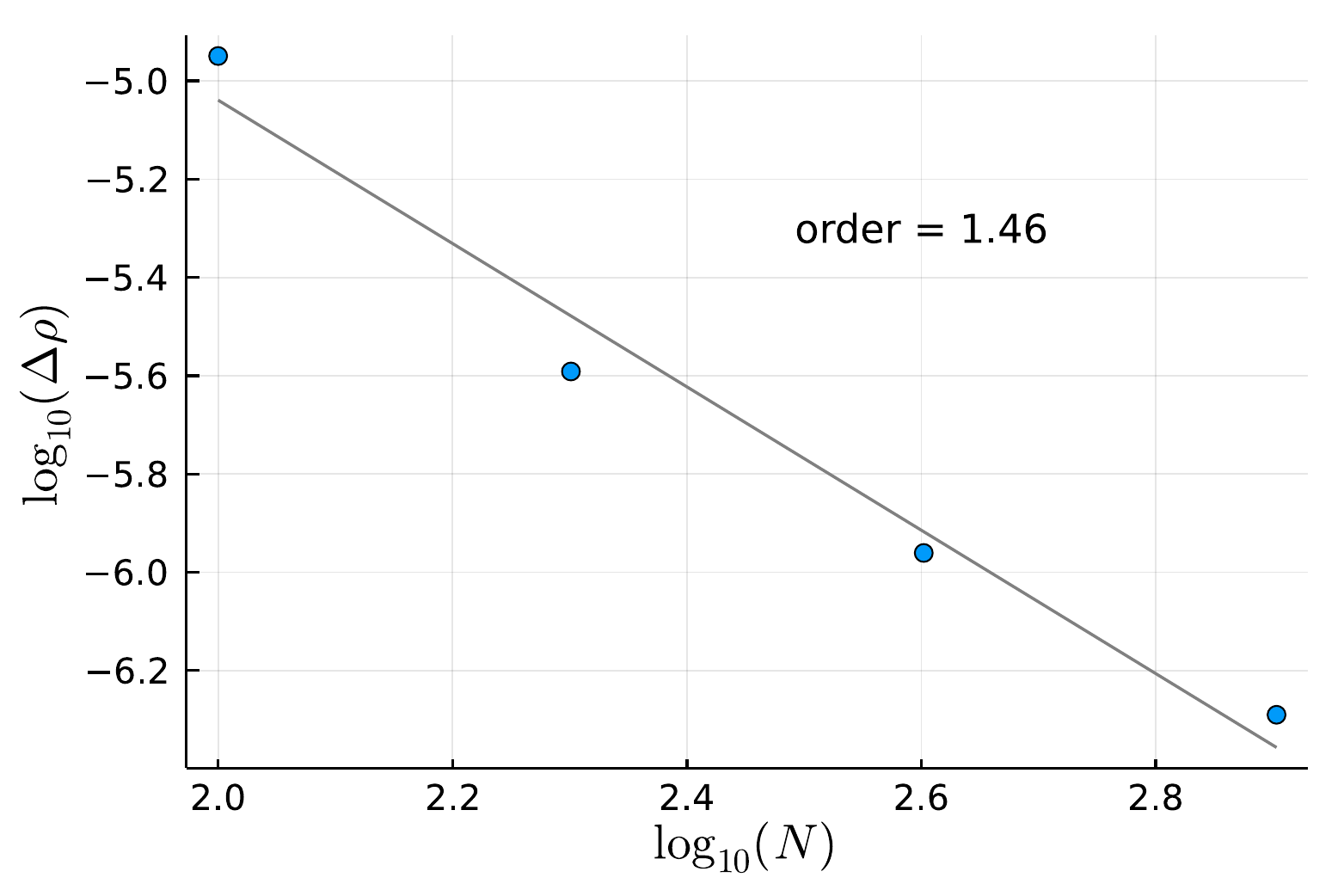}
\caption{Neutron star oscillations. On the top, the evolution of the central
rest mass density.
On the bottom, the residuals at the end of the evolution (20~ms)
for different resolutions and for $v_\mathrm{pert}=0$.}
\label{fig:osc}
\end{figure}

We show the results in Table~\ref{tab:osc} and Fig.~\ref{fig:osc}.  We obtain
the spectrum of the radial oscillations by Fourier-transforming the
cosine-tapered (with a 5\% taper ratio) signal obtained by interpolating (after
subtracting a linear fit) the central density on an evenly spaced time grid
with the same number of points of the original one.  As expected, (i) the
frequencies of the first 3 radial oscillation modes converge within the
uncertainty to the results from perturbation theory \citep{Baiotti09} with
increasing resolution, (ii) the gravitational mass is the same expected from the
spherical solution, (iii) the central rest mass density
oscillates around the initial value during
the simulation, and (iv) the maximal deviation from the initial total rest mass
decreases with increasing resolution.
In order to study how the residual changes with the resolution, we have set $v_\mathrm{pert}=0$
and we have computed the residual of the rest mass density at $t=\unit[20]{ms}$
with respect to the initial configuration. The convergence of the residual
is closer to the expected 2nd order than in the other test cases,
since the oscillation is a smooth problem.

\begin{table}
\centering
\caption{Frequencies of the first 3 radial modes, gravitational mass, and maximal variation of the absolute value of the baryonic mass
for the different resolutions, compared with the result from perturbation theory
(last row, see Tables~I and II of \citet{Baiotti09}). Since we are evolving the star for $t=\unit[20]{ms}$, the Fourier
spectrum has a resolution of $\delta f=\unit[50]{Hz}$.}
\label{tab:osc}
\begin{tabular}{cccccc}
N    & $f_0\unit{[Hz]}$ & $f_1\unit{[Hz]}$ & $f_2\unit{[Hz]}$ &
$M\unit{[M_\odot]}$ & $\mathrm{max} |\Delta M_b/M_b(t=0)|$ \\
\hline
100 & 1450 & 3850 & 5750 & 1.40 & $4.3\times10^{-8}$ \\
200 & 1450 & 3900 & 5850 & 1.40 & $4.8\times10^{-9}$ \\
400 & 1450 & 3950 & 5900 & 1.40 & $5.7\times10^{-10}$ \\
800 & 1450 & 3950 & 5900 & 1.40 & $6.9\times10^{-11}$ \\
\hline
-   & 1462 & 3938 & 5928 & 1.40 & -
\end{tabular}
\end{table}

\subsection{Migration from unstable to stable branch}
\label{ssec:test-migration}

For this test we use the same hydrodynamic equation and EOS of
Sec.~\ref{ssec:test-oscillations}, but the initial central rest mass density is
$\rho_0=7.993\times10^{-3}$, corresponding to a gravitational mass $M\simeq 1.45$
and an initial circumferential radius $R\simeq 5.8$ \citep{Font02, Bernuzzi10,
Thierfelder11}.

\begin{figure}
\includegraphics[width=\columnwidth]{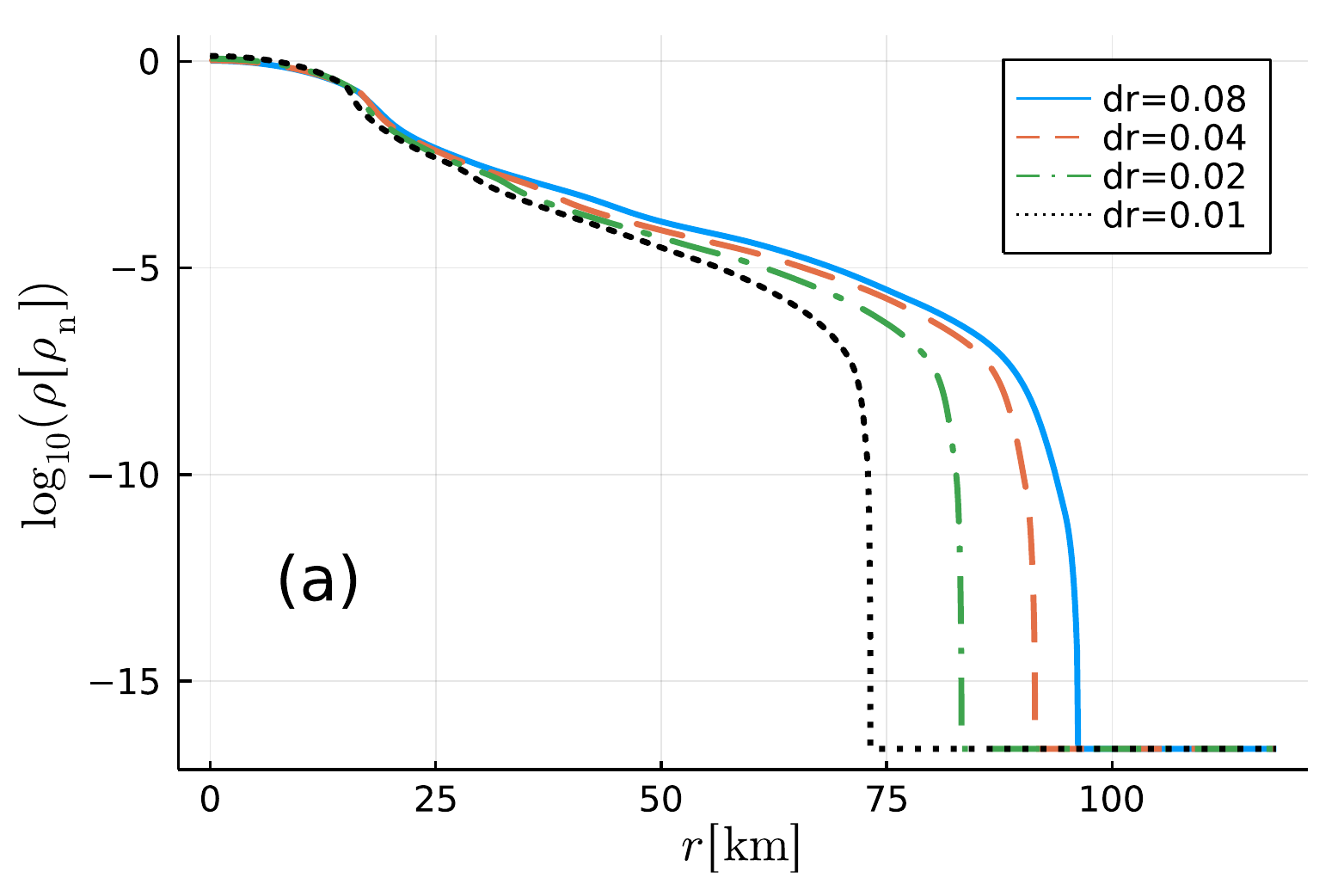}
\includegraphics[width=\columnwidth]{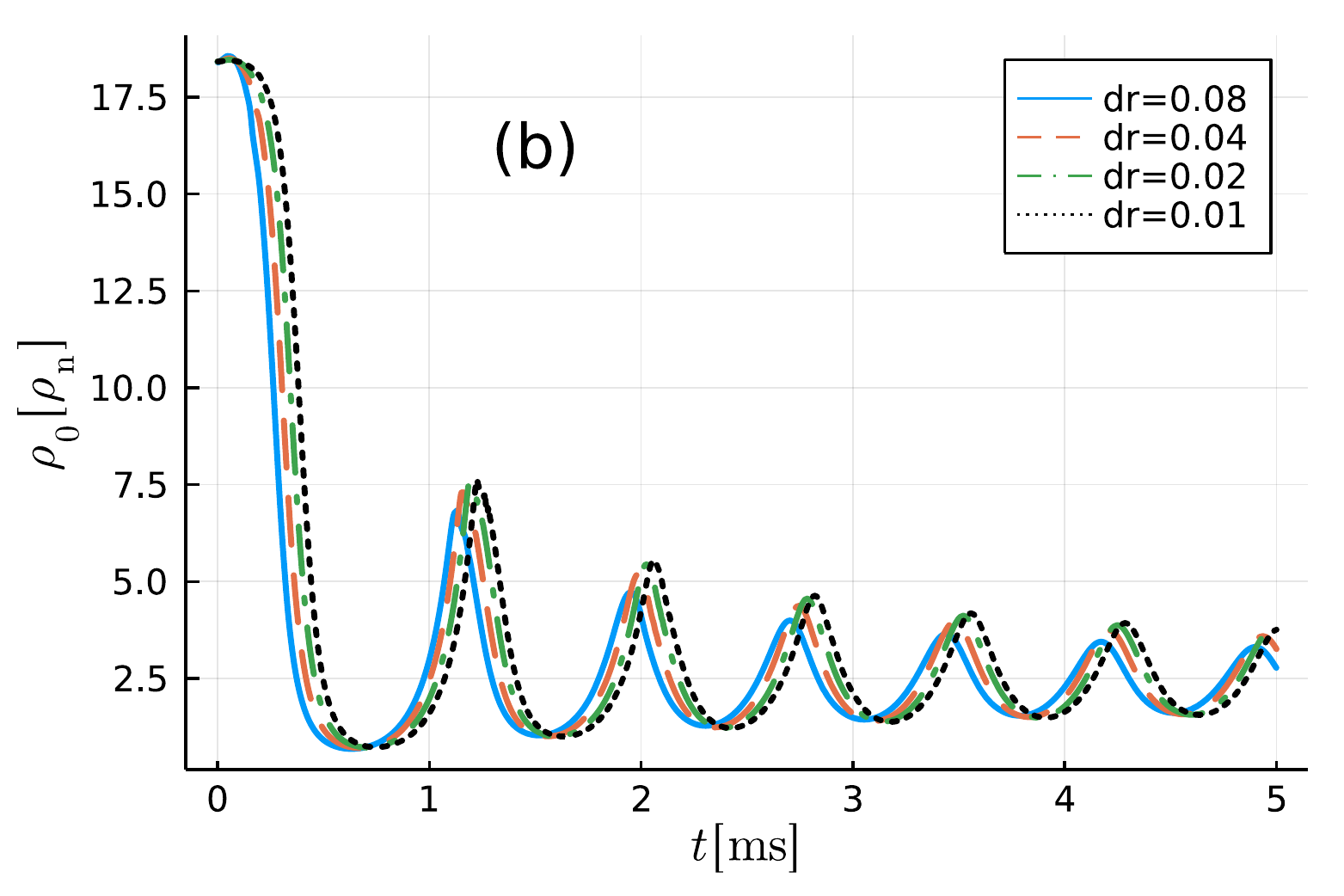}
\includegraphics[width=\columnwidth]{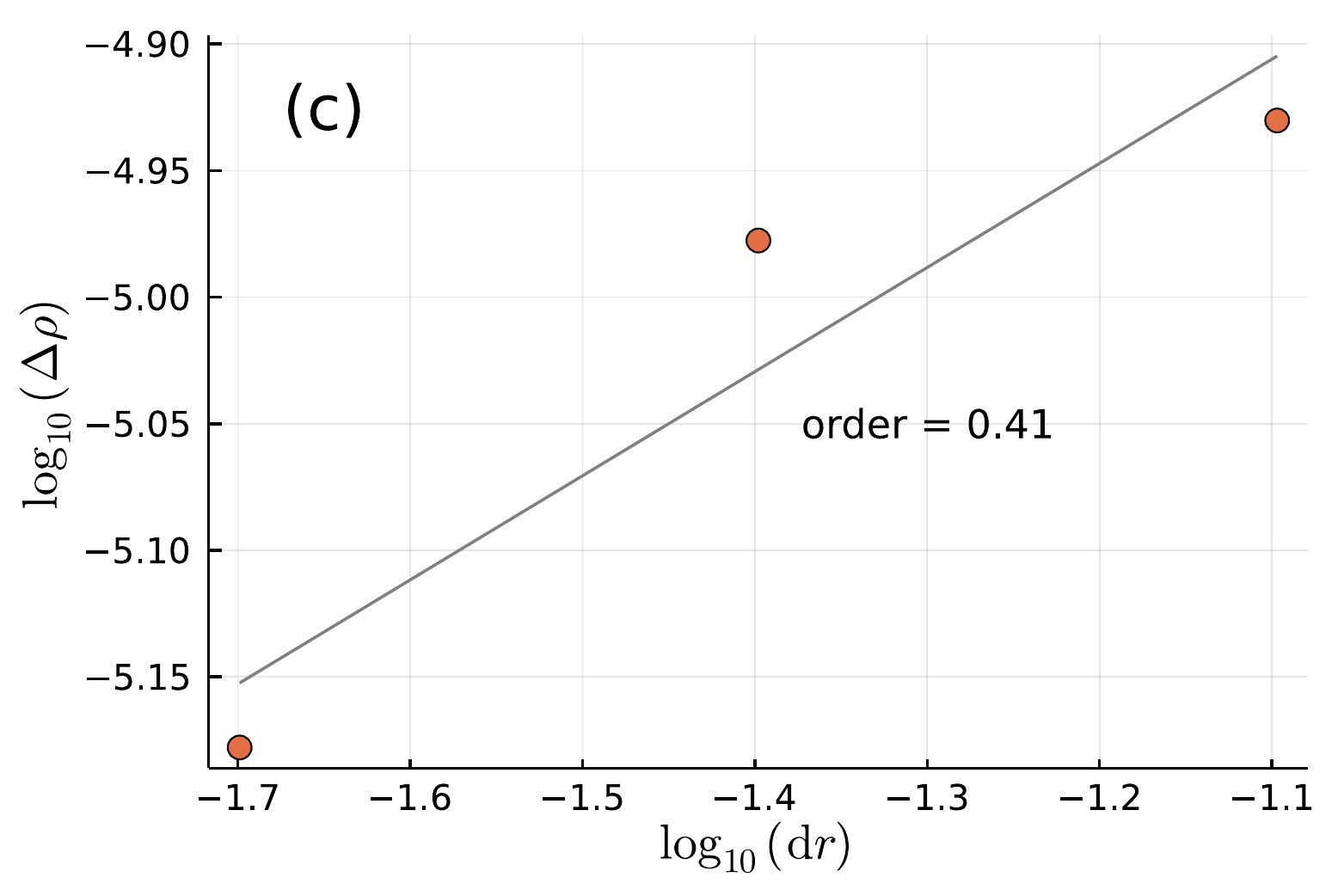}
\includegraphics[width=\columnwidth]{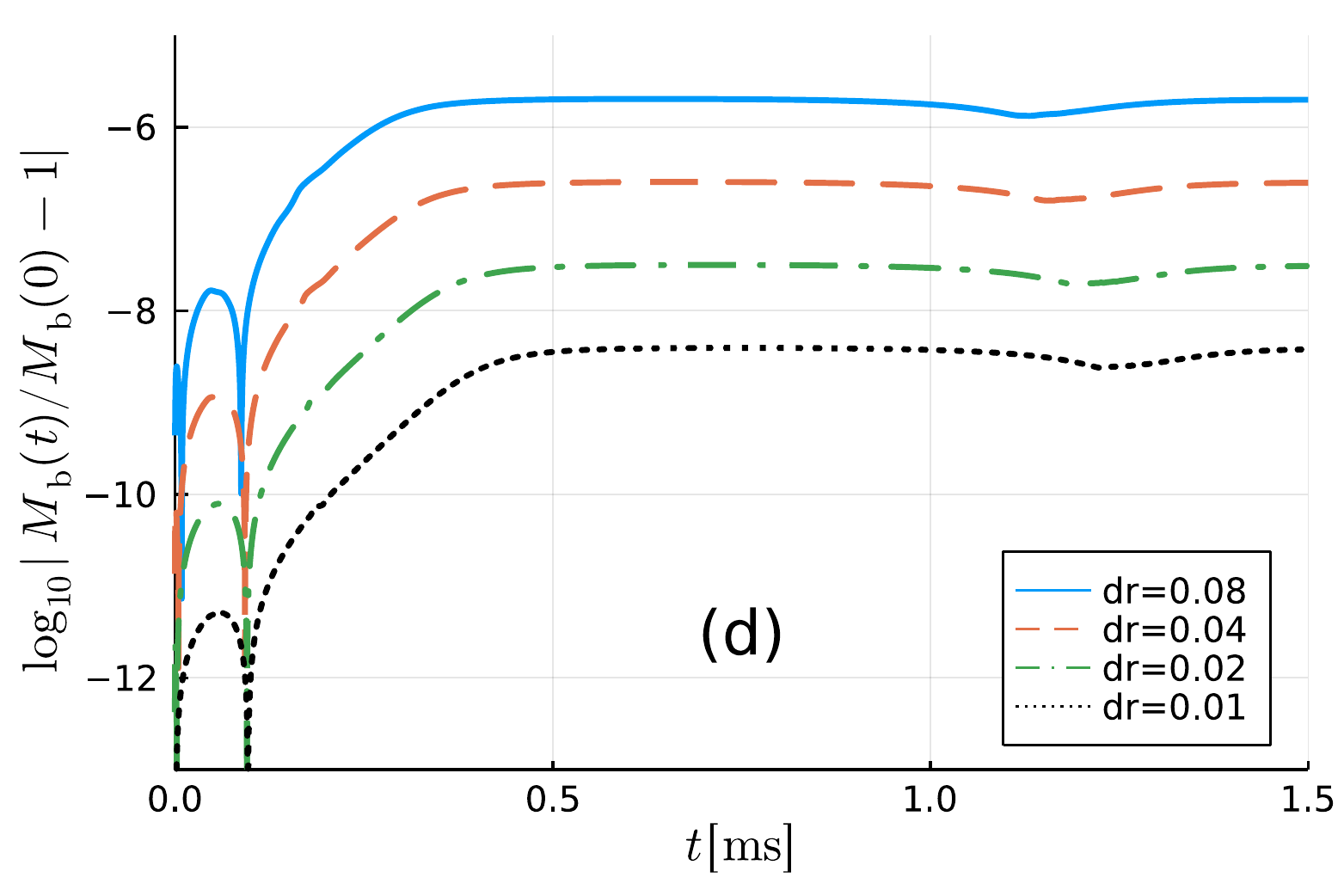}
\caption{Migration from unstable to stable branch. From top to bottom: (a)
profile of the rest mass density at 1.5~ms, (b) central rest mass density
evolution up to 5~ms, (c) residual at 1.5~ms with respect to the maximum resolution, and (d)
baryon mass conservation up to 1.5~ms.}
\label{fig:migration}
\end{figure}

We set $R_\mathrm{max}=80$ and consider different grid resolutions $\mathrm dr=\{0.08,0.04,0.02,0.01\}$,
which are equivalent to $N=\{1001,2001,4001,8001\}$. We
show the results in Fig.~\ref{fig:migration}.  The migration causes a series of
shock waves at the surface of the neutron star that eject material that
eventually reach the computational grid at a time greater than $\sim$1.6~ms.
For this reason, in Fig.~\ref{fig:migration} we show the
profile and the residuals at 1.5~ms, and the mass conservation is considered
only up to 1.5~ms, while the central density oscillations are shown up to 5~ms
since they are less affected by what happens close to the external grid border
(at 5~ms the density at the external grid point is of the order $\sim10^{-8}$).
The final result of the oscillation is the configuration considered in Appendix~\ref{ssec:test-oscillations},
but (i) part of the gravitational mass is converted into kinetic energy
and (ii) due to the shocks at the surface part of the material is ejected.
The code converges with the resolution with an order smaller than 1 because there
are strong shocks at the surface of the neutron star.

\section{Damping timescale}
\label{sec:tdamp}

In this appendix we derive the relativistic linearized damping time of an
oscillation without assuming a specific regime (i.e., without assuming that the
reaction timescale is much larger or smaller than the perturbation timescale)
and discuss its implications on our simulations with direct and modified Urca
reactions.
Our derivation follows Sec.~3.3 of \citet{Alford20}, but it has been refined to
properly take into account relativistic corrections.
The resulting formula can be seen as a generalization of Eq.~(4.9) of
\citet{Kovtun19} beyond the Navier-Stokes regime.

In the linear regime, the second law of thermodynamics can be expressed in
terms of the information current $E^\mu$ as follows: $\nabla_\mu
E^\mu=-\sigma$, where $\sigma$ is the entropy production rate. Working at a
scale where the background state can be approximated as homogeneous (in the equilibrium
local rest frame \cite{MTW_book}), we can integrate the second law in space,
obtaining
\begin{equation}\label{eq:gavas1}
\dfrac{d}{dt} \int E^0 d^3x = -\int \sigma d^3 x .
\end{equation}
Treating viscous effects as small corrections, the information density $E^0$
(averaged over the oscillations in space) can be approximated to be the perfect
fluid one, which for a sound wave traveling, say, in the $x^1$ direction can be
expressed as \cite{Gavassino21stable}:
\begin{equation}\label{eq:gavas2}
\left<E^0\right> \approx \dfrac{\epsilon+p^\req}{2T^\req} (\delta \tilde{u}^1)^2 ,
\end{equation}
where $\delta \tilde{u}^1(t)$ is the amplitude of the velocity perturbation,
i.e.~$\delta u^1{=}\delta \tilde{u}^1(t) e^{i(kx^1-\omega t)}$ ($\delta u^1$
coincides with the velocity in the equilibrium rest frame), where $k,\omega
\in \mathbb{R}$ are the perturbation wavelength and angular velocity, respectively.
The entropy production rate is (Eq.~(8) of \citet{Camelio22a}) $\sigma=\delta \Pi^2/(\zeta T^\req)
$, where $\delta \Pi$ is the perturbation to the bulk-viscous stress. From the
Maxwell-Cattaneo equation (Eq.~(14) of \citet{Camelio22a}), we find that $\delta \Pi=-(1-i\omega \tau)^{-1}\zeta
ik\delta u^1$, where $\tau$ is the reaction timescale. Hence, averaging over space, we obtain (assuming $\omega=c_s k$)
\begin{equation}\label{eq:gavas3}
\left<\sigma\right> \approx \dfrac{\zeta \omega^2 (\delta \tilde{u}^1)^2}{2T^\req c_s^2(1+\omega^2 \tau^2)} .
\end{equation}
Inserting \eqref{eq:gavas2} and \eqref{eq:gavas3} into \eqref{eq:gavas1}, we
obtain an ordinary differential equation for $\delta \tilde{u}^1(t)$, whose solution
is in the form $\delta \tilde{u}^1(t)=\delta \tilde{u}^1(0)e^{-t/t_{\text{damp}}}$, where
the damping timescale is
\begin{equation}
\label{eq:tdamp}
t_{\text{damp}}= \dfrac{2c_s^2(\epsilon+p^\req)(1+\omega^2 \tau^2)}{\zeta \omega^2} .
\end{equation}
Note that:
\begin{itemize}
\item the quantity $c_s$ is the \textit{phase} velocity of the wave that is
being damped. Hence, it should be either $c_\mathrm{s,ir}$ or $c_\mathrm{s,uv}$ (or something
in between) depending on whether the fluid is in the Navier-Stokes (i.e., quasi-stationary or `parabolic' regime, see
Sec.~II of \citet{Gavassino21bulk} or Sec.~II of \citet{Camelio22a}) or in
the frozen regime.
\item in the limit of $\omega\ll\tau$, namely in the Navier-Stokes regime,
Eq.~\eqref{eq:tdamp} tends, as expected, to Eq.~(4.9) of \citet{Kovtun19}.
\item although in the derivation we made use of the Maxwell-Cattaneo equation,
the result still holds unchanged within the Hiscock-Lindblom theory, because
the latter is equivalent to the Maxwell-Cattaneo theory in the linear regime.
\end{itemize}

\begin{figure}
\includegraphics[width=\columnwidth]{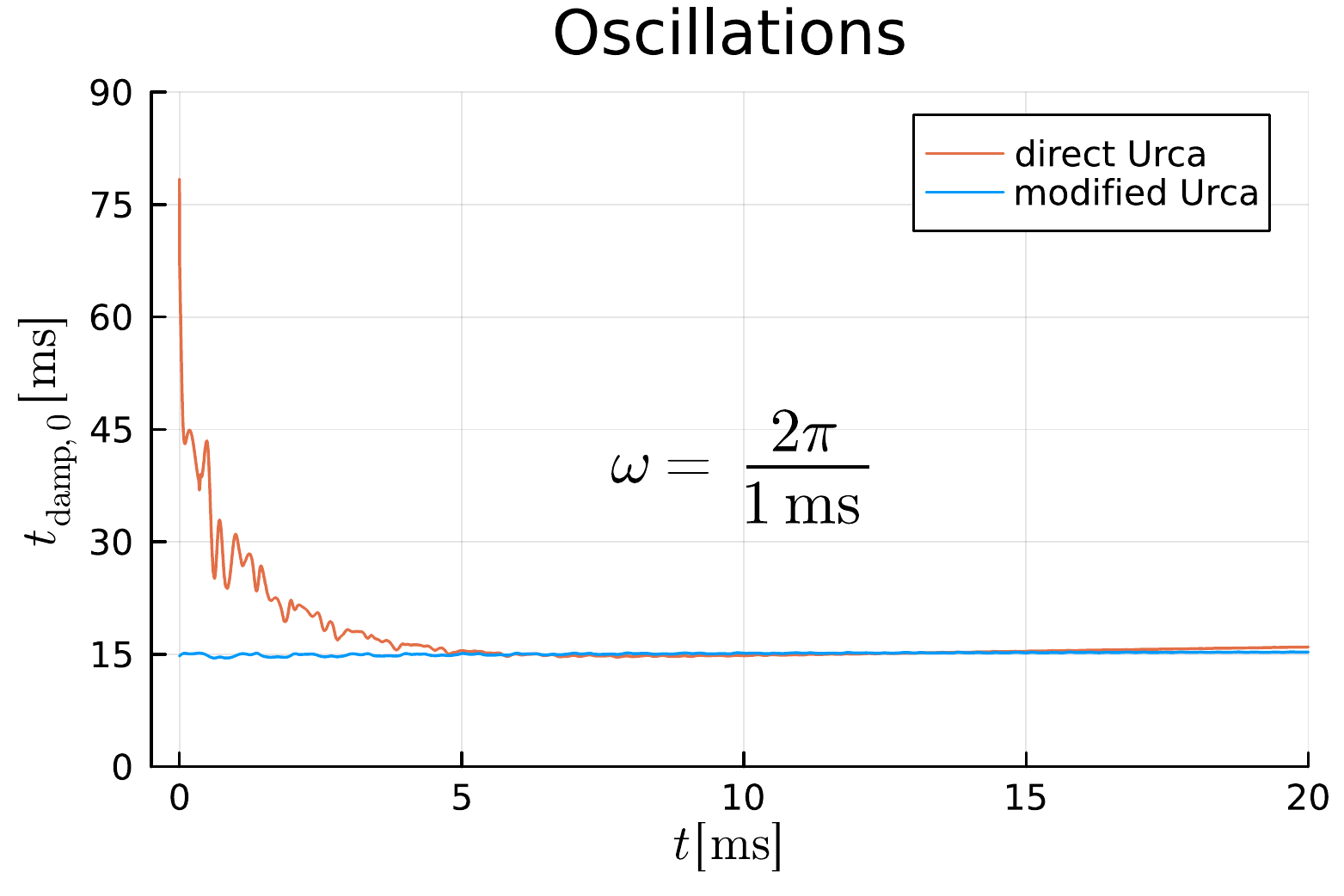}
\caption{Central linearized damping time for the oscillating MF model with
direct and modified reactions.  We use
Eq.~\eqref{eq:tdamp} assuming a constant hydrodynamic timescale of
$t_\mathrm{pert}=\unit[1]{ms}$ and $c_\mathrm{s}=c_\mathrm{s,uv}$
(the difference between the `ultraviolet' and the `infrared' speed of sound is
of the order of 2\% and does not affect our conclusions).}
\label{fig:osc-tdamp}
\end{figure}

In Fig.~\ref{fig:osc-tdamp} we compare the evolution of the linearized damping
time in the center of the star for the oscillation case, for direct and
modified Urca reactions, for the MF model.
From Plots~(a) of Figs.~\ref{fig:rho0} and \ref{fig:rho-murca}, we estimate an
oscillation period of $t_\mathrm{pert} = 2\pi/\omega\simeq \unit[1]{ms}$ in
both cases.
As can be seen in Fig.~\ref{fig:osc-tdamp}, the initial central damping time for
the direct Urca simulation is much larger than that of the modified Urca one.
This is due to the fact that we set (indirectly, by setting the initial uniform
entropy in the star) an initial central reaction timescale of $\tau \simeq
\unit[1.68]{ms} \approx t_\mathrm{pert}$ for the direct Urca reactions (i.e.,
$\omega^2\tau^2 \simeq 111$), while we set $\tau \simeq\unit[0.16]{ms} \approx
1/\omega$ for the modified Urca ones.
However, while the modified Urca damping time remains of the order of
$\unit[15]{ms}$ (which is very similar to the duration of the simulation,
i.e.~20\,ms) during the whole simulation, the direct Urca one starts at
approximately $\unit[80]{ms}$ and then reaches the same damping timescale of
the modified Urca case.
This large decrease in the damping time for the direct Urca case is due to the
fact that if $\omega\tau\gg1$, then
$t_\mathrm{damp}\propto\tau^2/\zeta\propto\Xi^{-1}\propto T^{-4}$, where the
last proportionality holds for the direct Urca reactions [Eq.~\eqref{eq:R}].
Moreover, the entropy production rate is generally inversely proportional to
the temperature (Eq.~(8) of \citet{Camelio22a}).
In our case, the initial temperature for the direct Urca reactions is small
enough for causing a huge increase of the entropy and as a consequence also of
the temperature, see Plot~(a) of Fig.~\ref{fig:temp}, which in turn causes a
huge decrease of the initial damping time.
On the other hand, for the modified Urca case, the initial temperature is much
larger and therefore, even accounting for the different order in the dependence
on the temperature, it is not enough to cause a significant increase of the
entropy and as a consequence also of the temperature, see top plot in
Fig.~\ref{fig:temp-murca}.
This is the reason why, in order to see the damping of the oscillation during
the simulation with modified Urca reactions, we had to choose an initial
entropy such that $\tau\omega\approx 1$ instead of $\tau\approx t_\mathrm{pert}$
as with the direct Urca reactions.

\end{document}